\documentclass[onecolumn]{svjour3}       
\usepackage{amsmath,bm,amssymb}
\usepackage{graphicx}

\usepackage{amsbsy}
\usepackage{setspace}
\usepackage{ulem}
\let\isout\sout \renewcommand{\sout}[1]{\ifmmode\text{\isout{\ensuremath{#1}}}\else\isout{#1}\fi}
\usepackage{url}
\usepackage{xcolor}
\usepackage{epstopdf, epsfig}

\newcommand{\DETAILS}[1]{}
\newcommand{\I}{\mathrm{i}}
\newcommand\D{\mbox{d}}

\newcommand{\R}{\mathbb R}
\newcommand{\N}{\mathbb N}
\newcommand{\C}{\mathbb C}
\newcommand{\e}{\eqref}

\newcommand{\rev}[1]{#1}  
\newcommand{\revc}[1]{{\color{red}\sout{}}}  
\newcommand{\revmike}[1]{{\color{cyan}}} 

\newcommand\bx{\mathbf{x}}
\newcommand\by{\mathbf{y}}
\newcommand\bu{\mathbf{u}}
\newcommand\bom{\boldsymbol{\omega}}

\begin{document}

\title{Collapse vs. blow up and global existence in the generalized Constantin-Lax-Majda equation}
\titlerunning{Collapse vs. blow up}        

\author{Pavel M. Lushnikov \and Denis A. Silantyev \and Michael Siegel
}

\institute{Pavel M. Lushnikov
\at Department of Mathematics and Statistics, University of New Mexico, Albuquerque, MSC01 1115, NM, 87131, USA \\
\email{plushnik@math.unm.edu}
\and
Denis A. Silantyev
\at
Courant Institute of Mathematical Sciences, New York University, 251 Mercer Street New
York, NY 10012-1110, USA
\and
Michael Siegel
\at
Department of Mathematical Sciences and Center for Applied Mathematics and Statistics, New Jersey Institute of Technology, Newark, NJ 07102, USA
}
\authorrunning{P.M. Lushnikov, D.A. Silantyev and  M. Siegel} 


\date{
\today}

\maketitle

\begin{abstract}
The question of finite time singularity formation vs. global existence for solutions to the generalized Constantin-Lax-Majda equation is studied, with particular emphasis on the influence of a parameter
$a$ which controls the strength of advection. For solutions on the
infinite domain we find a new critical value
$a_c=0.6890665337007457\ldots$ below which there is  finite time singularity formation 
that has a form of self-similar collapse, with the spatial extent of blow-up
shrinking to  zero.
\rev{We prove the existence of a leading-order power-law complex singularity for general values of $a$ in the analytical continuation of the solution from the real spatial coordinate into the complex plane, and identify the power-law exponent. This singularity controls the leading order behaviour of the collapsing solution.   We prove that this singularity can persist over time,  without other singularity types present, provided $a=0$ or $1/2$.  This enables the construction of exact analytical solutions for these values of $a$. For other values of $a$, this leading-order singularity must coexist with other singularity types over any nonzero interval of time. }
For $a_c<a\leq1$, 
we find a blow-up solution
in which the spatial extent of the blow-up region expands infinitely fast at the singularity time.
For  $a \gtrsim 1.3$
, we find that the solution exists globally with exponential-like
growth of the solution amplitude in time. We also consider the case of periodic
boundary conditions. We identify collapsing solutions for $a<a_c$ which
are similar to the real line case. For $a_c<a\le0.95$, we find new blow-up solutions which are   neither expanding nor collapsing. For $ a\ge 1,$ we identify a
global existence of solutions.

\end{abstract}

\section{Introduction} \label{sec:intro}

In this paper we investigate finite-time singularity formation in the generalized Constantin-Lax-Majda (CLM) equation \cite{ConstantinLaxMajda,DeGregorio,Okamoto2008}
\begin{equation}\label{CLM01}
\begin{split}
&\omega_t=-au\omega_x +\omega u_x,\ \quad \omega, x\in \R, t>0,
\\
&u_x= {\mathcal H}\omega,
\end{split}
\end{equation}
which is a  1D model for the advection and stretching of vorticity in
a 3D incompressible Euler fluid.  Here  $\omega$ and $u$  are a scalar vorticity and velocity, respectively, $a \in \mathbb{R}$ is a parameter, and ${\mathcal H}$ is the Hilbert transform,
\begin{equation} \label{HilbertHdef}
\quad {\mathcal H} \omega(x):= \frac{1}{\pi} \text{p.v.} \int^{+\infty}_{-\infty}\frac{\omega(x')}{x-x'}\D x'.
\end{equation}
This equation, with $a=0$, was first introduced by Constantin, Lax and Majda \cite{ConstantinLaxMajda}
as a simplified  model to study the possible formation of finite-time singularities in the  3D incompressible Euler equations. It was later generalized by DeGregorio \cite{DeGregorio} to include an advection term $u \omega_x$,
and by Okamoto, Sakajo and Wensch \cite{Okamoto2008}, who introduced the real parameter $a$
to give different relative weights to advection and  vortex stretching, $u_x \omega$. In addition to its relationship to the 3D Euler equation,  (\ref{CLM01}) has a direct connection to the surface quasi-geostrophic (SQG) equation \cite{Elgindi2020}.

The 3D incompressible Euler equations can be written as
\begin{align}
\partial_t \bom &+ \bu \cdot \nabla \bom = \bom \cdot \nabla \bu,~~~ \bx \in \mathbb{R}^3,~t>0, \label{eq:vorticity}\\
\bu&= \nabla \times  (- \Delta)^{-1} \bom. \label{eq:bs}
\end{align}
The second equation above is the Biot-Savart law, which in free-space has an equivalent representation as a convolution integral
\begin{equation} \label{eq:bsint}
\bu(\bx,t)= \frac{1}{4 \pi} \int_{\mathbb{R}^3} \frac{(\bx-\by) \times \bom(\by,t)}{|\bx-\by|^3} \ d\by.
\end{equation}
The term $\bom \cdot \nabla \bu$ on the right-hand side (r.h.s.) of
(\ref{eq:vorticity}), where $\nabla \bu=S(\bom)$ is a matrix of singular
integrals,   is known as the vortex stretching term.  Standard estimates
from the theory of singular integral operators \cite{Stein} show that $\|
\bom \|_{L^p} \leq \| \nabla \bu \|_{L^p}  \leq c_p \| \bom \|_{L^p}$ for
$1<p<\infty$, which formally implies that the vortex stretching term
scales quadratically in the vorticity, i.e., $S(\bom) \bom \approx
\bom^2$. This term is therefore destabilizing and has the potential to
generate singular behavior.  However, analysis of the regularity of Eqs.
(\ref{eq:vorticity}),  (\ref{eq:bs}) is greatly complicated by the
nonlocal and matrix structure of $S$, and remains an outstanding open
question
\rev{(see \cite{Elgindi2019arxiv}, \cite{Elgindi2019finite} for recent developments)}.

In contrast to the vortex stretching term, the advection term $\bu \cdot \nabla \bom$ does not cause any growth of vorticity.  As a result, it has historically been thought to play an unimportant role in the regularity of the incompressible Euler and Navier-Stokes equations. Recent studies, however, show that advection-type terms can have an unexpected smoothing effect.
 For example, Hou and Lei \cite{HouLei} present numerical evidence that a finite-time singularity forms from smooth data in solutions to a reformulated version of the  Navier-Stokes equations for axisymmetric flow with swirl, when the so-called convection terms $u_r \partial_r (\omega_\theta/r)+u_z \partial_z (\omega_\theta/r)$ and $u_r \partial_r (u_\theta/r)+u_z \partial_z (u_\theta/r)$ are omitted. Here $(u_r, u_\theta, u_z)$ and $\omega_\theta$ are velocity and vorticity components  in  cylindrical coordinates $(r, \theta, z)$. Adding the convection back  is found to suppress  a finite-time  singularity formation.
Related work  on the  smoothing effect of advection/convection in the Euler and Navier-Stokes equations is given in \cite{hou2018potential,hou2014finite,HouLi2008,HouLi2006,hou2012singularity,OkamotoOkhitani}.

The generalized CLM equation (\ref{CLM01}) (also called the Okamoto-Sakajo-Wunsch model in Ref. \cite{Elgindi2020}) is obtained from the 3D Euler equations by replacing the advection term $\bu \cdot \nabla \bom$ with $u \omega_x$ and the vortex stretching term $S(\bom) \bom$   by its 1D analogue ${\mathcal H}( \omega) \omega$. The Hilbert transform ${\mathcal H}$ is the  unique singular
integral operator in 1D that preserves certain important properties of $S(\bom)$, \rev{namely, it commutes with translations
and dilations} \cite{ConstantinLaxMajda}.
In addition,  the 1D vortex stretching term ${\mathcal H}( \omega) \omega$ preserves the quadratic scaling of the vortex stretching term $S(\bom)\bom$ in the 3D problem. The resulting equation (\ref{CLM01}) provides a simplified setting to understand the competition  between the stabilizing effect of  advection and destabilizing effect of vortex stretching.
In this work we focus on  smooth (analytic or $C^\infty$) initial data which we consider as the most physically relevant. There are also a number of results  on singularity formation for (\ref{CLM01})  in the case of Holder continuous initial data, see Refs.
\cite{ChenHouHuang,Elgindi2020} for recent reviews.

We summarize some of the known results, concentrating on those which apply to smooth (analytic or $C^\infty$) initial data.  In the case $a=0$, Constantin, Lax and Majda  \cite{ConstantinLaxMajda} obtained a closed-form exact solution to  the initial value problem
for (\ref{CLM01}) which
develops a self-similar finite-time singularity for a  class of analytic initial data.   When $a \ne 0$,  the simplifications that enable a closed-form solution no longer hold, and various analytical and numerical methods have been applied to investigate singularity formation.  Castro and Cordoba \cite{CastroCordoba} proved finite-time blow-up for  $a<0$  using a Lyapunov-type argument. In this case,  advection and vortex stretching act together to produce a singularity. In contrast, for $a>0$ the stabilizing effect of advection competes with the destabilizing effect of vortex stretching.   For  $\epsilon-$small values of $a>0$,  vortex stretching dominates and Elgindi and Jeong \cite{Elgindi2020}  proved the existence of  self-similar finite-time singularities in the form%
\begin{equation}\label{self-similar1}
\omega=\frac{1}{\tau}f\left ( \xi \right ), \ \xi=\frac{x}{\tau^\alpha}, \ \tau=t_c-t,
\end{equation}
where $t_c>0$\ is the singularity time and $\alpha$ depends on $a$, approaching $\alpha=1$ in the limit $a\to 0.$
  Also, $f(\xi)$ is an odd function, i.e. $f(-\xi)=-f(\xi),\ \xi\in \mathbb{R}$. The proof of \cite{Elgindi2020} is based on a continuation argument in a small neighborhood of the exact solution at $a=0$.
Chen, Hou and Huang \cite{ChenHouHuang} proved a similar result using a different method.

The special case of $a=1$ of Eq. (\ref{CLM01}) was first considered by De Gregorio \cite{DeGregorio} and has been the subject of extensive numerical computations in the periodic geometry by
Okamoto, Sakajo and Wensch \cite{Okamoto2008}.  These suggest that singularities do not occur in finite-time from
smooth initial data on a periodic domain.   Okamoto et al. \cite{Okamoto2008} use a least squares fit to the decay of Fourier modes to track the distance $\delta(t)$ from the real line to the nearest singularity in the complex-$x$ plane.
They find that  $\delta(t)$ decays exponentially in time, which is consistent with global existence.
Global existence for $a=1$ in the specific case of non-negative (or non-positive) initial vorticity is
proven by Lei et al. \cite{lei2019constantin}.

The above analytical and numerical results might suggest  the existence of
a threshold value $a=a_{threshold}$ below which finite-time singularities
occur for smooth initial data,  and at/above which the solution exists
globally in time.  Okamoto et al. conjecture that $a_{threshold} =1$.
However, for this value  $a=1$, Chen et al. \cite{ChenHouHuang}   recently
proved the existence of an  ``expanding" self-similar solution
\e{self-similar1} for the problem on $x\in\mathbb{R}$.  In this solution
$f(\xi)$ is an odd function with  finite support and  $\alpha=-1$. It
implies that $\omega(x,t)\to f'(0)x$ as $t\to t_c$ for any finite value
of $x\in\mathbb{R}$ while the boundary of compact support expands
infinitely fast in the spatial coordinate $x$  as $t\to t_c.$  \rev{We compute this solution numerically, and demonstrate that analytic initial data
converges to the expanding self-similar solution.} The form of this solution
is apparently incompatible with the periodic geometry,  and thus does not
rule out the  possibility of global existence of the solution in that
geometry when $a=1$.

We are not aware of any theory or simulation which consider solutions to (\ref{CLM01})  over  a wide range of the parameter $a $ as well as any simulation on $x\in\mathbb{R}$    addressing even the particular case $a=1$. The main goal of this paper is to fill this gap by presenting theory and highly accurate  computations to assess singularity formation for a wide range of $a$ for both the  periodic geometry and $x\in\mathbb{R}$.

\rev{We obtain two main analytical results (\rev{Theorems 1 and 3} below). The first one (Theorem 1) establishes the specific form of the leading-order complex singularity of $f(\xi)$ in \e{self-similar1} and determines its dependence on $a$, when that singularity is of power-law type.
{We show that this singularity can persist over time, without other singularity types present, provided $a=0$
or $1/2$. This enables the construction of exact analytical solutions for these values of $a$. }
The second main analytical result (Theorem 3)  proves that the exact solutions, consisting only of leading order power-law singularities,  is impossible beyond the particular cases $a=0$ and $1/2.$ It implies that  for any value of $a$, beyond $a=0$ and $1/2,$  the leading-order power-law singularity must coexist with other singularities for any   nonzero duration of time. If the initial condition contains only these leading order singularities, then other singularities must appear in arbitrarily small time to be consistent with
equation (\ref{CLM01}).}

Our spectrally accurate numerical simulations address all real values of $a$. We use a variable numerical precision, beyond the standard double precision, to mitigate loss of accuracy when computing poles and branch points in the complex plane, and employ fully resolved  spatial Fourier spectra on an adaptive grid with 8th order adaptive time stepping. Computations are performed both for  periodic boundary conditions (BC)
as well as on the real line  $x\in\mathbb{R}$ with the decaying BC%
\begin{equation}\label{omegadecay}
\omega(x,t)\to 0 \ \text{for} \ x\to\pm\infty.
\end{equation}
For the problem on $\mathbb{R}$,  we  reformulate Eq. (\ref{CLM01}) in a new spatial variable $q$ using a conformal mapping from Ref. \cite{LushnikovDyachenkoSilantyevProcRoySocA2017} between the real line   $x\in\mathbb{R}$ and $q\in (-\pi,\pi).$   Then our spectral simulations with a uniform spatial grid for $q\in (-\pi,\pi)$ ensure  spectral precision on the corresponding highly non-uniform grid for   $x\in\mathbb{R.}$

Our results make use of two distinct types of numerical simulation. The first type is time-dependent simulation which allows us to establish the convergence of generic initial conditions to the self-similar solution \e{self-similar1}. As a by-product of such simulations, we obtain values of $\alpha$ and the functional form of  $f(\xi).$  The second type of simulation directly solves the nonlinear eigenvalue problem for $\alpha$ to obtain the similarity solution  \e{self-similar1} of Eq. (\ref{CLM01}) for each value of $a.$ We solve that nonlinear eigenvalue problem by iteration on the real line    $x\in\mathbb{R}$  using a version of  the generalized  Petviashvili method (GPM)~\cite{Petviashvili1976,LushnikovOL2001,LY2007,PelinovskyStepanyantsSIAMNumerAnal2004,DLK2013}.
In {Theorem 4} we show that there exists a nonstable eigenvalue for the linearization of the original Petviashvili method \cite{Petviashvili1976} which prevents its convergence. However, the version of GPM employed here avoids that instability.

The results of the first and the second type of simulation are in excellent agreement  with {Theorems 1-3, and the exact similarity solutions.} The first major result of these simulations is the discovery of a critical value

\begin{equation}\label{acvalues}
a=a_c=0.6890665337007457\ldots
\end{equation}
below which (i.e., for $a<a_c$) there is finite-time singularity formation, but at which point (i.e., for $a=a_c$) the singularity transitions or changes character.
For  $a<a_c$ the value of $\alpha$ is positive with $f(\xi)$
an analytic function in a strip in the complex plane of $\xi$
containing the real line.
The positive values of $\alpha$ ensure,
in accordance with Eq. \e{self-similar1},  that the solution shrinks in
$x$ as $t\to t_c$ while the solution amplitude diverges in that limit.  This type of shrinking self-similar solution is compatible with both kinds of boundary conditions  (i.e., periodic and decaying on $\mathbb{R}$), and our simulations reveal  the same type of singularity formation at
$t\to t_c.$
The shrinking and divergence of amplitude is qualitatively reminiscent of
the collapse in both  the nonlinear Schr\"odinger equation and the
Patlak-Keller-Segel equation, see e.g. Refs.
\cite{ZakharovJETP1972,ChPe1981,SulemSulem1999,BrennerConstantinKadanoff1999,KuznetsovZakharov2007,LushnikovDyachenkoVladimirovaNLSloglogPRA2013}.
The terminology  ``collapse'' or ``wave collapse'' was first introduced
in Ref.  \cite{ZakharovJETP1972} in analogy with  gravitational
collapse and has been widely used ever since.
The   singularity
formation found  for $a<a_c$ is therefore of collapse type. We also find that $\alpha=0$
at the critical  value $a=a_c$.

The second major result of our simulations is the uncovering of a qualitatively different type of self-similar singularity formation
for $a_c< a\le 1$, in which the spatial scale of the solution does not shrink. We refer to this type of singularity as ``blow up."
An additional finding in the aforementioned range of  the parameter $a$ is that the blow-up solution on the real line  $x\in\mathbb{R}$ and the blow-up solution for  periodic BC are qualitatively different. In the  case   $x\in\mathbb{R}$ we find that $-1\le\alpha<0$ with $\alpha=-1$ only for $a=1.$ Thus Eq. \e{self-similar1}
  corresponds to an expanding self-similar solution. In particular, at $a=1$, we find that $\alpha=-1$ in agreement with the results of Ref. \cite{ChenHouHuang}. A Taylor series expansion of Eq. \e{self-similar1} at $x=0$ results in $\omega(x,t)=\tau^{-1-\alpha}xf'(0)+O(\tau^{-1-2\alpha}x^2).$  It shows that the linear slope $\propto x$  increases to infinity as $t\to t_c$ for $a_c\le a<1$, while it remains constant for  $a=1.$$ $ Time-dependent simulations for $x\in\mathbb{R} $ with analytic initial conditions and $a_c\le a\le1$ demonstrate  convergence of the solution at  $t\to t_c$  to Eq. \e{self-similar1} with $f(\xi)$ being of finite support.
  This extends the results of Ref. \cite{ChenHouHuang}
  from $a=1$ to $a_c\le a\le 1$.

The third major result of our simulations concerns  periodic BC. While
the collapse case $a<a_c$ is similar for both  $x\in\mathbb{R}$
and periodic BC, as mentioned the case  $a_c< a\le 1$ is qualitatively different.
Indeed, the spatial expansion or blow up observed  for $a_c\le a\le 1$  and $x \in \mathbb{R}$ would contradict the
periodic BC as $t$ approaches $t_c.$ Instead, we find a new self-similar
blow-up solution
\begin{equation}\label{self-similar1periodic}
\omega(x,t)=\frac{1}{t_c-t}f(x),
\end{equation}
which is valid for  $a_c< a\le 0.95. $   Formally, we can interpret Eq. \e{self-similar1periodic} as Eq. \e{self-similar1} with $\alpha=0$. However, periodic BC are qualitatively different from the finite support solution  of Eq.  \e{self-similar1} because of the nonlocality of the Hilbert transform in Eq. (\ref{CLM01}).  \rev{We find that $f(x)$ in Eq. \e{self-similar1periodic} has a discontinuity  in a high-order (or $n$th-order) derivative at the periodic boundary, i.e., at $x=\pm \pi$ when the domain is centered about  the point $x=0$ where the singularity occurs.  In addition,  $n\to \infty$  in the limit $a\to a_c^+,$  i.e. $f(x)$ approaches  a $C^\infty$ function in that limit. A complex singularity is also present in $f(x)$ on the imaginary axis away from the real line, the  form of which obeys Theorem 1.}

In the range    $0.95< a< 1, $ our simulations are inconclusive regarding whether  blow up occurs.   The value $a=1$ is a special case for the  periodic BC, with no blow up observed in our simulations for generic initial conditions. Instead, the solution exists globally  with the first spatial derivative remaining bounded,  while the second derivative grows exponentially in time. This agrees with the result on global existence for the particular case $a=1$ investigated in  Ref. \cite{Okamoto2008}.

For $a\ge1,$ we find that  the solution exists globally for all initial conditions considered in the case of periodic BC, while for the solution on the real line the situation is not conclusive. In the latter case, the maximum of $|\omega|$ initially grows with time but this growth saturates at larger times at least for $a\gtrsim1.3$, so we expect the global existence of solutions in this parameter range. In the intermediate range $1<a\lesssim1.3,$ our simulations catastrophically lose precision  at sufficiently large times, and a conclusive determination between  blow up and global existence of solutions is not possible.

We also find from the simulations that the kinetic energy on the infinite line $x\in \R$, %
\begin{equation}\label{KinetiEdef}
E_K:=\int \limits ^\infty_{-\infty} u^2(x,t)\D x,
\end{equation}
 with an initially finite value approaches a constant as $t \rightarrow t_c$  when $a < 0.265 \pm 0.001$, while it tends to infinity  for  $ 0.265 \pm 0.001<a\leq1$. In the case $a \gtrsim 1.3$  corresponding to global existence, the kinetic energy tends to infinity as $t \rightarrow \infty$.
On the periodic domain $x\in [-\pi,\pi],$ we find the same behaviour of the kinetic energy up to  $a=0.95$.   For $a \ge 1$ (when there is global existence),
$E_K$ approaches a non-zero constant as $t\to \infty$ ($a=1$)  or tends to zero ($a>1$).

Solutions with  finite energy are of interest by analogy with the fundamental question
on  global regularity of the 3D Euler and Navier-Stokes equations with smooth initial
data,  see Refs. \cite{FeffermanMilleniumprize2006,GibbonPhysD2008}.

To reveal the structure of singularities of  $\omega(x,t)$ and $f(\xi)$ in the complex plane of $x$ and $\xi,$   we use both a  fitting of the Fourier spectrum similar to Ref. \cite{Okamoto2008} (see also Refs. \cite{CarrierKrookPearson1966,DyachenkoLushnikovKorotkevichJETPLett2014,DyachenkoLushnikovKorotkevichPartIStudApplMath2016,SulemSulemFrischJCompPhys1983} for more detail),  and more general methods of analytical continuation by rational interpolants (see Refs.  \cite{AGH2000,DyachenkoLushnikovKorotkevichPartIStudApplMath2016,DyachenkoDyachenkoLushnikovZakharovJFM2019,TrefethenAAA}). As time evolves, these singularities approach the real line in agreement with Eq. \e{self-similar1}. We have formulated a system of ordinary differential equations
(ODEs) describing the motion of such singularities.  Fourier fitting allows us to track only singularities which  are nearest to the real axis, while rational interpolants go beyond this, by giving information on singularities other than the closest one. In particular, it reveals that for $a\ne 0,1/2$ with $a<a_c$, there are generically branch points beyond the leading order singularities, consistent with Theorem 3. The exceptional cases are $a=0,1/2$ and $ 2/3$ where the nearest singularities are poles of the first, second, and third order, respectively. However, already for $a=2/3$, the third order pole coexists with additional branch points.  For other values of $a$, the nearest singularities are branch points.    We find that for $a_c<a\le1,$ the singularities approach the real line as $t\to t_c$ in the spatial regions near the boundary of the support of $f(\xi).$

The rest of this paper is organized as follows.
Section \ref{sec:main} establishes Theorem 1, which describes the leading-order complex singularity and determines its dependence on  $a$.
Section \ref{sec:a0} reinterprets the results of Ref. \cite{ConstantinLaxMajda} for $a=0$  in terms  of moving complex poles and  the self-similar solution \e{self-similar1}.
In Section \ref{sec:a1p2},  we derive  an  exact blow-up solution for $a=1/2$ (Theorem 2), and  transform that exact solution to  the self-similar form \e{self-similar1}.
Section \ref{sec:generala} considers  solutions for  general values of $a$ and establishes in Theorem 3 that, except for $a=0,1/2$, the leading order singularity cannot fully characterize the exact solution.
Two preliminary steps for computations on $x \in \mathbb{R}$ are developed in Sections \ref{sec:self-similar} and \ref{sec:transform}.  In particular,  Section \ref{sec:self-similar} reformulates  Eq.\e{CLM01} as a nonlinear eigenvalue problem for the self-similar solution   \e{self-similar1}, and
Section \ref{sec:transform} rewrites Eq. \e{CLM01} in an auxiliary variable $q$ mapping the real line into the finite interval.
Section \ref{sec:numerics} then describes the results of time-dependent numerical simulations for $x\in\R$, and
Section \ref{sec:e_value_problem} presents self-similar solutions of the type \e{self-similar1} 
via numerical solution of the nonlinear eigenvalue problem using a generalized
Petviashvili method. Section \ref{sec:analyticalcontinuation} addresses the analytical continuation into the complex plane of $x$ by rational approximation and uses it to study the structure of singularities. Section
\ref{sec:numerics2} describes the results of both time-dependent numerical
simulations and the generalized
Petviashvili method for periodic BC. Section \ref{sec:conclusions} provides a
summary of the results and discusses future directions. Appendix
\ref{sec:AppendixTransformationHilberttransform} gives a derivation for
the form of the Hilbert transform over $x$ in variable $q.$

\section{Leading order spatial singularity} \label{sec:main}

We assume that $\omega(x,t)$ is an analytic function in the open strip containing $x\in\R$ in the complex  plane $x\in\mathbb C$ decaying at $x\to\pm\infty$.  Then we can represent $\omega$ as %
\begin{equation}\label{omegapm}
\omega=\omega^++\omega^-,
\end{equation}
where $\omega^+(x.t)$ is analytic in the upper complex half-plane $x\in\C^+$ and  $\omega^-(x.t)$ is analytic in the lower complex half-plane $x\in\C^-$.

The Hilbert transform \e{HilbertHdef} implies that %
\begin{equation}\label{Hpm}
{\mathcal H}\omega=-\I(\omega^+-\omega^-).
\end{equation}

Assume that the solution exhibits a leading order singularity of power $\gamma>0$ in the complex plane $x$ for  $\omega$ at $x=\pm\I v_c, \ v_c>0,$ so that
\begin{equation}\label{omegagamma}
\omega(x,t)=\frac{\omega_{-\gamma}(t)}{[x-\I v_c(t)]^\gamma}+\frac{\bar \omega_{-\gamma}(t)}{[x+\I v_c(t)]^\gamma}+l.s.t.,
\end{equation}
where $l.s.t$ designates less singular terms at  $x=\pm\I v_c$, i.e.%
\begin{equation}\label{limit1}
\lim_{x\to\pm\I v_c}[x\mp\I v_c(t)]^\gamma l.s.t.=0.
\end{equation}
If we additionally assume that  $\omega(-x)=-\omega(x),$ for $x\in \R$, then Eq. \e{omegagamma} implies that
\begin{equation}\label{omegagammainteger}
\frac{\omega_{-\gamma}(t)}{[x-\I v_c(t)]^\gamma}+\frac{\bar \omega_{-\gamma}(t)}{[x+\I v_c(t)]^\gamma}=-\frac{\omega_{-\gamma}(t)}{[-x-\I v_c(t)]^\gamma}-\frac{\bar \omega_{-\gamma}(t)}{[-x+\I v_c(t)]^\gamma},
\end{equation}
i.e., $\bar \omega_{-\gamma}(t)(-1)^{\gamma+1}=\omega_{-\gamma}(t). $  Then we can define %
\begin{equation}\label{omegagammadef}
\omega_{-\gamma}(t):=-\I e^{-\I\pi \gamma/2} \tilde \omega_{-\gamma}(t), \quad \tilde \omega_{-\gamma}(t)\in \R
\end{equation}
so that Eq. \e{omegagamma} takes the following form %
\begin{equation}\label{omegagamma2}
\omega(x,t)=-\I\tilde \omega_{-\gamma}(t)\left (\frac{ e^{-\I\pi \gamma/2} }{[x-\I v_c(t)]^\gamma}-\frac{  e^{\I\pi \gamma/2}  }{[x+\I v_c(t)]^\gamma}\right )+l.s.t..
\end{equation}

Using Eqs. \e{CLM01},\e{Hpm} and \e{omegagamma2}
we obtain  that %
\begin{equation}\label{uxgamma1}
u_x={\mathcal H}\omega=\tilde \omega_{-\gamma}(t)\left (\frac{ e^{-\I\pi \gamma/2} }{[x-\I v_c(t)]^\gamma}+\frac{  e^{\I\pi \gamma/2}  }{[x+\I v_c(t)]^\gamma}\right )+l.s.t.,
\end{equation}

and
\begin{equation}\label{ugamma1}
u:=u^++u^-=-\frac{\tilde \omega_{-\gamma}(t)}{(\gamma-1)}\left (\frac{ e^{-\I\pi \gamma/2} }{[x-\I v_c(t)]^{\gamma-1}}+\frac{  e^{\I\pi \gamma/2}  }{[x+\I v_c(t)]^{\gamma-1}}\right )+l.s.t.,
\end{equation}
where we have additionally assumed that $\gamma\ne1$.

Plugging Eqs. \e{omegagamma2}-\e{ugamma1} into Eq. \e{CLM01} and collecting the most singular terms $\propto[x-\I v_c(t)]^{-2\gamma}$ at $x=\I v_c(t)$ on the right-hand side of Eq. \e{CLM01} gives %
\begin{equation}\label{twogamma}
\frac{\I e^{-\I\pi \gamma} \tilde\omega_{-\gamma}^2(t) }{[x-\I v_c(t)]^{2\gamma}}\left (\frac{a\gamma}{\gamma-1}-1 \right )=0.
\end{equation}
By assumption $\omega_{-\gamma}(t)\ne 0$. Then Eq. \e{twogamma} implies that%
\begin{equation}\label{gammaa}
\gamma=\frac{1}{1-a}.
\end{equation}

Thus we have proved the following:

  {\it Theorem} 1. \rev{If a solution $\omega(x,t)$ of Eq. \e{CLM01}  is  (i) analytic in an open strip of ${\mathbb C}$ containing $\mathbb{R}$, (ii) tends to zero as as $x \rightarrow \pm \infty$, and (iii)  has a complex conjugate pair of   power law singularities located
at $x=\pm\I v_c$ for $\ v_c>0$  given by  Eqs. \e{limit1}, \e{omegagamma2}
  with   $\gamma>0,$ then $\gamma$ is determined by  Eq. \e{gammaa}.}

{\it Remark 1}. The condition $\gamma>0$ is essential in Theorem 1. If we assume $\gamma<0$, then the leading order term in Eq. \e{CLM01} at $x=\pm\I v_c \ $is  $\propto[x-\I v_c(t)]^{0}.$

{\it Remark 2}. Eq. \e{gammaa} is in excellent agreement with the simulations of Section \ref{sec:numerics}. The singularities with  $\gamma<0$ in our simulations are always located further away from the real axis than the leading order singularities given by
Eq. \e{gammaa}. These more remote singularities 
provide  a smaller contribution to the solution near the origin.

Eq. \e{gammaa} with $a=0$ results in $\gamma=1.$ Also $\gamma \to \infty$ for $a\to1-$. For the particular values %
\begin{equation}\label{nnm1}
a=\frac{n-1}{n}, \quad n=1,2,3,\dots,
\end{equation}
 we obtain the integer values $\gamma=n$ resulting in  complex pole singularities of order $n$ in Eq. \e{omegagamma2} while  the other values of $a\in(0,1)$ result in the branch points at $x=\pm\I v_c(t).$

\section{Exact blow-up solution for $a=0$}\label{sec:a0}

The particular value  of the parameter $a=0$ implies from Eq. \e{gammaa} that $\gamma=1$. This  case recovers the results of Ref. \cite{ConstantinLaxMajda}. The general solution of Eq. \e{CLM01} is immediately obtained by noticing that  Eqs. \e{CLM01}, \e{Hpm} result in %
\begin{equation}\label{CLMexplicit}
\omega_t=\omega^+_t+\omega^-_t=-\I(\omega^+)^2 +\I(\omega^-)^2
\end{equation}
which  decouples into two independent ODEs%
\begin{equation}\label{ODEomegapm}
\omega^+_t=-\I(\omega^+)^2, \quad \omega^-_t=\I(\omega^-)^2.
\end{equation}
The solutions of these ODEs with the generic initial conditions  $\omega^+(x,t)|_{t=0}=\omega^+_0(x)$ and $\omega^-(x,t)|_{t=0}=\omega^-_0(x)$ are given by %
\begin{equation}\label{omegapmsol}
\omega^+(x,t)=\frac{\omega^+_0(x)}{1+\I t\omega^+_0(x)} \quad \text{and} \quad
\omega^-(x,t)=\frac{\omega^-_0(x)}{1-\I t\omega^-_0(x)}.
\end{equation}
Eqs. \e{omegapm}, \e{Hpm} and \e{omegapmsol} lead to  the solution of Constantin-Lax-Majda equation found in Ref. \cite{ConstantinLaxMajda}%
\begin{equation}\label{omegaCLM}
\omega(x,t)=\frac{4\omega_0(x)}{[2-t {\mathcal H}\omega_0(x)]^2+t^2\omega_0^2(x)}
\end{equation}
for the generic initial condition $\omega(x,t)|_{t=0}=\omega_0(x)=\omega^+_0(x)+\omega^-_0(x)$.
Also Eqs.
\e{Hpm} and  \e{omegapmsol} imply that (as in Ref. \cite{ConstantinLaxMajda})
\begin{equation}\label{HomegaCLM}
{\mathcal H}\omega(x,t)=\frac{2{\mathcal H}\omega_0(x)[2-t{\mathcal H}\omega_0(x)]-2t\omega_0^2(x)}{[2-t {\mathcal H}\omega_0(x)]^2+t^2\omega_0^2(x)}.
\end{equation}

Assume that there exists an  $x_0\in\R$ such that $\omega_0(x_0)= 0$ and  $ {\mathcal H}\omega_0(x_0)>0$.   Then Eq. \e{omegaCLM} implies a singularity in the solution at the time $t_c:=2/ {\mathcal H}\omega_0(x_0)>0.$ If there are multiple points  $x\in\R$ such that $\omega_0(x)= 0$ and  $ {\mathcal H}\omega_0(x)>0$ then  $t_c:=2/\text{sup}\{ {\mathcal H}\omega_0(x)|\omega_0(x)=0\}>0$ \cite{ConstantinLaxMajda}. Below we assume that $x_0$ corresponds to the singularity at the earliest time $t=t_c.$   A particular example is any odd function $\omega_0(x)$ with respect to $x=x_0$ (implying that $\omega_0(x_0)=0$)  which is strictly positive for $x>x_0$ and decays at $x\to \infty$.

A  series expansion of Eq. \e{omegaCLM} at $x\to x_0$ and $t\to t_c^-$ implies that%
\begin{equation}\label{omegaser1}
\omega(x,t)=\frac{1}{t_c-t}\frac{4\xi\omega_0'(x_0)[ {\mathcal H}\omega_0(x_0)]^2}{\left([ {\mathcal H}\omega_0(x_0)]^2-2\xi {\mathcal H}\omega'_0(x_0)\right )^2+4\xi^2[\omega'_0(x_0)]^2}+O((t_c-t)^0),
\end{equation}
where%
\begin{equation}\label{xidef0}
\xi:=\frac{x-x_0}{t_c-t}
\end{equation}
is the self-similar variable.
Eqs. \e{omegaser1} and \e{xidef0}
provide a universal profile of the solution at  $t\to t_c^-$ in a spatial neighborhood of $x\to x_0$ after we neglect the correction term $O((t_c-t)^0)$.
That profile has the form of a sum of two complex poles at complex conjugate points $\xi=\xi_\pm$ as follows
\begin{equation}\label{omegaser1poles}
\omega(x,t)=\frac{\I}{t_c-t}\left(\frac{\xi_+}{\xi-\xi_+}-\frac{\xi_-}{\xi-\xi_-}\right ),
\end{equation}
where %
\begin{equation}\label{xipm}
\xi_\pm=\frac{[ {\mathcal H}\omega_0(x_0)]^2}{2[ {\mathcal H}\omega'_0(x_0)\pm\I\omega'_0(x_0)]}
\end{equation}
are positions of poles in the complex plane of $\xi.$

Eqs. \e{omegaser1poles} and \e{xipm} provide the exact solution of Eq. \e{CLM01} for $\omega'_0(x_0)<0$ as can be immediately verified by  direct substitution into Eq. \e{CLM01}. Here the condition  $\omega'_0(x_0)<0$  ensures that $\xi_+\in \C^+.$ This solution is asymptotically stable with respect to  perturbations of the initial condition as follows from Eq. \e{omegaser1}. The only trivial change due to  the perturbation of the initial condition is  a shift of both $x_0$ and $t_c.$

One can also recover from the solution \e{omegaser1poles}  the representation  \e{omegagamma2}
with $\gamma=1$\ which gives the exact solution %
\begin{align}\label{omegagamma2a0}
\omega(x,t)=-\tilde v_c\left (\frac{ 1 }{x-x_0-\I \tilde v_c(t_c-t)}+\frac{  1  }{x-x_0+\I\tilde  v_c(t_c-t)}\right )\nonumber \\=-\frac{\tilde v_c}{t_c-t}\left (\frac{ 1 }{\xi-\I \tilde v_c}+\frac{  1  }{\xi +\I\tilde  v_c}\right )
\end{align}
 of Eq. \e{CLM01} for any values of the real constants $t_c, \ \tilde v_c>0$ and $x_0.$ %
Here without loss of generality we have shifted the origin
 in the real direction compared with the solution \e{omegaser1poles}.

\section{Exact blow-up solution for $a=1/2$}\label{sec:a1p2}

The particular value  of the parameter $a=1/2$ implies from Eq. \e{gammaa} that $\gamma=2$. In this section we look for the solution to Eq. \e{CLM01} in the form  \e{omegagamma2}
 assuming that the  $l.s.t.$ are identically zero, i.e.,  %
\begin{equation} \label{omegagamma2a1p2}
\omega(x,t)=\I\tilde \omega_{-2}(t)\left (\frac{ 1 }{[x-x_0-\I v_c(t)]^2}-\frac{  1  }{[x-x_0+\I v_c(t)]^2}\right ),
\end{equation}
where for  generality we have also allowed a shift of the origin by  introducing the arbitrary real constant $x_0.$
Eq. \e{ugamma1} then becomes  %
\begin{equation}\label{ugamma11p2}
u=\tilde \omega_{-2}(t)\left (\frac{ 1 }{x-x_0-\I v_c(t)}+\frac{  1  }{x-x_0+\I v_c(t)}\right )=\frac{2 \tilde \omega_{-2}(t)(x-x_0) }{(x-x_0)^2+ v_c(t)^2}.
\end{equation}
Plugging  Eqs. \e{omegagamma2a1p2} and \e{ugamma11p2} into Eq. \e{CLM01}, we find the latter equation  is identically satisfied provided %
\begin{equation}\label{vcom1p2a}
\begin{split}
& \frac{d v_c(t)}{dt}=- \frac{\tilde \omega_{-2}(t)}{4 v_c(t)}, \\
\end{split}
\end{equation}
and
\begin{equation}\label{vcom1p2b}
\begin{split}
& \frac{d\tilde \omega_{-2}(t)}{dt}= \frac{\tilde \omega_{-2}^2(t)}{4 v_c^2(t)}.
\end{split}
\end{equation}
Solving the system of  ordinary differential equations
(ODEs)  \e{vcom1p2a} and \e{vcom1p2b}  results in %
\begin{equation}\label{vcomt1p2power}
v_c(t)=(t_c-t)^{1/3}\tilde v_c , \quad \tilde \omega_{-2}(t)=\frac{4\tilde v^2_c }{3(t_c-t)^{1/3} },
\end{equation}
where $\tilde v_c>0$ and $t_c$ are two arbitrary real constants.
Assuming the initial condition is given at $t=0$ and that $t_c>0$, we obtain that $t=t_c$ is the time of singularity formation.

Section \ref{sec:numerics} below 
shows the convergence during the
evolution in time $t$ of the solution of Eq. \e{CLM01} to  the exact
solution given by Eqs.  \e{omegagamma2a1p2} and \e{vcomt1p2power}. The
spatial extent of the solution  shrinks while the maximum amplitude
increases until  the singularity is reached  at $t=t_c$.

One can rewrite the solution \e{omegagamma2a1p2}, \e{vcomt1p2power} in the self-similar form as follows%
\begin{equation}\label{omegasera1p2}
\omega(x,t)=\frac{1}{t_c-t}\frac{4\I\tilde v^2_c }{3 }  \left (\frac{ 1 }{[\xi-\I\tilde  v_c]^2}-\frac{  1  }{[\xi+\I \tilde v_c]^2}\right )=\frac{1}{t_c-t}\frac{16\tilde v^3_c\xi }{3(\xi^2+\tilde  v_c^2)^2},
\end{equation}
where%
\begin{equation}\label{xidef0p5}
\xi:=\frac{x-x_0}{(t_c-t)^{1/3}}
\end{equation}
is the self-similar variable.

\rev{ {\it Note.} After our arXiv preprint submission \cite{LushnikovSilantyevSiegelarXiv2020} we learned that the self-similar solution (\ref{omegasera1p2}) was recently discovered by Jiajie Chen in \cite{Chen2020Singularity}. The result presented here was found independently via the complex singularity approach, and has a somewhat more general form by including the additional real parameter $\tilde v_c$.}


To summarize, this section proves the following theorem:

{\it Theorem} 2.  Eqs. \e{omegasera1p2} and \e{xidef0p5}
provide an exact solution of Eq. \e{CLM01} for $a=1/2$ for any value of the real constants $t_c, \ \tilde v_c>0$ and $x_0.$

{\it Remark 3}. The decay of $u(x,t)$ in  Eq.  \e{ugamma11p2} as $x \rightarrow \pm \infty$ ensures that the kinetic energy
\e{KinetiEdef} has a  finite  value for $t<t_c$. In contrast, $E_K$ for the solution
\e{omegagamma2a0} at $a=0$  is  infinite.

\section{The solution for  general values of $a$}\label{sec:generala}

The explicit self-similar solutions \e{xidef0}-\e{xipm} and  \e{omegasera1p2}, \e{xidef0p5} (corresponding to the values  $a=0, 1/2$) represent the particular situation where the leading order singularity in Eqs. \e{omegagamma2} and \e{gammaa}   
 provides the exact solution with  identically zero  $l.s.t.$.
All other values of $a$ are addressed in the following theorem:

  {\it Theorem} 3. \rev{A solution \e{omegagamma2} and \e{gammaa} of Eq.   
 \e{CLM01} which satisfies assumptions (i) and (ii) of Theorem 1
 requires  $l.s.t.$  which are not identically zero for any $a\in \R$ except $a=0$ and $a=1/2.$}

\begin{proof}

The case $a\ge 1$ is trivial because $a=1$ corresponds to the singular value of $\gamma$ as follows from Eq.  \e{gammaa}, while $a>1$ implies that $\gamma<0$, contradicting the assumption of Theorem 3 that  $\omega$ at $x\to \pm \infty$ .  Thus below we assume that $a<1$ which implies that $\gamma>0$.

We assume by contradiction that   $l.s.t.$  in Eq.  \e{omegagamma2} are identically zero. Then we plug  Eq.  \e{omegagamma2} into Eq. \e{CLM01} and collect terms with different powers of $x-\I v_c(t).$ The most singular term  $\propto[x-\I v_c(t)]^{-2\gamma}$ is identically zero by Eq. \e{gammaa} as
follows from the proof of Theorem 1. Collecting the next most singular terms   $\propto[x-\I v_c(t)]^{-1-\gamma}$ we obtain that  %
\begin{equation}\label{vcom1p2agamma}
\begin{split}
\frac{d v_c(t)}{dt}=- \frac{2^{1-\gamma}\tilde \omega_{-\gamma}(t)}{ v_c^{\gamma-1}(t)\gamma}, \end{split}
\end{equation}
which generalizes  Eq. \e{vcom1p2a}  to arbitrary values of $\gamma.$ We note that there is no overlap between terms of different orders  in this proof except in the case $\gamma=1$, for which $ -2\gamma=-\gamma-1$. However, this case is fully considered in Section \ref{sec:a0} and excluded by  assumption in the statement of  {Theorem 3} because it corresponds to $a=0.$

Collecting the terms   $\propto[x-\I v_c(t)]^{-\gamma}$ we obtain that   \begin{equation}\label{vcom1p2bgamma}
\begin{split}
\frac{d\tilde \omega_{-\gamma}(t)}{dt}= \frac{2^{-\gamma} (\gamma-1)\tilde \omega_{-\gamma}^2(t)}{ v_c^\gamma(t)} \\
\end{split}
\end{equation}
which generalizes  Eq. \e{vcom1p2b}  to arbitrary values of $\gamma.$

However, at the next order, collecting terms    $\propto[x-\I v_c(t)]^{-\gamma+1}$
leads to  %
\begin{equation} 
\frac{2^{-\gamma-2}(\gamma-2)(\gamma+1)\I e^{-\I\pi \gamma/2} \tilde\omega^2_{-\gamma}(t) }{ v_c^{\gamma+1}(t)}=0,
\end{equation}
which cannot be satisfied by any nontrivial solution $\tilde\omega_{-\gamma}(t)\not\equiv 0$   except if $\gamma=2$, i.e. $a=1/2.$ This contradiction completes the proof of {Theorem 3}.

\end{proof}

\begin{figure}
\centering
    \includegraphics[width=0.5\linewidth]{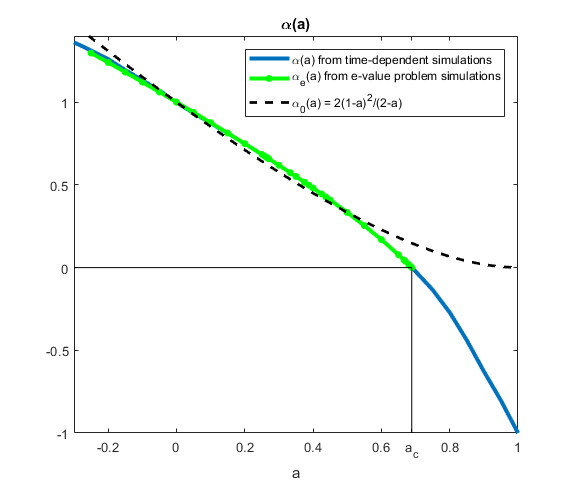}
    \caption{Dependence of $\alpha(a)$ on $a$,  obtained via time-dependent simulations 
of Section \ref{sec:numerics}
and via nonlinear eigenvalue problem  of  Section \ref{sec:e_value_problem}. The green curve terminates at $a=a_c$ since the  iteration used to solve the nonlinear eigenvalue problem for $x \in \mathbb{R}$ does not converge for $a>a_c$. Also included for comparison is an  approximation to $\alpha(a)$ from Eq. \e{xidef0pgamma}, $\alpha_0(a)=\frac{2}{\gamma(a)(\gamma(a)+1)}=\frac{2(1-a)^2}{(2-a)}$. \label{fig:alpha(a)}}
\end{figure}

{\it Remark 4}. The ODE system \e{vcom1p2agamma} and \e{vcom1p2bgamma} can be immediately solved for any $\gamma$ resulting in
\begin{equation}\label{vcomgamma}
\begin{split}
&v_c(t)=\tilde v_c \,(t_c-t)^{\frac{2}{{\gamma} ({\gamma}+1)}}, \\
&\tilde \omega_{-\gamma}(t)=\frac{2^{{\gamma}} {\tilde v_c}^{\gamma}}{\gamma+1}(t_c-t)^{\frac{1-\gamma}{{\gamma}+1}},
\end{split}
\end{equation}
where  $\tilde v_c$ and $t_c$ are arbitrary real constants.
Then neglecting  $l.s.t.$, we obtain from Eqs.
\e{omegagamma2} and \e{vcomgamma} the following self-similar ``solution"%
\begin{equation}\label{omegaselfsimilargammaincorrect}
\omega(x,t)=-\frac{\I}{t_c-t}\frac{2^{{\gamma}} {\tilde v_c}^{\gamma}}{\gamma+1}\left (\frac{ e^{-\I\pi \gamma/2} }{[\xi-\I\tilde  v_c]^\gamma}-\frac{  e^{\I\pi \gamma/2}  }{[\xi+\I\tilde  v_c]^\gamma}\right ),
\end{equation}
where%
\begin{equation}\label{xidef0pgamma}
\xi:=\frac{x-x_0}{(t_c-t)^{\alpha_0}}, \quad  \alpha_0={\frac{2}{{\gamma} ({\gamma}+1)}},
\end{equation}
is the self-similar variable. For $\gamma=1 (a=0)$ and $\gamma=2 (a=1/2)$,
Eqs. \e{omegaselfsimilargammaincorrect} and \e{xidef0pgamma} recover Eqs.
\e{xidef0}, \e{omegagamma2a0} and \e{omegasera1p2},  \e{xidef0p5},
respectively. However, {Theorem 3} ensures that Eqs.
\e{omegaselfsimilargammaincorrect} and \e{xidef0pgamma} are not the exact
solution for $\gamma\ne 1,2$. One may hope that even if  $\gamma\ne 1,2$, the
self-similar solution is well approximated by Eqs.
\e{omegaselfsimilargammaincorrect} and \e{xidef0pgamma} because
\e{omegagamma2} is the leading order singularity of the solution. However,
we find below in Section \ref{sec:numerics} (see also Fig. \ref{fig:alpha(a)})
that the numerically computed self-similar solution has a different power scaling
for $\xi=\frac{x-x_0}{(t_c-t)^{\alpha}}$ than in Eq.  \e{xidef0pgamma}, i.e.
$\alpha_0\ne \alpha$ for $\gamma\ne 1,2$.  This implies that the  $l.s.t$, neglected in \e{xidef0pgamma},
lead to a
non-trivial modification of $\alpha$ compared with $\alpha_0.$

\revc{
{\it Discussion.} To address $l.s.t. $ in Eq.  \e{omegagamma2}, one can naively write a solution as a  formal expansion  in    powers of $x\pm\I v_c(t)$ as follows }


\revc{
 where we have assumed integer values of $\gamma$ as given by Eq. \e{nnm1}. For example, if $a=2/3$ and, respectively, $\gamma=3$, such a series has the  form
}


\revc{
 For non-integer values of $\gamma$ we can write a more general Puiseux series involving multiple powers of $n\gamma+m,$ $n,m\in \N$, however, this is beyond the scope of the current paper.

A limitation of the formal series  \e{omegagamma2integer} and \e{omegagamma3p3} is that they simultaneously involve expansions of $\omega^+(x,t)$ at $w=-\I v_c$ and $\omega^-(x,t)$ at $w=\I v_c$. If the radiuses  of convergence $r_c$ of both series exceed $v_c$, then there is overlap of the corresponding disks of convergence, e.g., at $x=0.$ The simplest situation is when both radiuses of convergence  exceed $2v_c,$ which would turn these formal series into proper Laurent series at $x=\pm\I v_c $ for integer values of $\gamma$, while allowing a straightforward evaluation of the Hilbert transform by Eq. \e{Hpm}.    However, we find from the  simulations of Section \ref{sec:analyticalcontinuation} that additional singularities  exist  in the complex plane of $x$ beyond $x=\pm v_c$, which implies that  $v_c<r_c<2v_c$  for  $\gamma\ne 1,2$. Thus for integer values of $\gamma\ne 1,2,$ one has  to use a more general form for the
Laurent series  at $x=\pm\I v_c $ which is not necessarily equivalent to \e{omegagamma2integer}. For example,  in the case $\gamma=3$ one has to replace Eq. \e{omegagamma3p3} by the Laurent series
}

\revc{
 at $x=\I v_c$, where the series coefficients are the same for the singular part  in Eq. \e{omegagamma3p3}, i.e.    $ \tilde{\tilde \omega}_{j}(t)\equiv{\tilde \omega}_{j}(t)$ for $j=-1,-2,-3$, while for $j\ge0$ these coefficients are different  than ${\tilde \omega}_{j}(t)$. One can also prove that   ${\tilde \omega}_{-2}(t)\equiv{\tilde \omega}_{-1}(t)\equiv 0$. The disadvantage of using the Laurent series \e{omegagamma3p3Laurent} compared with \e{omegagamma3p3} is that one cannot easily find ${\mathcal H}\omega$ from \e{omegagamma3p3Laurent}  because both $\omega^+(x,t)$ and $\omega^-(x,t)$ contribute to the coefficients    $ \tilde{\tilde \omega}_{j}(t)$. A technique from Ref. \cite{DyachenkoDyachenkoLushnikovZakharovJFM2019} can be used to to obtain  partial information on ${\mathcal H}\omega$, however a full solution (i.e. a knowledge of all coefficients)  requires global information.
}

\section{Self-similar solution and nonlinear eigenvalue problem} \label{sec:self-similar}

The results of Sections \ref{sec:a0}-\ref{sec:generala} suggest looking for a solution of Eq. \e{CLM01}
in the general self-similar form \e{self-similar1}.
Substitution of the ansatz \e{self-similar1} into  Eq. \e{CLM01}
reduces it to %
\begin{equation}\label{CLM01selfsimilar}{\mathcal
 M}f:=f+\alpha\xi f_\xi=-a(\partial^{-1}_\xi{\mathcal H}f)f_\xi+f{\mathcal H}f,
\end{equation}
where ${\mathcal M} $ is a  linear operator. One can also rewrite Eq. \e{CLM01selfsimilar} as the system %
\begin{equation}\label{CLM01selfsimilar2}
f+\alpha\xi f_\xi=-agf_\xi+fg_\xi, \quad g=\partial^{-1}_\xi{\mathcal H}f,
\end{equation}
where
\begin{equation}\label{uselfsimilar}
u={\tau}^{\alpha-1}g\left ( \xi \right ).
\end{equation}

\revc{We note that the series \e{omegagamma2integer} or \e{omegagamma3p3Laurent} are reduced to   self-similar form provided we add the restriction that $v_c(t)=(t_c-t)^\alpha \tilde v_c $ and     $\tilde \omega_{-\delta}(t)=(t_c-t)^{-1+\alpha\delta}     \tilde{\tilde {\omega}}_{-\delta}$, where $\tilde{\tilde {\omega}}_{-\delta}$ is an arbitrary constant and the subscript/power $\delta$ represents a general subscript/power  in the series starting from $\delta=\gamma$. }

We can iterate Eq. \e{CLM01selfsimilar}
for different values of $\alpha$ to find the optimal  $\alpha$ which realizes the dominant collapse regime. To do this we have to invert  the operator ${\mathcal M}$ in Eq. \e{CLM01selfsimilar} at each iteration.
The equation ${\mathcal M}f=0$ has a general solution %
\begin{equation}\label{fasymp}
f\propto|\xi|^{-\frac{1}{\alpha}}
\end{equation}
 for $\alpha\ne 0$ and $f\equiv 0$ for $\alpha=0.$ Depending on the sign on $\alpha$, this solution is singular either at $x\to 0$ or $x\to\pm \infty$. Thus the operator $\mathcal\ M$ is invertible
for the class of smooth solutions decaying at  $x\to\pm \infty$ which we use below in Section \ref{sec:e_value_problem}.

The  condition that the solution of  Eq. \e{CLM01selfsimilar} decays at both $x\to\pm\infty$ requires a specific choice of $\alpha$ for each $a$. It forms a version of nonlinear eigenvalue problem for $\alpha(a).$
Section \ref{sec:e_value_problem}  finds $\alpha(a)$ by iterating Eq. \e{CLM01selfsimilar} numerically.

{\it\ Asymptotics for $\xi\to \pm \infty$.} \rev{ If we assume smooth (e.g., power law)  decay in $f$ and it's derivative as $\xi \rightarrow  \pm \infty$, then in this limit  the quadratically nonlinear r.h.s. of  \e{CLM01selfsimilar2} will be subdominant to the  linear terms on the left hand side. This implies that
Eq. \e{fasymp} describes the decay of $f$  for  $\xi\rightarrow{\pm\infty}$ provided $\alpha>0$, in agreement with the exact results of Section  \ref{sec:a0} (Eq. \e{omegaser1poles}) and Section \ref{sec:a1p2} (Eq. \e{omegasera1p2}) for $\alpha=1$ and $\alpha=1/3$, respectively. For $\alpha<0$, 
the assumed smooth decay of $f$ as $\xi \rightarrow \pm \infty$ is inconsistent with \e{fasymp}. This suggests that
\begin{equation}\label{fanympalphanegative}
f(\xi)\equiv 0 \ \text{at} \ \xi\to \pm \infty \ \text{for} \ \alpha<0,
\end{equation}
so  that $f(\xi)$ has the finite support for $\alpha<0$. This is consistent with Ref. \cite{ChenHouHuang} which considers the particular case $\alpha=-1$. }

Eq. \e{CLM01selfsimilar} is invariant under a stretching of the self-similar coordinate $\xi$,
\begin{equation}\label{stretching}
\xi\to A\xi, \ A=const\in \R.
\end{equation}
 i.e., if  $f(\xi)$ is a solution for Eq. \e{CLM01selfsimilar} then $f({A\xi}{})$ is also a solution of the same equation.
 Therefore if one finds a solution of Eq. \e{CLM01selfsimilar} then it immediately implies an infinite family of solutions from the stretching \e{stretching}. Despite this nonuniqueness, we find that the version of GPM employed here converges to a solution of Eqs. \e{CLM01selfsimilar2}, \e{fanympalphanegative}. Further details are given in Section \ref{sec:e_value_problem}.

\section{Transformed version of the equation} \label{sec:transform}

The analysis of previous sections assumes  the solution exists on the real line $x\in(-\infty, \infty)$ with the decaying BC \e{omegadecay}.  To address this infinite domain  in simulations, we use the auxiliary (computational) variable  $q$ defined by  \begin{equation}\label{transform}
x=\tan \left ( \frac{q}{2} \right ).
\end{equation}
Eq. \e{transform} maps the segment of the real line  $(-\pi,\pi)$  of $q$ onto the real line $(-\infty, \infty) $  of $x.$ Extending both $x$\ and $q$ into the complex plane, we find that Eq. \e{transform} maps the infinite strip   $-\pi <Re(q)<\pi$ onto the complex plane $x\in\C$, except for the half-lines $(-\I\infty,-\I)$ and $(\I,+\I\infty)$, with the upper half-strip being mapped onto the upper half-plane $\C^+$ and the lower half-strip being mapped onto the lower half-plane $\C^-$. Also the boundaries of the strip, $Re(q)=\pm\pi$ are mapped onto $(-\I\infty,-\I)$ and $(\I,+\I\infty)$, see e.g., Refs.
\cite{DyachenkoLushnikovKorotkevichPartIStudApplMath2016,LushnikovDyachenkoSilantyevProcRoySocA2017} for details of this mapping.
Here and below we abuse notation and use the  same symbols for functions of either  $x$ or $q$. For example,      we  assume that  $\tilde f(q):= f(x(q)) $ and remove the $\tilde ~$ sign.

Using the Jacobian of the mapping \e{transform},
\begin{equation}\label{transform_jacobian}
\frac{dx}{dq}=\frac{1}{2\cos^2(\frac{q}{2})}=\frac{1}{1+\cos{q}},
\end{equation}
and the results of Appendix \ref{sec:AppendixTransformationHilberttransform}, we rewrite
Eqs. \e{CLM01}-\e{HilbertHdef}  for independent variables $q$ and $t$ as
\begin{equation}\label{CLM_transformed}
\begin{split}
&\omega_t=-a (1+\cos{q}) u\omega_q +\omega[ {\mathcal H}^{2\pi}\omega+ C_\omega^{2\pi}],   \quad q\in (-\pi,\pi), \\
&(1+\cos{q})u_q=[ {\mathcal H}^{2\pi}\omega+ C_\omega^{2\pi}],
\end{split}
\end{equation}
where the Hilbert transform $ {\mathcal H}^{2\pi}$ on the interval $(-\pi,\pi)$ is defined by (see also Appendix \ref{sec:AppendixTransformationHilberttransform}) %
\begin{equation} \label{HilbertHdef_periodic}
{\mathcal H}^{2\pi} f(q):= \frac{1}{2\pi} \text{p.v.} \int^{\pi}_{-\pi}\frac{f(q')}{\tan(\frac{q-q'}{2})}\D q' ,
\end{equation}
and the constant $ C_\omega^{2\pi}$ is determined by%
\begin{equation}\label{C2pidef}
 C_\omega^{2\pi}=-\frac{1}{2\pi}\int^{\pi}_{-\pi}\omega(q')\tan(\frac{q'}{2})\D q'.
\end{equation}

We call Eq. \e{CLM_transformed} the transformed CLM equation.
 Note that Eq. \e{HilbertHdef_periodic} is the reduction of Eq. \e{HilbertHdef} to the class of $2\pi$-periodic functions, see Appendix \ref{sec:AppendixTransformationHilberttransform}. The decaying BC  \e{omegadecay}  allows a $2\pi$-periodic extension of $\omega(q,t)$  with $\omega(q,t)|_{q=\pi+2\pi n}=0, n\in \N.$  It enables us to work with  $\omega(q,t)$ in terms of a  Fourier series over $q.$

\section{Results of time dependent simulations on the real line} \label{sec:numerics}

Based on the results of Section \ref{sec:transform},  we numerically solve Eq. \e{CLM_transformed} on the real line $x\in \R$ with a pseudo-spectral Fourier method by representing the  $2\pi$-periodic solution $\omega(q,t)$ as a sum of $2N$ Fourier modes $\hat{\omega}_k(t)$ as
\begin{equation}\label{solution_Fourier}
\omega(q,t)= \sum_{k=-N}^{k=N-1}{\hat{\omega}_k (t)e^{\I kq}}.
\end{equation}
We use $2N$ uniformly spaced grid points in $q$ from $-\pi$ to $\pi-\Delta q$, where $\Delta q=\pi/N$. The  Fast Fourier transform (FFT) allows us to efficiently find numerical values of  $\hat{\omega}_k(t)$  from values of   $\omega(q,t)$  on that grid. The resolution  $N$ is chosen depending on the initial condition (IC) and adaptively adjusted throughout the computation so that the spectrum $\hat{\omega}_k$  is fully resolved with the desired precision.  This  means that  $|\hat{\omega}_k|$  decays  by 16-17 orders of magnitude at $|k| \sim N$  compared to $\max\limits_{-N\le k\le N-1}|\hat{\omega}_k|$, down to the round-off floor of the error for  double precision. For the  multi-precision simulations which were performed, this decay is further enhanced (or equivalently, the round-off is reduced)  by any desired number of orders. Below we focus on the description of double precision simulations while noting that  higher precision simulations were also extensively performed.

The decay of the Fourier spectrum $\hat{\omega}_k$  is checked at the end of every time step. If $|\hat{\omega}_k|$ is larger than the numerical round-off at $|k| \sim N$ at the given time-step, then the simulation is `rewound' for one time-step backwards with $N$  increased by factor of 2, and the time-stepping is continued. Amplitudes of the new extra Fourier modes are set to $0$, which is equivalent to performing a spectral interpolation of the solution at the newly inserted grid points in $q$ space.  Rewinding is done to avoid accumulation of error due to the tails of the spectrum not being fully resolved at the time-step before the grid refinement. For time-marching we use 11-stage explicit Runge-Kutta method of $8^{th}$ order \cite{RK8_CooperVerner72} with the adaptive time-step $\Delta t$ determined by the condition $\Delta t=\text{CFL}\cdot\min\{\Delta q/(a\max\limits_q|(1+\cos{q})u|),1/\max\limits_q|(1+\cos{q})u_q|\}$, where the numerical constant $\text{CFL}$ is typically chosen as $\text{CFL}=1/4, 1/8 \text{ or } 1/16$ to achieve  numerical stability in the  time-stepping and ensure that the error of the method is near round-off level.
Also, the scaling of $\Delta t$ with $\max\limits_q|(1+\cos{q})u|$ and $\max\limits_q|(1+\cos{q})u_q|$ ensures numerical stability of the method during possible singularity formation events. We additionally  enforced the real-valuedness of $\omega(q)$  at each time-step to avoid numerical instability, since the FFT and inverse FFT  lead to accumulation of a small imaginary part at the level of round-off, which can be amplified during time evolution.

Typically, we used the following two types of initial conditions (ICs):
\begin{eqnarray}
\text{IC1:} \quad \omega_0(q) &=& -(\sin(q)+0.1\sin(2q)), \label{IC1} \\
\text{IC2:} \quad \omega_0(q) &=&\I \frac{ 4V_c^2}{3T_c} \left (\frac{1}{(\tan(\frac{q}{2})-\I V_c)^2} -\frac{1}{(\tan(\frac{q}{2})+\I V_c)^2} \right), \label{IC2} 
\end{eqnarray}
where the real-line IC1 is similar in form  to the periodic IC in Ref. \cite{Okamoto2008} except for
an opposite sign.  In IC2, $V_c$ and $T_c$ are real numbers and in most of our simulations we used
$V_c=1$, $T_c=1$, for which  IC2 reduces to
\begin{equation}\label{IC2a}
\omega_0(q)=-\frac{4}{3}(\sin(q)+0.5\sin(2q)).
\end{equation}
  Note the first two derivatives of (\ref{IC2a})
are zero at $q=\pm \pi$, i.e., $\omega_0^{(n)}(q=\pm \pi)=0 \text{ for }
n=0,1,2$. Both ICs \e{IC1} and \e{IC2} are real-valued odd functions with a negative slope at
$q=0$, and lead to the formation of a singularity   at $q=0$ at some moment
in time  for $a<a_c$ (see Eq. \e{acvalues} for the definition of $a_c$)
while $\omega(q,t)$ stays real-valued and odd.
The function $\omega_0(q)$ in IC1 is an entire function, and that in IC2 has two double poles at $x=\tan \left (
\frac{q}{2} \right )=\pm \I V_c$ in $x$-space or at $q=\pm \I q_c$ in
$q$-space, where $q_c=2\, \text{arctanh}(V_c).$ Note that  IC2 corresponds to
the   exact solution for the case $a=1/2$ with a collapse at $t=T_c$ (see
Eq. \e{omegasera1p2}), while for other values of the parameter $a,$ it is not
an exact solution but qualitatively resembles one on the real interval
$[-\pi,\pi]$ and serves as a good IC to obtain collapsing
solutions.

Computation of the $2\pi$-periodic Hilbert transform ${\mathcal H}^{2\pi}$ (see Appendix \ref{sec:AppendixTransformationHilberttransform} for the definition of  ${\mathcal H}^{2\pi}$) is easily done in Fourier space as
\begin{equation}\label{HilbertHdef_periodic_k}
\hat{{\mathcal H}}^{2\pi}_k = -\I \, \text{sign}(k), 
\end{equation}
where $ \text{sign}(k)=1$ for $k>0$, $\text{sign}(k)=0$ for $k=0$ and $\text{sign}(k)=-1$ for $k<0$.
 Also the constant $C_\omega^{2\pi}$ \e{C2pidef} in Eq. \e{CLM_transformed} is computed  from the condition that  ${\mathcal H}^{2\pi} \omega(q=-\pi) + C_\omega^{2\pi} =0$, i.e. $ -\I \sum\limits_{k=-N}^{k=N-1}{\hat{\omega}_k (-1)^{-k}\text{sign}(k) + C_\omega^{2\pi}= 0}$.

While computing the values of $u_q$ from the second equation in
\e{CLM_transformed}, one has to take special care at the point $q=-\pi$.
Expanding both the left-hand side (l.h.s.) and r.h.s. of that equation in a
Taylor series at the point $q=-\pi,$ we obtain that $u_q(q=-\pi)={\mathcal
H}^{2\pi}_{qq}\omega(q=-\pi)$, which can also be computed using
$\hat{\omega}_k$. The term with ${\mathcal H}^{2\pi}_q$ in the Taylor
series of the r.h.s. vanishes since ${\mathcal
H}^{2\pi}_q\omega(q=-\pi)=\sum_k{|k|\hat{\omega}_k}=0$ for the real-valued
odd function $\omega(q)$
\rev{with $\hat{\omega}_{-k}=-\hat{\omega}_k$.}

For each simulation  we made a  least squares fit of the Fourier spectrum $|\hat{\omega}_k|$ at time $t$ to the asymptotic decay model  %
\begin{equation}\label{deltakmodel}
|\hat{\omega}_k(t)| \approx C(t)\frac{ e^{-\delta(t)|k|}}{|k|^{p(t)}}
\end{equation}
for  $|k|\gg1$ \cite{CarrierKrookPearson1966}, where $C(t), \, \delta(t)$ and $p(t)$ are the fitting parameters for each value of $t$.
 This allows us to obtain both $\delta(t)>0$ and $p(t)$ as functions of $t$.
The value of $\delta(t)$ indicates the distance  of the closest singularity of $\omega(q)$ from the real line in the complex $q$-plane, and the value of $p(t)$ is related to
the type or power  of  that complex singularity, see  Refs.
\cite{Okamoto2008,DyachenkoLushnikovKorotkevichJETPLett2014,DyachenkoLushnikovKorotkevichPartIStudApplMath2016,SulemSulemFrischJCompPhys1983} for more details.
In particular, if the singularity in the solution is of a power law type  $\omega(q) \sim (q-\I q_c)^{-\gamma}$ then using complex contour integration one obtains
(see e.g. Ref. \cite{CarrierKrookPearson1966})  that
$|\hat{\omega}_k| \approx C e^{-q_c |k|}/|k|^{1-\gamma}$, meaning that $\delta=q_c$ and%
\begin{equation}\label{pgamma}
p=1-\gamma
\end{equation}
 which follows from Eq. \e{deltakmodel}. According to Eq. \e{transform}, the distance $\delta_x$ from the closest singularity to the real line in the complex $x$-plane is  $\delta_x=\tanh{\left(\frac{\delta}{2}\right)}$. It implies that  $\delta_x = \frac{\delta}{2}+O(\delta^3)$ for  $\delta\ll 1$.

\begin{figure}
\includegraphics[width=1\textwidth]{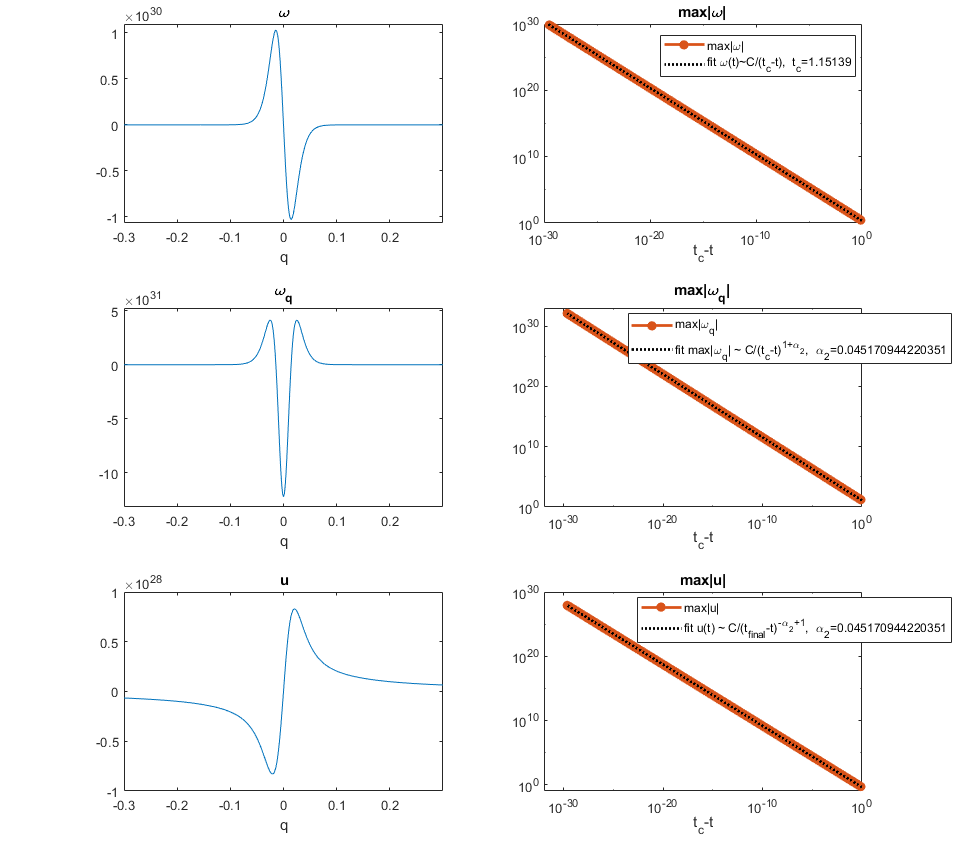}
\caption{Results of the simulation of Eqs. \e{CLM_transformed}-\e{HilbertHdef_periodic} with $a=2/3$ and initial condition IC2 \e{IC2a}. Left panels:  the  solution $\omega(q,t)$, its derivative $\omega_q(q,t)$ and $u(q,t)$ for $t=1.15139$. Right panels: the time dependence of maximum values of these functions. Dashed lines show the prediction of Eq.  \e{self-similar1}  with $\alpha_2$ extracted from the simulations as explained in the text. The collapse time $t_c$ is extracted from the fit (by extrapolation) to $\max|\omega(x,t)| \propto \frac{1}{(t_c-t)}$.  
} \label{fig:a2p3_6pics}
\end{figure}

\begin{figure}
\includegraphics[width=0.328\textwidth]{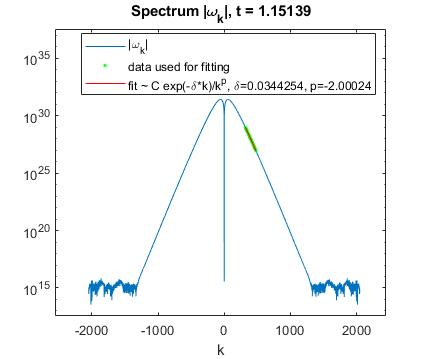}
\includegraphics[width=0.328\textwidth]{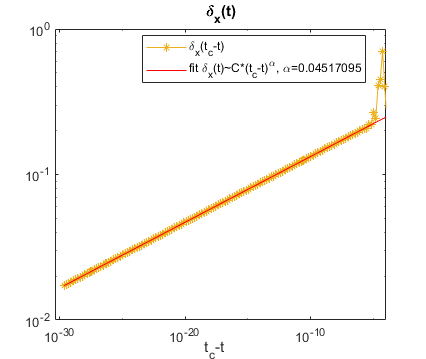}
\includegraphics[width=0.328\textwidth]{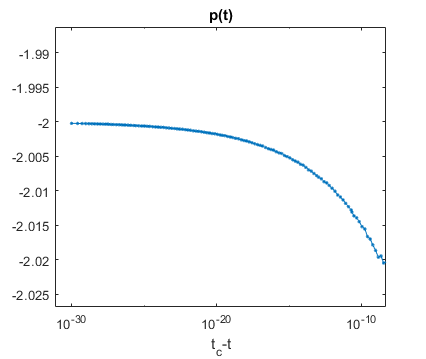}
\caption{Left panel: The Fourier spectrum $|\hat{\omega}_k|$ at a
particular time $t=1.15139$ from the same simulation as in Fig.
\ref{fig:a2p3_6pics} with  $a=2/3$. The red line represents a fit to the
model \e{deltakmodel} with green line showing the portion of the
$|\hat{\omega}_k|$ used for the least-squares  fit. Center and right
panels: time dependence of
$\delta_x(t)=\tanh{\left(\frac{\delta(t)}{2}\right)}$ and  $p(t)$ recovered from the fit of the spectrum to Eq.  \e{deltakmodel}  at different times. The red solid line at the center panel represents a fit to the model $\delta_x(t) \sim (t_c-t)^\alpha.$  
} \label{fig:a2p3_spectrum}
\end{figure}

Results of a simulation with the parameter value $a=2/3$ and IC2 with  $V_c=1$, $T_c=1$ (i.e. Eq. \e{IC2a}) are provided in Figs.  \ref{fig:a2p3_6pics} and \ref{fig:a2p3_spectrum}. The maximal value  $\max\limits_q|\omega(q,t)|$  of the numerical solution increases from an initial value $\sim 1$ up to $\sim 10^{30}$ at the final simulation time.
Fig. \ref{fig:a2p3_spectrum} shows the spectrum $|\hat{\omega}_k|$ and its fit to the model \e{deltakmodel}. This fit provides numerically extracted values of both $\delta(t)$ and $p(t)$. Then $\delta_x(t)=\tanh{\left(\frac{\delta(t)}{2}\right)}$ is computed from $\delta(t)$ and fitted to $\delta_x(t) \propto (t_c-t)^\alpha$, per Eq. \e{self-similar1}, to determine $\alpha$. \rev{ We first obtain an estimate for $t_c$  from a fit to $\max\limits_x|\omega(x,t)| \propto \frac{1}{(t_c-t)}$ by extrapolating the numerical solution up to $t=t_c$.}  From these fits  we obtain that $\alpha \approx 0.04517095$, giving the temporal rate of singularity approach to the real line in  complex $x$-space. The algebraic decay rate $p(t)$ appears to stabilize at the value $-2$ as $t$ approaches the singularity time $t_c$. 
An initial transient is not included in the  data used for the $\delta_x(t)$ fit, since  $\delta(t)$ and $p(t)$ cannot be determined  accurately at these times due to the spectrum $|\hat{\omega}_k|$ being oscillatory. These oscillations quickly die out as the self-similar regime is approached.

\revmike{Addressing Ref. 2 comments 5 and 6, we removed the paragraph below starting with `In order', but added a short explanation of the estimate of $t_c$ to the paragraph above.}

\revc{
In order to extract $\alpha$ from the fit $\delta_x(t) \propto (t_c-t)^\alpha$ with maximum accuracy, we need an accurate estimate of $t_c$. We obtain this estimate from the fit $\max\limits_x|\omega(x,t)| \propto \frac{1}{(t_c-t)}$ by extrapolating the numerical solution up to $t=t_c$. Using this $t_c$ for the fit instead of $t_{final}$ (that one may naively use instead of the correct $t_c$)
to perform the fit $\delta_x(t) \propto (t_c-t)^\alpha$
provides considerably more accurate values of $\alpha$, for example,  giving 7 significant digits instead of 4 when $a=2/3$.
Having a more accurate estimate of $t_c$ is especially important for simulations with $a<0.3$ where the spectrum widens more quickly than it grows in amplitude, so there is less of a useful range of data for fitting.
}

We find that we get the best accuracy for $\delta$ and $p$ from the fit of $|\hat{\omega}_k|$ to the model \e{deltakmodel} if we confine the least square fit to a window of data between 1/4 and 1/3 of the total effective width of the spectrum (shown on the left part of Fig. \ref{fig:a2p3_spectrum} with a green color). This is due to an increase in the relative error of the spectrum data at the tails, as the round-off floor is approached. Moreover,  the model \e{deltakmodel} is accurate only asymptotically as $|k|\to \infty$ so we cannot use too small values of $|k|$. \revmike{Some very minor edits to this paragraph.}

\begin{figure}
\includegraphics[width=0.5\textwidth]{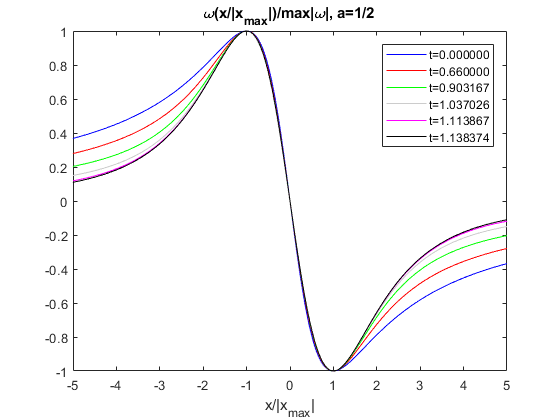}
\includegraphics[width=0.5\textwidth]{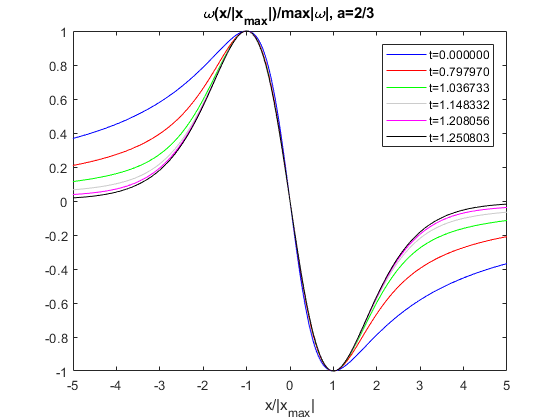}
\caption{ Convergence of time-dependent numerical solution of Eqs.
\e{CLM_transformed}-\e{HilbertHdef_periodic} with $a=1/2$ (left panel) and
$a=2/3$ (right panel) to the self-similar solution  \e{self-similar1}. In
both cases we used IC1 \e{IC1}. Solutions  shrink horizontally and
increase in amplitude  vertically until collapse  occurs at $t=t_c$,
$t_c\approx1.180602237542$ (left panel) and $t_c\approx1.272876000077$
(right panel). Solutions are plotted in $x$-space, where
$x=\tan(\frac{q}{2})$. Horizontal and vertical scales are dynamically
changed in both panels to exactly match the positions and amplitudes of
the local maximum at $x=x_{max}$ and minimum at $x=-x_{max}$. }
\label{fig:selfsimilarconvergence}
\end{figure}

\begin{figure}
\includegraphics[width=0.5\textwidth]{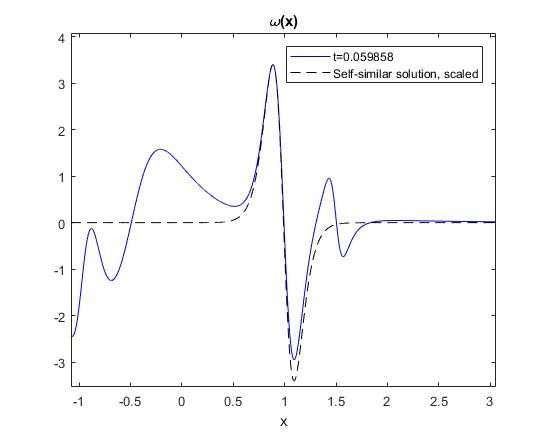}
\includegraphics[width=0.5\textwidth]{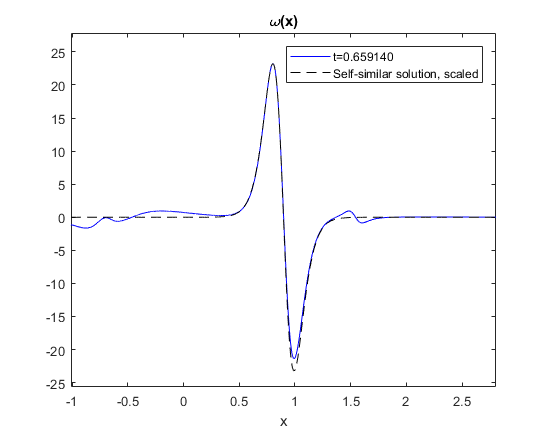}
\caption{ Convergence of the  time-dependent numerical solution of Eqs.
\e{CLM_transformed}-\e{HilbertHdef_periodic} to the self-similar profile
\e{self-similar1} as $t \rightarrow t_c$. Here $a=2/3$ and we use the
generic initial condition $\omega_0(x) = - \frac{1}{64}
\left(\frac{1}{(x-x_1^+)^3} + \frac{1}{(x-x_1^-)^3} \right) - \frac{\I}{3}
\left(\frac{1}{(x-x_2^+)^2} - \frac{1}{(x-x_2^-)^2} \right) + \frac{1}{32}
\left(\frac{1}{(x-x_3^+)^3} + \frac{1}{(x-x_3^-)^3} \right) +
\frac{\I}{96} \left(\frac{1}{(x-x_4^+)^2} - \frac{1}{(x-x_4^-)^2}
\right),$ where $x_1^\pm=-1 \pm \frac{\I}{4}, x_2^\pm=-\frac{1}{2} \pm
\frac{\I}{2}, x_3^\pm=1 \pm \frac{\I}{4}, x_4^\pm=\frac{3}{2} \pm
\frac{\I}{8}.$ The solution is shown at two different moments in time,
where for each time we overlaid the self-similar profile as in Fig.
\ref{fig:selfsimilarconvergence},  matching their corresponding maximum
and minimum positions horizontally and  vertically.}
\label{fig:evolution_generic_IC}
\end{figure}

\begin{figure}
\centering
    \includegraphics[width=0.5\linewidth]{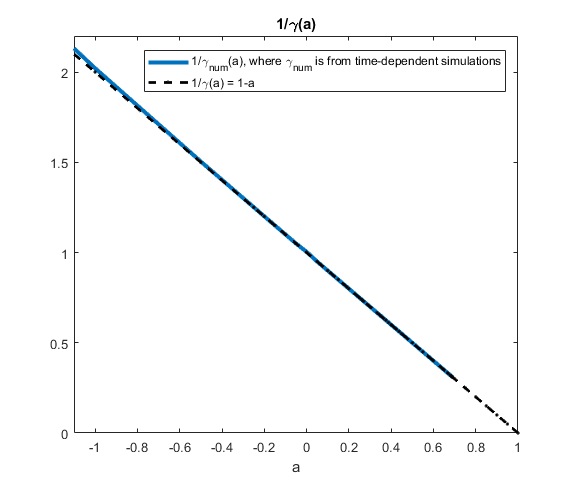}
    \caption{Dependence of $\gamma_{num}(a)=1-p(a)$ using $p(a)$ obtained via time-dependent simulations by the fit to Eq. \e{deltakmodel}. These data are also provided in Table \ref{Table1}. Also shown is  $\gamma(a)=\frac{1}{1-a}$ from Eq. \e{gammaa} for comparison. Here we plot $1/\gamma(a)$ instead of $\gamma(a)$ for the easier comparison. }  \label{fig:gamma(a)}
\end{figure}

For $0\leq a < a_c$ (with $a_c$ given by Eq. \e{acvalues}) and for both  IC1 \e{IC1} and IC2 \e{IC2}, we find that $\delta_x(t)$ evolves in time toward 0  while $p(t)$ approaches a constant value after a quick transient phase, see Fig. \ref{fig:a2p3_spectrum} (right panel). We  observe  spontaneous formation of a universal self-similar solution profile of the form \e{self-similar1} during time evolution
(see Fig. \ref{fig:selfsimilarconvergence}). These self-similar profiles, as well as the value of $\alpha$ in $\delta_x(t)$ and the terminal value of $p(t)$ as $t\rightarrow{t_c}$ are the same for a wide class of ICs (e.g. one can change  a power of singularity in IC2 from $-2$ to any negative number below $-2$ and/or change numerical values of both $V_c>0$\ and $T_c>0$). Thus, these self-similar profiles are only functions of the parameter $a$.
Table \ref{Table1} provides the universal values of $\alpha$ and $p$ vs $a$. Fig. \ref{fig:alpha(a)} shows the dependence of  $\alpha(a)$ on $a$.
However, one can also find particular IC  in which finite time singularities do not form. Two such choices are  -IC1  and -IC2, i.e. IC1 \e{IC1} and IC2 \e{IC2} taken with the opposite sign.  In these two cases we did not observe collapse or singularity formation  in finite time, but rather an algebraic-in-time approach of a singularity to the real line, $\delta_x(t) \sim 1/t^\mu, \mu>0$.
\revc{The low-order derivatives of $\omega(q)$  were found to either decay  algebraically in time or approach constant values while higher-order derivatives of $\omega(q)$ grow algebraically in time.}
Other smooth generic initial conditions that were tried were found to produce   blow up after an initial  transient, as exemplified in Fig. \ref{fig:evolution_generic_IC}. These transients made the simulation considerably slower (due to the need for more modes in the spectrum of  to resolve the solution  down to double precision round-off). However, in a space-time neighborhood of the singularity these solutions recover the same self-similar profile as shown in Fig. \ref{fig:selfsimilarconvergence}, see also  Fig. \ref{fig:evolution_generic_IC}.  We note that the velocity $u(x,t)$ evolves toward the self-similar profile \e{uselfsimilar} with $\max\limits_x|u|\to \infty$  for $0<a< a_c$.  Below we focus on  IC1 and IC2, but the reader should  but keep in mind that they appear  generic.

\begin{table} [h]
\centering
  \caption{Table of values of $\alpha$, $p$ and $\alpha_2$ extracted via fits to $\delta_x(t)$, $|\hat{\omega}_k|$ and $\max|\omega_x(x,t)|$  in time-dependent simulations of Eqs. \e{CLM_transformed}-\e{HilbertHdef_periodic} for various values of $a$. Also shown are  values of $\alpha_e$ and $\beta$ obtained from eigenvalue problem simulations of Eqs. \e{CLM01selfsimilar_transformed} and \e{HilbertHdef_periodic} described in Section \ref{sec:e_value_problem}. 
  Accuracy of $\alpha(a)$ (for $-1\leq a\leq0.689$) and $\alpha_2(a)$ (for $-1\leq a\leq0.689066533$) is 
  at least 3-4 digits of precision, whereas accuracy of $\alpha_e(a)$ is about 3-4 digits of precision for $a < 0.3$ and at least 5 digits of precision for $a \geq 0.3$, with more precision for $0.3\leq a \leq 0.6890665$.}
\resizebox{\textwidth}{!}{%
\begin{tabular}{|l|l|l|l|l|l|}
  \hline
  $a$ & $\alpha_e$ & $\beta$ & $p$ & $\alpha$ & $\alpha_2$ \\
  \hline
-5&-&-&0.855 & 7.495 &7.517 \\
-2&-&-&0.680 & 3.444 & 3.422  \\
-1&-&-&0.505 & 2.208 & 2.206 \\
-0.5&-&-&0.335120 & 1.603747 & 1.600222 \\
-0.25&1.296593455&-&0.200942 & 1.303708 & 1.302424  \\
-0.2&1.239824952&-&0.167139 &1.243558 & 1.242436 \\
-0.15&1.181358555&0.133308&0.130811  & 1.183300 & 1.182701  \\
-0.1&1.121312899&0.100401&0.091110 & 1.122630 & 1.122093 \\
-0.05&1.061051829&0.060633&0.047696  & 1.061617 & 1.061334\\
0&1&0&0.004 & 1.000243 & 1.000019 \\
0.05&0.938365701&-0.070205&-0.052759 & 0.938381 & 0.938288  \\
0.1 &0.876129662&-0.136336&-0.111326 & 0.876329 & 0.876309  \\
0.15&0.813179991&-0.240380&-0.176727 & 0.813219 & 0.813215  \\
0.2 &0.749369952&-0.338799&-0.250265 & 0.749519 & 0.749549  \\
0.25&0.684513621&-0.460507&-0.333582 & 0.684650 & 0.684671  \\
0.265&0.664818990&-0.500444&-0.360765 & 0.664827 & 0.664830  \\
0.3&0.618374677&-0.610349&-0.428762 & 0.618375 & 0.618377  \\
0.35&0.550648498&-0.787978 &-0.538583 & 0.550661 & 0.550655  \\
0.4  &0.480939257&-0.939823&-0.666732 & 0.4809431 & 0.4809429   \\
0.425&0.445184823&-0.97452&-0.739156  & 0.4451863 & 0.4451860   \\
0.4375&0.427049782&-0.993899&-0.777804& 0.4270512 & 0.4270508  \\
0.45&0.408728507&-1&-0.818193         & 0.40872820 & 0.40872838 \\
0.5&0.333333333&-1&-1.0000007         & 0.33333354 & 0.33333340 \\
0.55&0.253852136994&-1&-1.222218      & 0.25385226 & 0.25385213 \\
0.6&0.169098936470&-1&-1.4999991      & 0.16909915 & 0.1690989367 \\
0.65&0.077532635626630 &-1&-1.857141  & 0.07753269 & 0.07753263562662 \\
2/3&0.045170944220367&-1&-1.999997    & 0.04517096 & 0.04517094422035 \\
0.68&0.018526534283004&-1&-2.125013      & 0.01852675  & 0.01852653428270 \\
0.685&0.008351682345844 &-1&-2.175083    & 0.00835210  & 0.008351682345843  \\
0.689&0.000137203824593 &-1&-2.219165    & 0.00013724   & 0.000137203824603 \\
0.68905&  3.409705703117e-05&-1&-2.221589   & 3.4145e-05 & 3.4097057039e-05 \\
0.68906&  1.347443362884e-05&-1&-2.220924   & 1.3418e-05 & 1.3474433654e-05 \\
0.689066& 1.10065641e-06&-1&-2.221505       & 1.0808e-06 & 1.1006564176e-06 \\
0.6890665&6.950143e-08&-1&-2.223142&-                    & 6.9501438524e-08 \\
0.68906653&7.632094e-09&-1&-2.222128&-                   & 7.6321058379e-09 \\
0.689066533&1.445152e-09&-1&-2.220519&-                  & 1.4451679770e-09 \\
0.6890665335&4.13992e-10&-1&-2.205923&-                  & 4.1401557848e-10 \\
0.6890665337&1.537e-12&-1&-2.220897&-&1.5519e-12 \\
0.6890665337007&9.43093e-14&-1&-2.227272&-&1.1097e-13 \\
0.68906653370074&1.18169e-14&-1&-2.222533&-&2.7574e-14 \\
0.689066533700745&1.505397e-15&-1&-2.221208&-&1.4711e-14 \\
0.6890665337007457&6.169686e-17&-1&-&-&- \\
0.7&-&-&-&-&-0.02281 \\
0.75&-&-&-&-&-0.13435 \\
0.8&-&-&-&-&-0.26008 \\
0.85&-&-&-&-&-0.40384 \\
0.9&-&-&-&-&-0.57118 \\
0.95&-&-&-&-&-0.76643 \\
1&-&-&-&-&-1.000000056 \\
  \hline
\end{tabular}} \label{Table1}
\end{table}

Using the terminal values of $p$ extracted by fits  to Eq. \e{deltakmodel}
 with various $a$, and employing Eq. \e{pgamma} to recover $\gamma$ from $p$, we confirmed the formula $\gamma(a)=\frac{1}{1-a}$ (see Theorem
1 and Eq. \e{gammaa} in Section \ref{sec:main}) and the corresponding
formula $p(a)=\frac{-a}{1-a}$ within 0.5\% for $0\leq a < a_c$.  Fig.
\ref{fig:gamma(a)} shows the numerical approximation, $\gamma_{num}(a)=1-p(a)$ using values of $p(a)$
from Table \ref{Table1} as well as the theoretical value  $\gamma=\frac{1}{1-a}$ for
comparison. We note that  the plot of $1/\gamma_{num}(a)$ in Fig. \ref{fig:gamma(a)} stops at $a=a_c$, since it is difficult to obtain accurate values of $p(a)$ (and hence $\gamma_{num}(a)$) from time-dependent simulations when $a>a_c$. This is due to a transition that occurs at $a=a_c$, in which the fitted singularity for $a<a_c$ corresponding to collapse is no longer closest to the real-$x$ line when $a>a_c$.

In addition to Fourier fitting,  we also extract values of $\alpha$ in an alternative way (these values are called $\alpha_2$ below), using the spatial derivative of the self-similar solution \e{self-similar1} given by \begin{equation}\label{self-similar2}
\omega_x(x,t)=\frac{1}{(t_c-t)^{1+\alpha}}f'\left( \frac{x}{(t_c-t)^\alpha} \right).
\end{equation}
Using Eq. \e{self-similar2} we fit $\max\limits_x|\omega_x(x,t)|$ to the model $\max\limits_x|\omega_x(x,t)| \propto \frac{1}{(t_c-t)^{1+\alpha_2}}$ to find $\alpha_2.$
\revmike{We removed this highlighted portion in response to referee 2 comment 6 recommending to omit some of the details in describing the different methods for computing $\alpha$:}
\revc{The same scaling is valid for $\max\limits_q|\omega_q(q,t)|$ since the maximum of $|\omega_q|$ (or $|\omega_x|$) is at $x=q=0$ for an odd collapsing solution.  Fig. \ref{fig:a2p3_6pics} (center  right panels) shows such a fit for $a=2/3$, where  $\alpha_2\approx 0.045170944220351.$ For comparison,  $\alpha$ extracted via $\delta_x(t)$ is $\alpha \approx 0.045170954667478$ (Fig. \ref{fig:a2p3_spectrum}, center panel). These values of $\alpha$ and $\alpha_2$ have errors of only $1\times10^{-8}$ and $1.6\times10^{-14}$, respectively  (we compared them to $\alpha_e=0.04517094422036721851569165527172$ that was obtained using a different method described in Section \ref{sec:e_value_problem} together with 32 digit 
arithmetic).}
Values of $\alpha_2$ for various $a$ are also gathered in Table \ref{Table1} for comparison with values of $\alpha$.
We confirmed that $\alpha$ and $\alpha_2$ obtained using the above two methods for $0<a\lesssim0.689$ agree within  a relative error of $< 0.02\%$.

\begin{figure}
\includegraphics[width=1\textwidth]{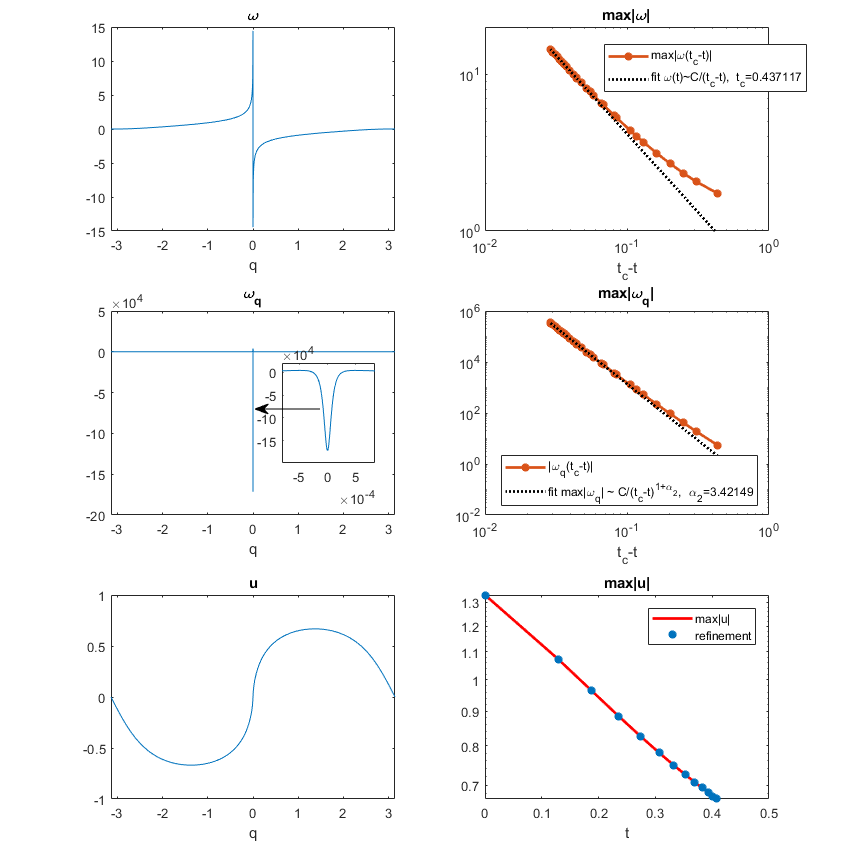}
\caption{ Results from  simulations of Eqs. \e{CLM_transformed}-\e{HilbertHdef_periodic} with $a=-2$ and initial condition IC2 \e{IC2a}. Left panels:  the  solution $\omega(q,t)$, its derivative $\omega_q(q,t)$, and $u(q,t)$ for $t=0.407228$. Right panels:  Time dependence of the  maximum values of these functions.   Dashed lines show the prediction of Eq.  \e{self-similar1}  with $\alpha_2$ extracted from simulations as explained in the text. 
}
  \label{fig:a-2_6pics}
\end{figure}

\begin{figure}
\includegraphics[width=0.328\textwidth]{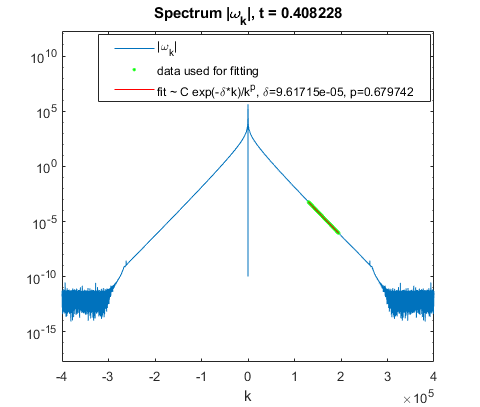}
\includegraphics[width=0.328\textwidth]{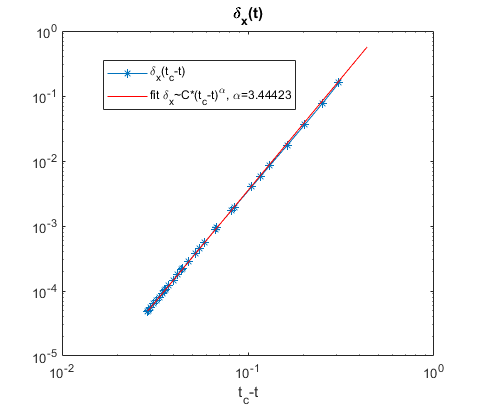}
\includegraphics[width=0.328\textwidth]{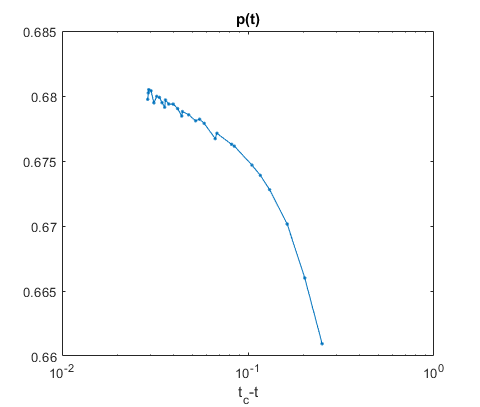}
\caption{Left panel: The Fourier spectrum $|\hat{\omega}_k|$ at  time
$t=0.407228$ from the same simulation as in Fig. \ref{fig:a-2_6pics}.  The
red line represents a fit to the model \e{deltakmodel} with green line
showing  portion of the  $|\hat{\omega}_k|$ used for the fit. Center and
right  panels: Time dependence of
$\delta_x(t)=\tanh{\left(\frac{\delta(t)}{2}\right)}$ and  $p(t)$
recovered from fit of the spectrum to Eq.  \e{deltakmodel}  at different
times. The red solid line at the center panel represents a fit to the
model $\delta_x(t) \propto (t_c-t)^\alpha.$ } \label{fig:a-2_spectrum}
\end{figure}

For $a < 0$ we observe a similar finite time blow up starting from both  IC1 and IC2 with $\max\limits_x|\omega| \rightarrow{\infty}$ as $t \rightarrow{t_c}$ according to the self-similar profile in Eq. \e{self-similar1}. %
 The extracted values of $\alpha$, $p$ and  $\alpha_2$ for $a < 0$ are also given in Table \ref{Table1}, see also Figs. \ref{fig:a-2_6pics} and \ref{fig:a-2_spectrum} for results of simulations with $a=-2$ and  IC2.
The velocity $u(x,t)$ during the temporal evolution approaches the self-similar profile \e{uselfsimilar} near the singularity location at $x=q=0$. 
A qualitative difference  for  $a < 0$  (in comparison with $0<a< a_c$) is that  the self-similar profile \e{uselfsimilar} approaches zero because $\alpha>1$ in the former case, while
away from the spatial singularity location 
the value of $u(x,t)$ is generally nonzero, even at $t\to t_c.$
This extends the result of \cite{CastroCordoba}, who proved that there is finite-time singularity formation for $a<0$ in the case of odd compactly supported data $\omega(x,0) \in C_c^\infty(\mathbb{R})$ with ${\mathcal H}\omega(0,0)>0$, to examples with analytic initial data.

\revc{The values of $\alpha(a)$ were challenging to obtain with more than 3-4 digits of accuracy using time-dependent simulations in double precision arithmetic, since it involves the fitting of several unknown coefficients, including $t_c$. Generally, fitting to $\max\limits_x|\omega_x(t)| \propto 1/(t_c-t)^{1+\alpha_2}$ provided more accurate results for $\alpha$ then obtaining $\alpha$ from fitting to $\delta_x(t)\propto (t_c-t)^\alpha$ (for $a<a_c$). This is especially true  for values of $a$ near $a_c$ since $\delta_x(t) \sim (t_c-t)^\alpha$ decays very slowly with $\alpha \rightarrow{0}$ as $a \rightarrow{a_c}$, while amplitudes satisfy  $\max\limits_x|\omega(t)| \propto 1/(t_c-t)$ and $\max\limits_x|\omega_x(t)| \propto 1/(t_c-t)^{1+\alpha_2}$. At $a=2/3$ ($\alpha=0.045\dots$) the quantity  $\delta_x(t)$ decays only by one order of magnitude over the simulation time (see Fig. \ref{fig:a2p3_spectrum}, center panel) even though $\max\limits_x|\omega(t)|$ grows from from the value $\sim 1$ up to $10^{30}$ (see Fig. \ref{fig:a2p3_6pics}, right panel, the first row).}

We  obtained much  more accurate values of $\alpha(a)$ (up to 14 digits of
precision) by numerically solving the nonlinear eigenvalue problem, Eq.
\e{CLM01selfsimilar2}, for a self-similar solution of  Eq. \e{CLM01}  (see
Section \ref{sec:e_value_problem}). In contrast, for $a_c$ we  were able to obtain 14 digits
of accuracy  using both time-dependent simulations and  the
nonlinear eigenvalue problem with double precision arithmetic. Another 3 digits of precision  are obtained (for a total of 17 digits of precision) if  quadruple precision arithmetic
is used in the nonlinear eigenvalue problem.

\begin{figure}
\includegraphics[width=0.328\textwidth]{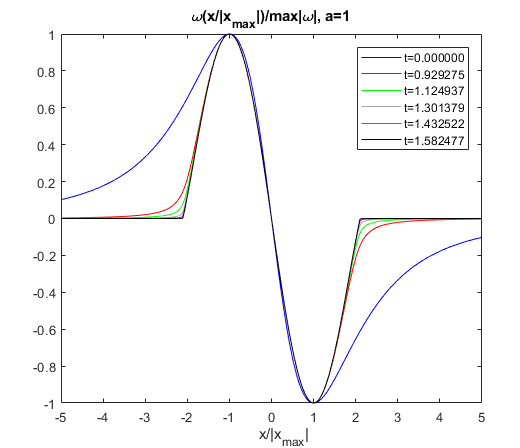}
\includegraphics[width=0.328\textwidth]{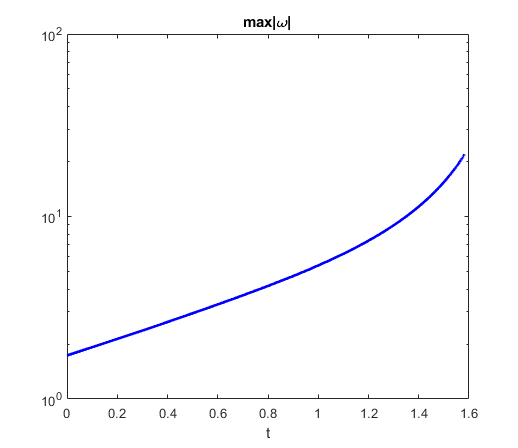}
\includegraphics[width=0.328\textwidth]{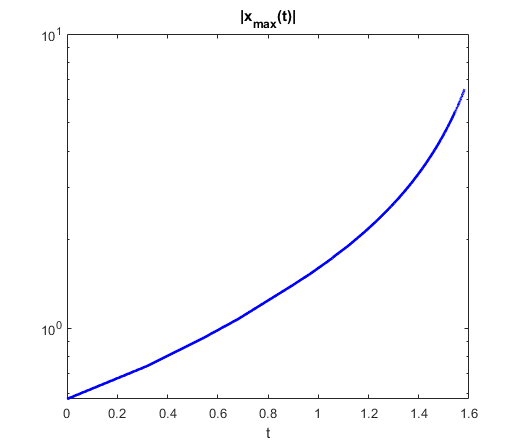}
\caption{Left panel: Convergence of  the time-dependent numerical solution
to Eqs. \e{CLM_transformed}-\e{HilbertHdef_periodic} with  $a=1$ and  IC2
\e{IC2a} to a self-similar profile with compact support.  The solution
expands horizontally and stretches vertically until blowing up at $t=t_c
\approx 1.77864$. The solution is plotted in $x$-space, where
$x=\tan(\frac{q}{2})$, and is scaled both horizontally and vertically to
exactly match the positions of the local maximum and minimum. Center and
right panels: The time dependencies of $\max\limits_x|\omega(x,t)|$ and
the absolute value its location $x_{max}(t)$ on $t$.}
\label{fig:a1_selfsimilarconvergence}
\end{figure}

\begin{figure}
\centering
\includegraphics[width=1\textwidth]{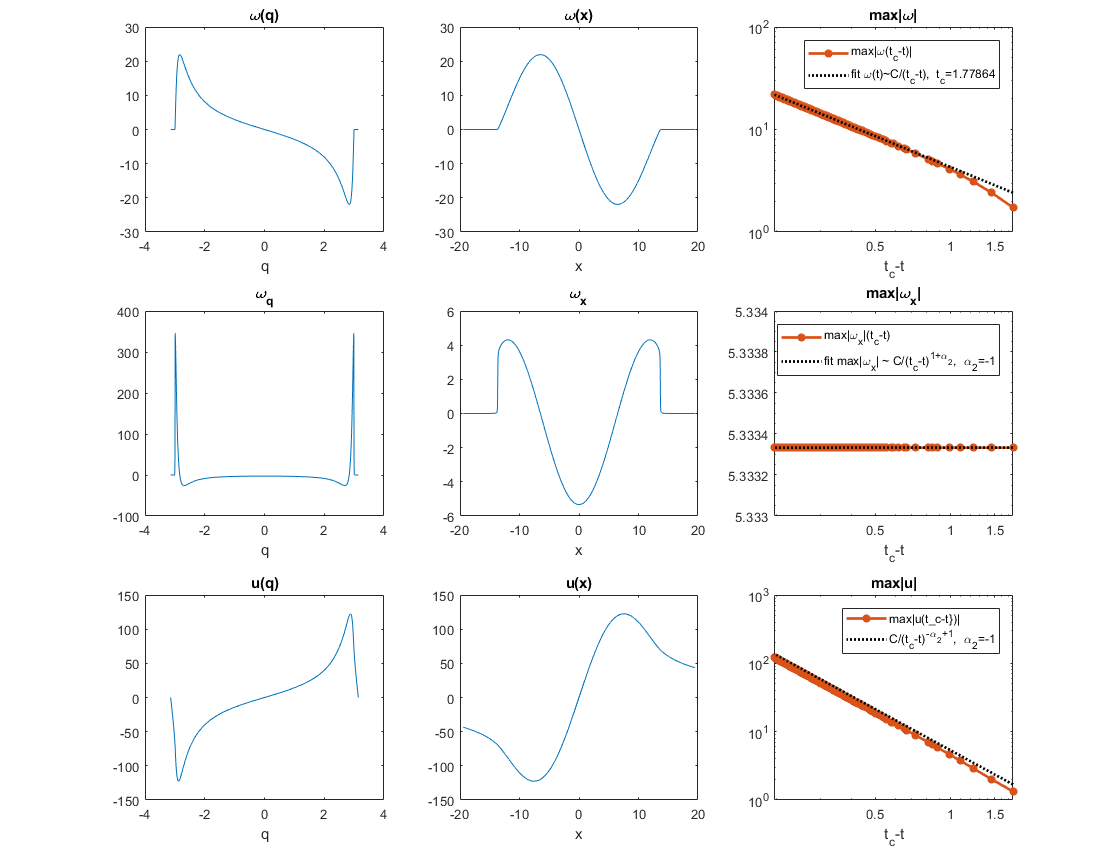}
\caption{Results of the same simulation as in Fig. \ref{fig:a1_selfsimilarconvergence} with  $a=1$  showing solution $\omega(q,t)$, $\omega_q(q,t)$ and $u(q,t)$ in $q$-space (left panels) as well as the same solution in $x$-space (center panels) at time $t=1.582477. $ Right panels show  the time dependence of their maximum values as functions of $(t_c-t)$, where $t_c$ is the blow-up time extracted from the fit to $\max|\omega(x,t)| \propto \frac{1}{(t_c-t)}$.} \label{fig:a1_9pics}
\end{figure}

\begin{figure}
\centering
\includegraphics[width=0.49\textwidth]{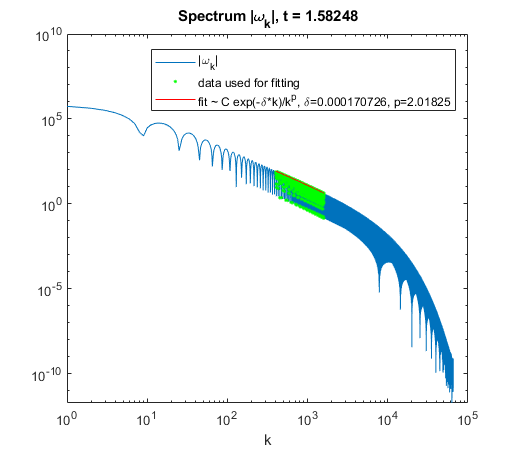}
\includegraphics[width=0.49\textwidth]{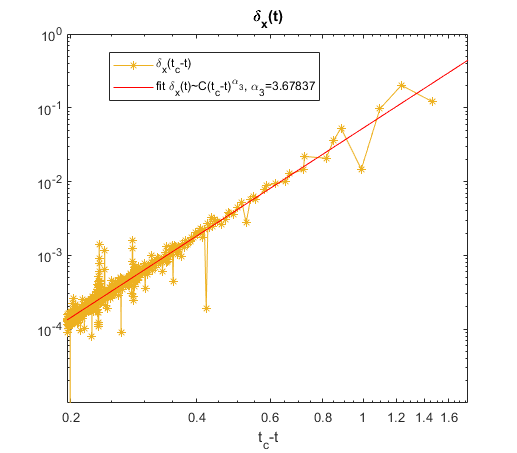}
\caption{Left panel: The Fourier spectrum $|\hat{\omega}_k|$ at time
$t=1.58248$ from the same simulation as in Fig. \ref{fig:a1_9pics} with
$a=1$.  The red line represents a fit to the model \e{deltakmodel} with
green line showing a portion of the  $|\hat{\omega}_k|$ used for the fit.
Right panel: Time dependence of
$\delta_x(t)=\tanh{\left(\frac{\delta(t)}{2}\right)}$ recovered from the
fit of the spectrum to Eq.  \e{deltakmodel}. The red solid line in the
right panel represents a fit to the model $\delta_x(t) \propto
(t_c-t)^{\alpha_3}.$  } \label{fig:a1_Spectrum}
\end{figure}

\begin{figure}
\includegraphics[width=0.328\textwidth]{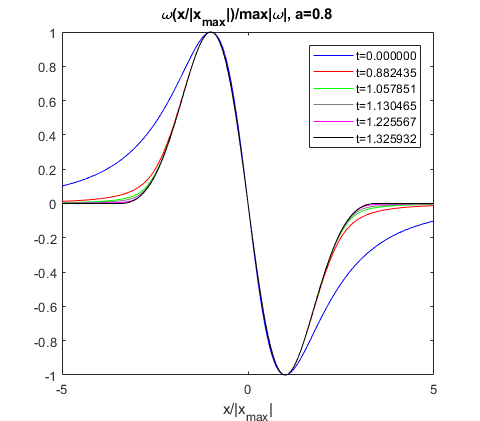}
\includegraphics[width=0.328\textwidth]{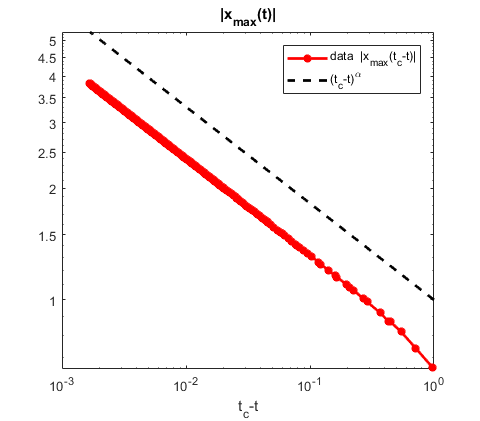}
\includegraphics[width=0.328\textwidth]{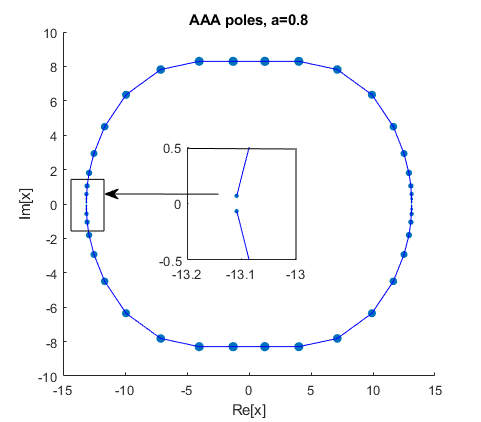}
\caption{Left panel: Convergence of time dependent numerical solution to Eqs. \e{CLM_transformed}-\e{HilbertHdef_periodic} with  $a=0.8$ and IC2 \e{IC2a} to a self-similar profile with  compact support. The solution expands horizontally and stretches vertically until blowing up at $t=t_c \approx 1.32761$. The solution is plotted in $x$-space, where $x=\tan(\frac{q}{2})$, and is scaled both horizontally and vertically to exactly match the positions of the local maximum and minimum.
Center panel: The time dependence  $x_{max}(t)$  of the location of $\max\limits_x\omega(x)$. The dashed lines shows that
it scales like $x_{max}(t)\propto (t_c-t)^\alpha$ with $\alpha \simeq-0.26008$ as $t\rightarrow{t_c}$.
Right panel: The structure of complex singularities  at $t=1.32593$ obtained using AAA-algorithm  (described in Section \ref{sec:analyticalcontinuation}) that approximates the solution by a set of simple poles, $\omega(x)\approx \omega_{AAA}(x)=\sum_{i=1}^{m-1}\frac{a_i}{x-b_i}$. The simple poles are shown as dots at locations $b_i$ with a size of dot scaled with $\log_{10}|a_i|$. The branch cuts are shown as lines connecting the dots, and form `U-shaped' curves in the upper and lower complex plane. The accumulation of poles approximates two pairs of branch points near the real line.
The location of these branch points scale as $x_{sing} \sim \pm (t_c-t)^\alpha x_b\pm \I y_0(t_c-t)^{\alpha_3}y_b$, where $x_0,y_0>0$,  $\alpha=-0.26008$ and $\alpha_3\approx0.908$.
} \label{fig:a0p8_selfsimilarconvergence}
\end{figure}

We have also performed simulations specifically with $a=1$ since this special
case was addressed in Chen et al. \cite{ChenHouHuang}, who proved
for this value of $a$ the existence of an ``expanding" self-similar solution of the type
\e{self-similar1} for the problem on $x\in\mathbb{R}$.  In this case
$f(\xi)$ is an odd function with a finite support and $\alpha=-1$. Their solution
implies that $\omega(x,t)\to f'(0)x$ as $t\to t_c$ for any finite value of
$x\in\mathbb{R}$ while the boundary of the compact support expands
infinitely fast into large $|x|$ as $t\to t_c.$
\revmike{Following referee 1 bullet 3, we removed comment on Ref. [7], and modified the following sentence:} \rev{Our numerical findings show an approach to this kind of expanding solution with compact support starting from a generic analytic initial condition, see Figs.
\ref{fig:a1_selfsimilarconvergence} and \ref{fig:a1_9pics}. This verifies that the similarity solution is attracting.} The solution
grows in amplitude and expands faster than exponentially in time, which is
demonstrated by semi-log plots of $\max\limits_x|\omega(x)|(t)$ and its
location $x_{max}(t)$ in the middle and right panels of Fig.
\ref{fig:a1_selfsimilarconvergence}. It obeys the self-similar
profile \e{self-similar1} and forms a finite time singularity at $t=t_c$.
Fig. \ref{fig:a1_9pics} (right panels) 
confirms the scales $\max\limits_x|\omega(x)| \propto 1/(t_c-t)$ and
$|\omega_x(x=0)| \propto 1/(t_c-t)^{1+\alpha}=const$ with $\alpha=-1$. One can also see (from the middle panel of Fig. \ref{fig:a1_9pics}) that  $\max\limits_x|\omega_{xx}(x)| \to \infty$ as $t \to t_c$. We
are able to simulate the growth in amplitude of $\omega(x)$ only by
about one order of magnitude with our spectral code,  since the
spectrum widens very quickly as $t\rightarrow{t_c}$
 and decays slowly, i.e.,  $|\hat{\omega}_k(x)| \sim k^{-2}$,
 as seen in  Fig.\ref{fig:a1_Spectrum} (left panel). The approach to a self-similar solution with  compact support
is expressed in the complex $x$-plane by the approach of  complex singularities (identified as branch points from our simulations) located at $x=x_{sing}$
to the real line near the boundaries of compact support. The small distances $|Im(x_{sing})|$ of these singularities to the real line for $t$ near $t_c$
means that the solution is ``almost of  compact support'' with ``almost a  jump" in the first derivative at the boundary of ``compact  support" in $x$-space.
The singularity locations  scale like%
\begin{equation}\label{xsing}
x_{sing} \simeq \pm (t_c-t)^{\alpha} x_b\pm \I\,(t_c-t)^{\alpha_3} y_b
\end{equation}
 (i.e. there are four symmetrically located singularities), where $\alpha=-1$ and $\alpha_3\approx3.68.   $ Here the real constants $t_c$, $x_b$\ and $y_b$ depend on the  IC. Note that
$\alpha_3$ is
different from $\alpha$ because it characterizes
the approach of the solution to the  compactly supported profile  \e{self-similar1}.   In contrast, the value $\alpha=-1$
is fully determined by Eq.  \e{self-similar1} and characterizes the self-similar behaviour of the central part of the solution. The nonzero value of $\alpha_3$  suggests that the ``almost compactly supported" solution turns into a truly compactly supported solution   at $t=t_c$,   with a jump in the first derivative.
Due to  oscillations in the spectrum, it is difficult to accurately extract the value of $\alpha_3$ from the fit to $\delta_x(t) \sim (t_c-t)^{\alpha_3}$.
However, using  rational approximation via the AAA algorithm (see details about AAA in  Section \ref{sec:analyticalcontinuation}) we can observe two pairs of branch cuts with branch points approach the real line near $x= \pm (t_c-t)^{\alpha}x_b$ as $t\rightarrow{t_c}$, similar to the case $a=0.8$. One can see from  Fig. \ref{fig:a0p8_selfsimilarconvergence} (right panel) that the structure of the singularity for $a=0.8$  is similar to the $a=1$ case.

\begin{figure}
\includegraphics[width=1\textwidth]{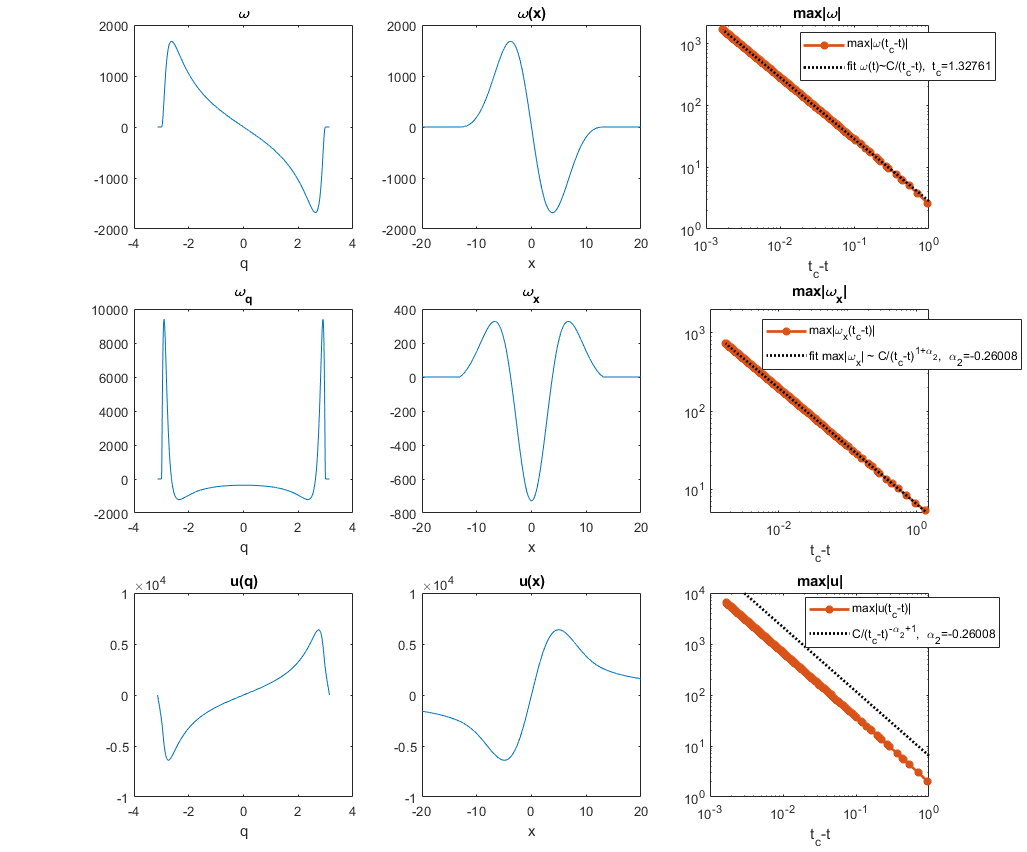}
\caption{Results of the same simulation as in Fig.  \ref{fig:a0p8_selfsimilarconvergence}  with $a=0.8$ showing  $\omega(q,t)$, $\omega_q(q,t)$, and $u(q,t)$ in $q$-space (left panels) as well as  in $x$-space (center panels) at time $t=1.32593. $ Right panels show  the time dependence of their maximum values as functions of $(t_c-t)$, where $t_c$ is the blow-up time extracted from the fit to $\max|\omega(x,t)| \propto \frac{1}{(t_c-t)}$.} \label{fig:a0p8_9pics}
\end{figure}

\begin{figure}
\centering
\includegraphics[width=0.48\textwidth]{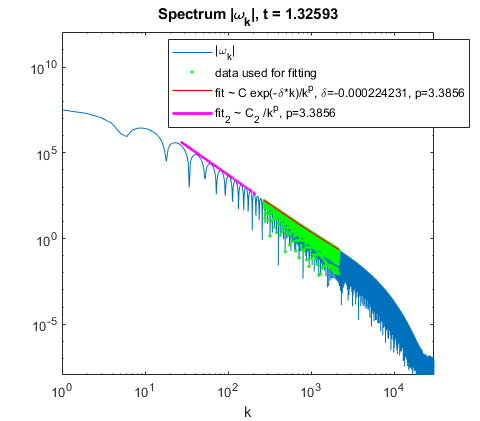}
\includegraphics[width=0.48\textwidth]{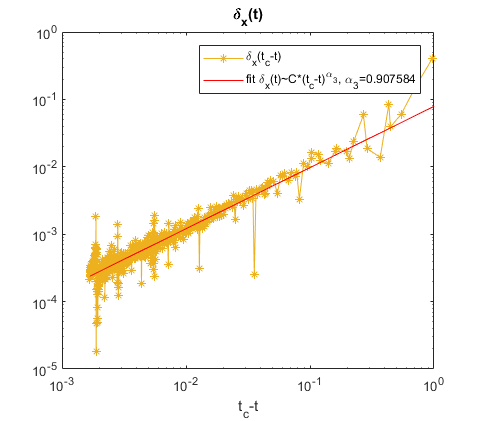}
\caption{ Left panel: The Fourier spectrum $|\hat{\omega}_k|$ at a
particular time $t=1.32593$ from the same simulation as in Fig.
\ref{fig:a0p8_selfsimilarconvergence} with $a=0.8$.  The red line
represents a fit to the model \e{deltakmodel} with the green line showing
the portion of the  $|\hat{\omega}_k|$ used for the fit. The purple line
shows  a fit to the rougher model  \e{deltakmodel} with $\delta=0$. Right
panel: Time dependence of
$\delta_x(t)=\tanh{\left(\frac{\delta(t)}{2}\right)}$ recovered from the
fit of the spectrum to Eq.  \e{deltakmodel}. The red solid line at the
right panel represents a fit to the model $\delta_x(t) \propto
(t_c-t)^{\alpha_3}.$  } \label{fig:a0p8_Spectrum}
\end{figure}

For $a_c < a < 1$ and both IC1 or IC2, we similary observe  finite time singularity formation with an expanding self-similar solution approaching a compactly supported profile (described again by Eq. \e{self-similar1}). This is qualitatively similar to the  $a=1$ case, but involves different values of $\alpha$. Another
difference compared to the  $a=1$ case is that there is  a discontinuity in a higher-order derivative  at the boundary of  "compact support", instead of a jump in the first derivative $\omega_x$ as occurs for $a=1$.  Figs. \ref{fig:a0p8_selfsimilarconvergence} - \ref{fig:a0p8_Spectrum} show the results of  simulations with the parameter $a=0.8$ and  IC2 \e{IC2a}. Here we find a jump in $\omega_{xx}$
forming at the boundary of  ``compact support".
\revc{In this case, we are able to simulate the growth of the amplitude of $\omega(x)$ only by a factor of  $\sim10^3$. This limitation is due to the rapid widening of the  spectrum with time so that it becomes almost algebraic, $|\hat{\omega}_k(x)| \sim k^{-p}, p \approx 3$
as $t\rightarrow{t_c}$.  This is again due to the solution achieving ``almost compact support" with a jump in the second derivative at the boundary of the ``compact support" in $x$-space.}
Fig. \ref{fig:a0p8_9pics} (right) shows the growth of both $\max\limits_x|\omega(x)|$ and  $\max\limits_x|\omega_x(x)|=|\omega_x(x=0)|$ as functions of $t_c-t$ confirming the scales $\max\limits_x|\omega(x)| \sim 1/(t_c-t)$ and $|\omega_x(x=0)| \sim 1/(t_c-t)^{1+\alpha}$ with $\alpha=-0.26008$.

Qualitatively similar to the case $a=1$, for $a_c < a < 1$ we again observe two pairs of branch cuts approach the real line as $t\rightarrow{t_c}$ according to Eq. \e{xsing}.
For example, when $a=0.8$ we find that  $\alpha=-0.26008$ and $\alpha_3\approx0.908$, see Fig. \ref{fig:a0p8_selfsimilarconvergence} (right panel).
 It was challenging to accurately extract values of $\delta(t)$ and $p(t)$ from a fit to Eq. \e{deltakmodel} due to the spectrum being oscillatory, see the left panel of Fig. \ref{fig:a0p8_Spectrum}. The right panel of Fig. \ref{fig:a0p8_Spectrum} provides the best fit which we were able to obtain for  $\delta(t)$.  The fitting parameter  $p(t)$ was more sensitive to the oscillations  and did not appear to stabilize at any particular value, so we do not provide a plot for it here.

 This type of oscillation in the spectrum occurs when there are two symmetric singularities that are equally close to the real line.  In this case, a more elaborate fitting procedure with additional parameters to account for the oscillation can yield improved results, see e.g. Ref. \cite{BakerCaflischSiegelJFM1993}. However, such fits are also more delicate to implement, and are beyond the scope of the current work.

Simulations with ICs  either of type  -IC1 or -IC2 and  $a_c < a \le 1$ resulted in monotonically decaying  $\max\limits_x|\omega(x,t)|$ and   $\max\limits_x|u(x,t)|.$  The maximum slope $\max\limits_x|\omega_x(t)|=|\omega_x(x=0,t)|$ is found to approach  a constant value for $a=1$ while it decays for $a<1. $ Also,  $\max\limits_x|\omega_{xx}(x,t)|$ grows algebraically as a function of $t$, 
 while $\delta_x(t)$ decays algebraically,  $\delta_x(t) \sim 1/t^\mu, \mu>0$. Since these ICs do not result in a finite-time singularity formation,  we do not discuss these cases in further detail.

For $a \gtrsim 1.3$ and for  both IC1 and IC2, we observe global existence of the solution. The vorticity $\omega$ has the form an an expanding self-similar function which  approaches a compactly supported profile (in the scaled variable $\xi$)   with  infinite slope at the boundary of the compact region,  so that $\max\limits_x|\omega|\rightarrow{0}$ and $\max\limits_x|\omega_x|, \max\limits_x|u| \rightarrow{\infty}$ as $t \rightarrow{\infty}$ (although $\omega_x(x=0) \rightarrow{0}$ as $t \rightarrow{\infty}$). The complex singularities approach the real line in infinite time with positions that scale like  $x_{sing} = \pm x_0 \exp{(\kappa_1 t^{\nu_1})}  \pm \I y_0 \exp{(-\kappa_2 t^{\nu_2})}$, where the constants $\kappa_1, \kappa_2, \nu_1, \nu_2>0$ depend on $a$. \revc{In particular, for $a=2$ we observe an  initial growth followed by the  eventual decay of $\max\limits_x|\omega|$. For $a=1.5,$ the amplitude of $\omega$ initially grows and then plateaus over the times we were able to reach in simulations.  For $a=1.3$ we observe a slowdown in the  growth of $\max\limits_x|\omega|$ so that it is below an exponential rate.}
For both -IC1 and -IC2 we again observe global existence of the solution with decay of $\omega$ and infinite growth of $\omega_x(x=0)$, with  an infinite slope forming at $x=0$ and a singularity approaching the real line like $x_{sing} = 0 \pm \I y_0 \exp{(-\kappa_2 t^{\nu_2})}$, where $y_0,\kappa_2,\nu_2>0.$

For $1<a\lesssim 1.3,$ we find from simulations that initially $\max\limits_x|\omega|$ grows. This period of initial growth is long, with the spectrum widening so quickly that it was challenging to distinguish between a finite time singularity  and  global existence when  $a$ is near 1, but we have numerical evidence of global existence for $a$ at least as small as 1.3, as described in the previous paragraph.

Here we summarize the  behaviour of solutions to  Eqs. \e{CLM_transformed}-\e{HilbertHdef_periodic} on $x\in\R$, and its  dependence on the parameter $a$, for quite generic smooth IC:
\begin{itemize}
\item $ a < a_c$ with $\alpha(a)>0$: Collapse  in  $\omega,$ i.e. $\max\limits_x|\omega|\to \infty$  at the finite time $t_c.$  As $t\to t_c$, solutions with generic IC  approach the shrinking universal self-similar profile  \e{self-similar1} near the spatial location of $\max\limits_x|\omega|$. As $t\to t_c,$ the profiles shrink to zero width.  The self-similar solution has  leading order complex singularities in agreement with Theorem 1 and Eq. \e{gammaa}. The location of these singularities approaches the real line as $x_{sing} = x_0 \pm \I\,\delta_x(t)$, where $\delta_x(t) \propto (t_c-t)^\alpha$, $\alpha=\alpha(a)>0$. In particular, $x_0=0$ for both IC1 or IC2. Also $u(x,t)$ near $x_0$ follows the self-similar profile \e{uselfsimilar} with $\max\limits_x|u|\to \infty$  for $0<a< a_c$.

\item $a_c < a \leq 1$ with $\alpha(a)<0$: Blow up in both $\omega$ and $u$ at the finite time $t_c.$ As $t\to t_c$, solutions with generic IC  approach the expanding self-similar profile Eq. \e{self-similar1} which has  compact support. As $t\to t_c$, the rate of expansion turns infinite. The complex  singularities closest to the real line correspond to the boundaries of compact support, and  they approach  the real line as $x_{sing} \sim \pm (t_c-t)^{\alpha} x_b\pm \I\,(t_c-t)^{\alpha_3}y_b$, where $\alpha=\alpha(a)<0$ and $\alpha_3(a)>0$,

\item $a\gtrsim1.3:$  global existence of solutions with $\max\limits_x|\omega|\rightarrow{0}$, $\max\limits_x|\omega_x|, \max\limits_x|u| \rightarrow{\infty}$ and $\omega_x(x=0) \rightarrow{0}$ as $t \rightarrow{\infty}$. The complex singularities approach the real line exponentially in time as $x_{sing} = \pm x_0 \exp{(\kappa_1 t^{\nu_1})} \pm \I y_0 \exp{(-\kappa_2 t^{\nu_2})}$, where $\kappa_1, \kappa_2, \nu_1, \nu_2>0$.
\end{itemize}

\section{Numerical solution of nonlinear eigenvalue problem  on the real line} \label{sec:e_value_problem}
Similar to the transformation of Eq. \e{CLM01} to Eqs.  \e{CLM_transformed}-\e{HilbertHdef_periodic} in Section \ref{sec:transform},
we obtain a transformed equation for self-similar solutions of Eq. \e{CLM01selfsimilar2} by mapping the  interval $(-\pi,\pi)$ of the auxiliary variable $q$  onto the real line $(-\infty, \infty)$ as
\begin{equation}\label{transform2}
\xi=\tan \left ( \frac{q}{2} \right ).
\end{equation}
With this mapping Eq. \e{CLM01selfsimilar2} turns into
\begin{equation}\label{CLM01selfsimilar_transformed}
\begin{split}
&{\mathcal M} f:=f+\alpha \sin{q} \,f_q =-a (1+\cos{q}) g f_q + f [ {\mathcal H}^{2\pi}f+ C_f^{2\pi}]:={\mathcal N} [f]f, \quad q\in [-\pi,\pi], \\
&(1+\cos{q})g_q= {\mathcal H}^{2\pi}f + C_f^{2\pi},
\end{split}
\end{equation}
where the $2\pi$-periodic Hilbert transform $ {\mathcal H}^{2\pi}$ and the constant $C_f^{2\pi}$ are defined in Eqs. \e{HilbertHdef_periodic},  \e{C2pidef}, and the linear operator $\mathcal M$\ is now defined in $q$ space by the l.h.s. of the first Eq. in \e{CLM01selfsimilar_transformed}.
We also define \rev{in equation \e{CLM01selfsimilar_transformed} the quadratically nonlinear operator $\mathcal N[f]$ such that ${\mathcal N}[f]f$} represents the  r.h.s. of the first Eq. in  \e{CLM01selfsimilar_transformed} with $g$ expressed through the second equation in  \e{CLM01selfsimilar_transformed} as%
\begin{equation}\label{gdef}
g=\partial_q^{-1}\left [\frac{ {\mathcal H}^{2\pi}f + C_f^{2\pi}}{(1+\cos{q})} \right ], \quad \partial_q^{-1}p:=\int\limits_{-\pi}^qp(q')dq'.
\end{equation}
 Then Eq. \e{CLM01selfsimilar_transformed} takes the following operator form%
\begin{equation}\label{MNdef}
{\mathcal M} f={\mathcal N}[f]f.
\end{equation}

A linearization of Eq. \e{MNdef} about $f$ together with Eqs. \e{CLM01selfsimilar_transformed} and \e{gdef} result in%
\begin{align}\label{MNlinear}
&{\mathcal L}[f]\delta f:=-{\mathcal M} \delta f-a (1+\cos{q})\partial_q^{-1}\left [\frac{ {\mathcal H}^{2\pi}\delta f + C_{\delta f}^{2\pi}}{(1+\cos{q})} \right ] f_q  \nonumber \\&-a (1+\cos{q})\partial_q^{-1}\left [\frac{ {\mathcal H}^{2\pi}f + C_f^{2\pi}}{(1+\cos{q})} \right ]\delta f_q \nonumber\\
&+ \delta f [{\mathcal H}^{2\pi}f + C_f^{2\pi}]+ f [{\mathcal H}^{2\pi}\delta f + C_{\delta f}^{2\pi}],
\end{align}
\rev{where ${\mathcal L}[f]$\ is the linearization operator and $\delta f$\ is the deviation from $f.$

Taking $\delta f=f$ in Eq. \e{MNlinear}
 and using Eqs. \e{CLM01selfsimilar_transformed}, \e{MNdef} to express the nonlinear terms in $f$ through the linear terms  proves the following theorem:}

{\it Theorem} {4.} The solution $f$ of Eq. \e{CLM01selfsimilar_transformed} satisfies the  relation%
\begin{equation}\label{MLeq}
{\mathcal L}[f] f={\mathcal M}  f.
\end{equation}

{\it Corollary 1.} The invertability of the operator $\mathcal M$ (see Section \ref{sec:self-similar}) and Eq. \e{MLeq} imply that the operator ${\mathcal\ M}^{-1}{\mathcal L}[f]$ has the eigenvalue $\lambda=1$ with  eigenfunction $f$, which is the same as the solution $f$ of Eq. \e{CLM01selfsimilar_transformed}.

Similar to Eq. \e{solution_Fourier}, we approximate a solution of Eq. \e{CLM01selfsimilar_transformed} as a truncated Fourier series
\begin{equation}\label{solution_Fourier2}
f(q)=\sum_{k=-N}^{k=N-1}{\hat{f}_k e^{\I kq}}.
\end{equation}
Then the discrete Fourier transform allows us to rewrite Eq. \e{CLM01selfsimilar_transformed} in  matrix form as \begin{equation}\label{CLM01selfsimilar_transformed_matrix}
{\bf M}\,{ \hat{\bf f}} =\widehat{{\mathcal N}[f]f}, \quad  
{\bf M}:=
\begin{pmatrix}
     1 & -\frac{\alpha k_2}{2}&   &   &   \\
     \frac{\alpha k_1}{2}& 1 & -\frac{\alpha k_3}{2}&   &   \\
      &  \frac{\alpha k_2}{2}& 1 &\dots&   \\
      &   &\dots& \dots  & -\frac{\alpha k_{2N}}{2}\\
      &   &   &  \frac{\alpha k_{2N-1}}{2}& 1  \\
\end{pmatrix},
\end{equation}
where  ${ \hat{\bf f}} =(\hat f_{k_1},\hat f_{k_2},\ldots,\hat f_{k_{2N}})^T$ is a column vector,  the tridiagonal matrix ${\bf M}\in \R^{2N\times 2N} $ represents the Fourier transform of the operator $\mathcal M$ and  $\widehat{{\mathcal N}[f]f}$ is the column vector of  Fourier coefficients of ${{\mathcal N}[f]f}$.  Also $k_1:=-N, \, k_2:=-N+1, \ldots , \, k_{2N}:=N-1.$  Note that the tridiagonal form of  ${\bf M}$ is a consequence of  the term    $\sin(q)=\frac{e^{\I q}-e^{-\I
q}}{2\I}$in the definition of $\mathcal M$ in Eq.  \e{CLM01selfsimilar_transformed}.

We solve Eq. \e{MLeq} in the truncated Fourier representation \e{CLM01selfsimilar_transformed_matrix} by iteration using the generalized Petviashvili method (GPM) \cite{LY2007} which relates the  $n+1$th iteration  $\hat{\bf f}^{n+1}$
to the $n$th iteration $\hat{\bf f}^{n}$ of $\hat{\bf f}$ as follows
\begin{equation}\label{Petviashvili}
\hat{\bf f}^{n+1}-\hat{\bf f}^{n}=\left( [-\hat{\bf f}^n + {\bf M}^{-1} \widehat{{\mathcal N}[f]f^n}]  - \left(1+\frac{1}{\Delta\tau}\right)\frac{\langle\hat{\bf f}^n,-{\bf M}\,{ \hat{\bf f}} ^n+\widehat{{\mathcal N}[f]f^n}\rangle}{\langle\hat{\bf f}^n,{\bf M}\,{ \hat{\bf f}} ^n\rangle} \hat{\bf f}^{n}  \right) \Delta\tau,
\end{equation}
where  superscripts give the iteration number,  $\langle{\bf a}\,,\,{\bf b}\rangle:=\sum_{k=-N}^{k=N-1}\bar a_k b_k$ is the complex dot product and $\Delta\tau$ is a parameter that controls the convergence rate of the iterations. 
At each iteration we need to solve Eq.
\e{CLM01selfsimilar_transformed_matrix} for $\hat{\bf f}$ (assuming  $\widehat{{\mathcal N}[f]f}$ is given) to effectively
compute $ {\bf M}^{-1} \widehat{{\mathcal N}[f]f^n}$. Since ${\bf M}$ is a tridiagonal matrix,
this is easily done in $O(N)$ numerical operations in Fourier space.  We note that if one tries to avoid the  FFT\ and iterate Eq.  \e{CLM01selfsimilar_transformed}
directly in $q$ space, then
the corresponding matrix $M$ on the l.h.s. of  Eq. \e{CLM01selfsimilar_transformed}
would be a full matrix and each iteration would require $O(N^2)$ numerical operations.

A fixed point of the iteration \e{Petviashvili} corresponds to the solution of Eq. \e{MLeq}.
The straightforward iteration of \e{MLeq} (instead of   \e{Petviashvili}) would diverge because of the positive eigenvalue $\lambda=1$ of Corollary 1 for the linearized operator ${\mathcal\ M}^{-1}{\mathcal L}[f]$. In contrast, Eq. \e{MLeq} ensures an approximate projection into the subspace orthogonal to the corresponding unstable eigenvector $f.$ The original Petviashvili method \cite{Petviashvili1976} is the nonlinear version of Eq.  \e{Petviashvili} for the particular value $\Delta \tau=1$ and  is often successful with both partial differential equations (PDEs) (see e.g. Refs. \cite{LY2007,YangBook2010}) and nonlocal PDEs (see e.g. Ref. \cite{LushnikovOL2001}).   However, the linear operator
${\mathcal\ M}^{-1}{\mathcal L}[f]$ generally has extra eigenvalues preventing the convergence   of the original Petviashvili method. GPM  however uses the freedom in choice of the parameter $\Delta \tau$ to achieve  convergence even with such extra eigenvalues, see Refs. \cite{DyachenkoLushnikovKorotkevichJETPLett2014,LY2007,YangBook2010} {\color{red}} for more discussion.


An additional complication that arises in our   Eq.  \e{CLM01selfsimilar_transformed}, compared with the straightforward use of GPM in general PDEs, is that we do not know $\alpha$ in advance. Instead, for each value of  $a$ there is a nonlinear eigenvalue $\alpha(a)$ to Eq. \e{CLM01selfsimilar_transformed} that we need to determine. If  we use a general value of $\alpha$, then the iteration  \e{Petviashvili} would not converge because the solution of Eq. \e{CLM01selfsimilar_transformed} does not exist for such general values of $\alpha.$

\begin{figure}
\includegraphics[width=0.5\linewidth]{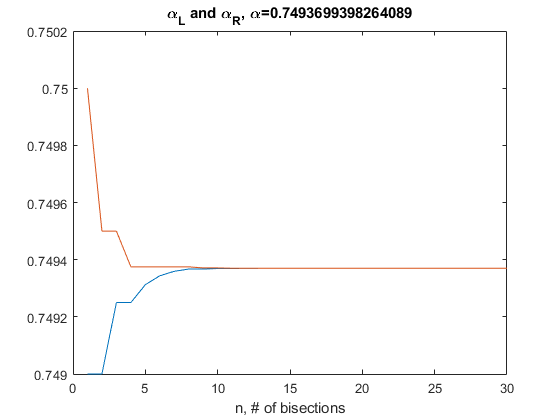}
\includegraphics[width=0.5\linewidth]{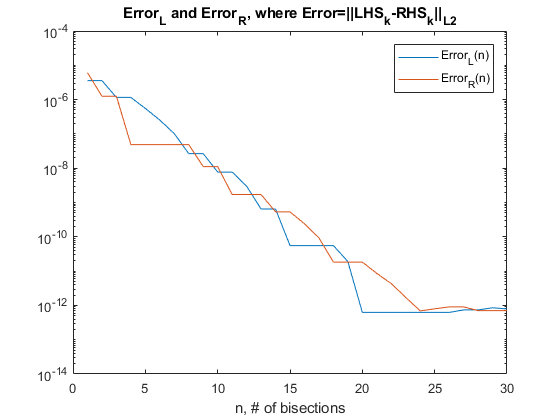}
\caption{ Convergence of the interval  $[\alpha_L, \alpha_R]$  to
$\alpha(a)$ (left panel) and  convergence of the residual of Eq.
\e{CLM01selfsimilar_transformed_matrix} (right panel) for the iteration
\e{Petviashvili} with $a=0.2.$ Here we used IC2 \e{IC2} with
$V_c=2^{-12}\approx2.44\cdot10^{-4}$ and $N=2^{18}$ as the zeroth
iteration.} \label{fig:alpha(n)Error(n)a0p}
\end{figure}

\begin{figure}
\centering
    \includegraphics[width=0.49\linewidth]{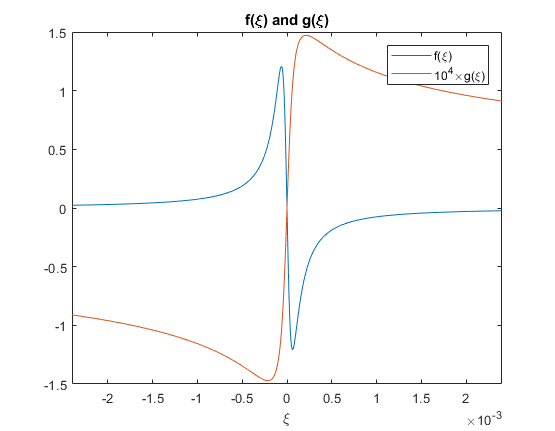}
    \includegraphics[width=0.49\linewidth]{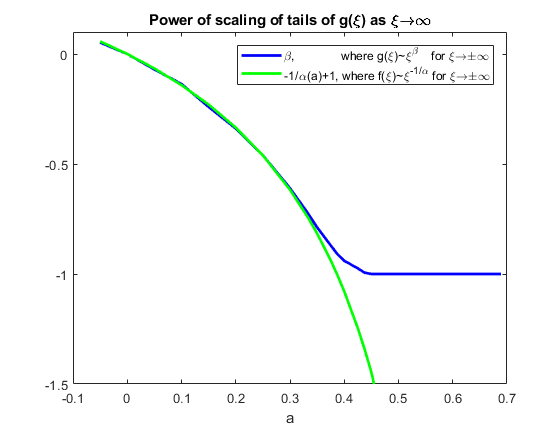}
    \caption{ Left panel: $a=0.2$. Functions $f(\xi)$ and scaled $g(\xi)$ obtained by the iteration \e{Petviashvili}. 
    Right panel: Power-law of scaling of the tails of $g(\xi)$ vs. $a$.}  \label{fig:f_g_a0p2} \label{fig:betta(a)}
\end{figure}

To address this additional complication, we make an initial guess of $\alpha=\alpha_{guess}$ for fixed $a$ and iterate Eq. \e{CLM01selfsimilar_transformed} for  $\alpha_{guess}$. If $\alpha_{guess} < \alpha(a)$ then the generalized Petviashvili iteration (after an initial transient) shrinks towards $q=0$.
If $\alpha_{guess} > \alpha(a)$ then the solution  expands away from $q=0$. We used the bisection method to determine $\alpha(a)$ for a given $a$. We start from a large enough interval $[\alpha_L, \alpha_R]$, so that $\alpha(a) \in [\alpha_L, \alpha_R]$. Then we try $\alpha_{guess}=(\alpha_L+\alpha_R)/2$ and based on the shrinking vs. expanding of iterations  for $\alpha_{guess},$ we obtain the updated values  $[\alpha_L, \alpha_R]$.  These updated values ensure  a factor  2 decrease of the length of the updated interval $[\alpha_L, \alpha_R]$, completing the first step of the bisection method. We continue such  bisection steps until convergence to $\alpha(a)$ (i.e., until the residual of Eq. \e{CLM01selfsimilar_transformed} decreases down to near round-off values and does not decrease anymore).
For each  updated $\alpha_{guess}$ we use the solution from the previous bisection step to speed-up the convergence.
We judged the expansion/shrinking of the solution by tracking the movement of its maximum point which was determined as a critical point of the function $f'(q)=\sum_{k=-N}^{k=N-1}{\I k \hat{f}_k e^{\I kq}}$ using spectral interpolation and a root-finding algorithm.
Also, in order to pass over the initial transient dynamics  (that depends on the initial guess of the solution) we  skip $10/\Delta\tau-20/\Delta\tau$ initial GPM iterations before judging the expansion/shrinking of the solution to classify the current $\alpha_{guess}$.
The larger $\Delta\tau$ we used, the less iterations were needed, but too large a $\Delta\tau$ leads to instability of the algorithm, so we need to keep it under a certain level.
For the initial guess of the solution we typically used IC2 from Eq. \e{IC2} with $V_c=1/2$ for $0.6<a<a_c$, and $N=64;$  $\Delta\tau$ was reduced from $0.1$ at $a=0.6$  to $10^{-4}$ near $a_c.$ For $a<0.6$ we used $\Delta\tau=0.1-1$ and progressively smaller  $V_c$ (down to $2^{-14}$) and larger  $N$ (up to $2^{22}$) because of the slowly decaying tails of the function $f(q)$ for small $a$ (see the next paragraph).
 Fig. \ref{fig:alpha(n)Error(n)a0p} illustrates the convergence of the $[\alpha_L, \alpha_R]$ interval to $\alpha(a)$ and  convergence of the residual of Eq. \e{CLM01selfsimilar_transformed} with bisection iterations for $a=0.2$, starting with an initial condition IC2 in \e{IC2} with $V_c=1/2^{12}\approx2.44\times10^{-4}$ (singularity is at $\xi=\I Vc$) and $N=2^{18}$. The converged solution is shown in Fig. \ref{fig:f_g_a0p2} (left panel) with a closest singularity at a distance $\xi_c=7.43\cdot10^{-5}$ from the real line in $\xi$-space and at a distance $q_c=1.49\cdot10^{-4}$  in $q$-space.

We note that the symmetry \e{stretching} implies that $\xi_c$  can be
stretched by an  arbitrary positive constant. The iteration
\e{Petviashvili} generally converges to different values of $\xi_c$ depending on
IC (i.e., the zeroth iteration). After that one can rescale any such solution in $\xi$ by any
fixed value of $\xi_c.$ This rescaling freedom can  also be seen
through the existence of the free parameter $\tilde v_c$ in the exact
solutions  \e{omegagamma2a0} and \e{omegasera1p2},  \e{xidef0p5}.

\begin{figure}
\centering
    \includegraphics[width=0.49\linewidth]{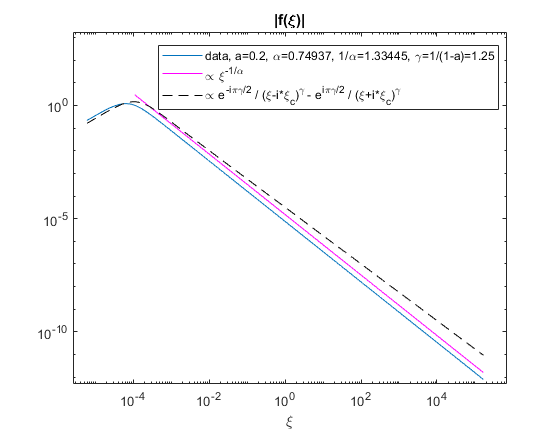}
    \includegraphics[width=0.49\linewidth]{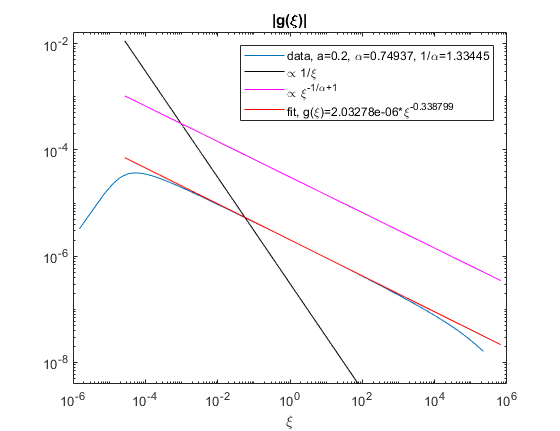}
    \caption{$a=0.2$. Left panel: Tail of $f(\xi)$ from Fig. \ref{fig:f_g_a0p2} (left panel). The dashed line shows the decay of $f(\xi)$ when it is approximated by its leading order singularities alone, as obtained from \e{omegagamma2}, neglecting the l.s.t.
    Right panel: Tail of $g(\xi)$ from Fig. \ref{fig:f_g_a0p2} (left panel) compared with different power laws. } \label{fig:f_g_tails_scaling}
\end{figure}

\begin{figure}
\centering
    \includegraphics[width=0.49\linewidth]{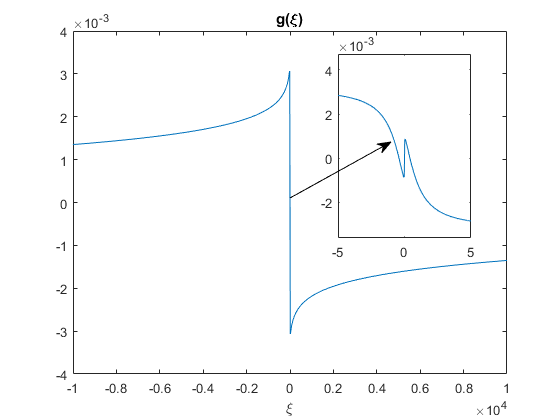}
    \includegraphics[width=0.5\linewidth]{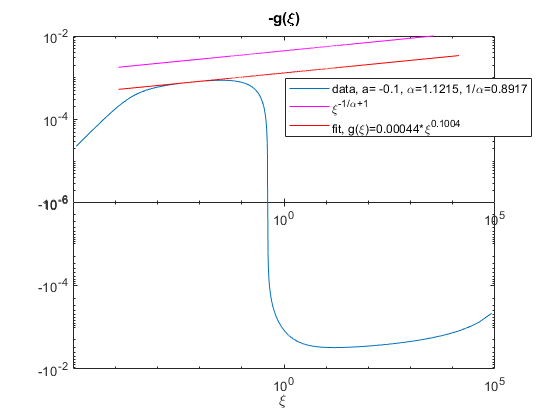}
    \caption{Plots of $g(\xi)$ for $a=-0.1$.  Left panel: Graph of $g(\xi)$ showing two extrema (one maximum and one minimum) in each half-space of $\xi$. The inset gives a
    magnified view showing extrema at small $\xi$.
    Right panel: Log-log plot of  $g(\xi)$  for positive $\xi$. Here $g(\xi_0)=0$ at $\xi_0\approx 0.41$. Solid lines show the scaling \e{gasymp} and a fit to power law \e{gpower}.}  \label{fig:g_tails_scaling_a-0.1}
\end{figure}

We computed  self-similar profiles $f(\xi)$ and $g(\xi)$ for various values  of $a<a_c$ to obtain  $\alpha(a)$  shown in Table \ref{Table1} as $\alpha_e(a).$ Additionally, we   make sure that the  $f(\xi)$ profile tails scale as in Eq. \e{fasymp} at $\xi\to\pm \infty$ and we also fit the $g(\xi)$ profile tails to the power law 
\begin{equation}\label{gpower}
g(\xi)\propto \xi^{\beta}.
\end{equation}
  Fig. \ref{fig:f_g_tails_scaling} show examples of such scaling and fit for $a=0.2$. Several other curves with different powers of $\xi$ are  present on the graphs for comparison.
The fitted values of  $\beta(a)$ are given in Table \ref{Table1} and Fig. \ref{fig:betta(a)} (right panel).
Ignoring for the moment the Hilbert transform, the integration operator $\partial_\xi^{-1}$ involved in determining $g(\xi)$ from $f(\xi)$ in  Eq.  \e{CLM01selfsimilar2} suggests that
\begin{equation}\label{gasymp}
g(\xi) \propto \xi^{-\frac{1}{\alpha}+1} \ \text{at} \ \xi\to\pm\infty,
\end{equation}
 which implies that %
\begin{equation}\label{betaa}
\beta=-\frac{1}{\alpha}+1.
\end{equation}
 However, the Hilbert transform in  Eq.  \e{CLM01selfsimilar2} can affect this scaling.   We find that \e{betaa} is valid for $0<a\lesssim 0.4$, while a transition to the constant scaling   $\beta=-1$ occurs  around $a\approx 0.45$ as seen in Table \ref{Table1} and Fig. \ref{fig:betta(a)} (right panel).
In particular, the exact analytical solution \e{ugamma11p2} for $a=1/2$ and $\alpha=1/3$  implies that $\beta=-1$  which is consistent with Table \ref{Table1} and Fig. \ref{fig:betta(a)} (right panel). One can see from comparison of Eqs. \e{omegagamma2a1p2} and \e{ugamma11p2} that the Hilbert transform indeed prevents the naive scaling \e{gasymp} in this particular case. In contrast, the scaling \e{fasymp} follows from the linear operator $\mathcal M$  as discussed in Section \ref{sec:self-similar}. That scaling was confirmed with  high precision in our simulations so we do not show it in Table  \ref{Table1}. For $a<0$ we find that $g(\xi)$ has two regions with two different scalings, see  Fig. \ref{fig:g_tails_scaling_a-0.1} for $a=-0.1$. While the tail of $g(\xi)$ still decays as $\xi\to\pm \infty$, there is an  intermediate scaling regime which approximately obeys \e{betaa} as seen in Fig.  \ref{fig:g_tails_scaling_a-0.1} (left panel). We are able to observe this intermediate scaling for $-0.2\leq a<0$. Going below $a=-0.2$ is difficult for the GPM method as the tails of $f(\xi)$ and $g(\xi)$ decay very slowly and it requires more than $10^6$ grid points to achieve good accuracy. For $a<0$ the values of $\beta$ in Table \ref{Table1} and in Fig. \ref{fig:betta(a)} (right panel) are from this intermediate scaling.

We estimate that our iteration procedure  provides at least 5-8 digits of precision of in $\alpha(a)$ and 2-3 digits of precision in $\beta(a)$ for $a\geq0.3$, when the spectrum of $f(q)$ is fully resolved.  The values of $\alpha(a)$ and $\beta(a)$ were challenging to obtain with more than 3-4 and $\sim$2 digits of accuracy, respectively, for $a\lesssim 0.2$ (corresponding to $\alpha\gtrsim0.75$) and especially for $a<0$ ($\alpha>1$) since we could not resolve the Fourier spectrum  $|\hat{f}_k|$ down to round-off level  $10^{-16}$, even with $N=2^{22}$ modes.
At its root, this is due to the slow decay of $f(\xi) \sim |\xi|^{-1/\alpha}$ for $|\xi| \rightarrow \infty$ and relatively large $\alpha$.

The numerical values of $\beta$ in the scaling \e{gpower} are important to distinguish between solutions with  infinite and finite energy $E_K$ \e{KinetiEdef}, which as mentioned is of interest in analogy with the question of singularity formation  in  the 3D Euler and Navier-Stokes equations.
Assuming that the solution is close to the self-similar profile \e{self-similar1}, changing the variable from $x$ to $\xi$ in \e{KinetiEdef} and using the self-similar profile \e{uselfsimilar} of the velocity $u(x,t)$ we obtain that
\begin{equation}\label{KinetiE}
E_K = E_K^{selfsim} + E_K^{rest},
\end{equation}
where
\begin{equation}\label{Eselfsim}
E_K^{selfsim}=\int \limits ^{x^{}_b}_{-x^{}_b} u^2(x)\D x  \sim \tau^{3\alpha-2} \int \limits ^{\xi^{}_b}_{-\xi^{}_b} g^2(\xi)\D \xi, \ \xi_b=\frac{x_b}{\tau^\alpha},
\end{equation}
is the kinetic energy of the approximately self-similar part of the solution located at $x \in [-x^{}_b, x^{}_b]$  and $E_K^{rest}$ is the kinetic energy of the numerical solution outside of
this interval.
Here we define the
cutoff value $x=x_b$ as the spatial location where the numerical solution deviates from the self-similar profile \e{self-similar1}  by $5\%$, while inside of the interval $[-x^{}_b, x^{}_b]$ the relative deviation is less than  $5\%$. We determine the variable $\xi$ by the same type of procedure as in Fig. \ref{fig:selfsimilarconvergence}. Then $x_b$ is determined by $5\%$ criterion above. We find from simulations with $a<a_c$  that %
\begin{equation}\label{xbtau}
x^{}_b(t) \approx const\sim \tau^0.
\end{equation}
 Such behaviour
 is typical for collapsing self-similar solutions, see e.g. Ref.  \cite{SulemSulem1999,KuznetsovZakharov2007,DyachenkoLushnikovVladimirovaKellerSegelNonlinearity2013,LushnikovDyachenkoVladimirovaNLSloglogPRA2013}.
 It  implies that $\xi_b\to \infty$ as $t\to t_c$.

There is no qualitative  difference between integrals $I_{g,\xi_b}:= \int \limits ^{\xi^{}_b}_{-\xi^{}_b} g^2(\xi)\D \xi $ and $I_{g,\infty }=\int \limits ^\infty_{-\infty} g^2(\xi)\D \xi$ provided $I_{g,\infty }<\infty$. The finiteness of $I_{g,\infty}$ requires that $\beta<-\frac{1}{2}$ for the scaling of the tails of $g(\xi)$ in \e{gpower}. Using equation \e{betaa} we obtain that $\beta=-\frac{1}{2}$ implies $\alpha= \frac{2}{3}$, i.e.  $\beta<-\frac{1}{2}$ for $\alpha < \frac{2}{3}$.
From the interpolation of the data of Table  \ref{Table1} we find that  $\alpha= \frac{2}{3}$ corresponds to $a=0.265 \pm 0.001$. Therefore for a self-similar profile, $I_{g,\infty }<\infty$ for $a>0.265 \pm 0.001$ and $I_{g,\infty }=\infty$ for $a<0.265 \pm 0.001$.

\begin{figure}
\centering
    \includegraphics[width=0.327\linewidth]{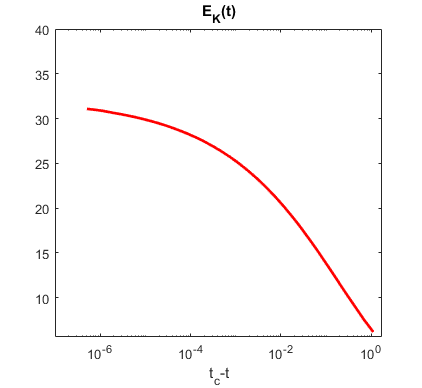}
    \includegraphics[width=0.327\linewidth]{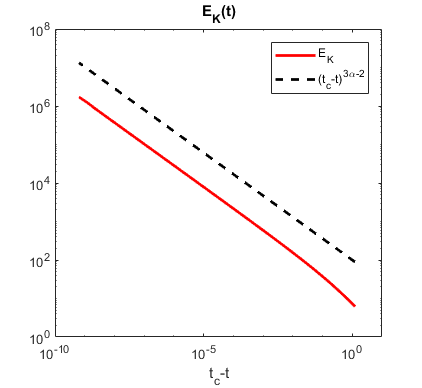}
    \includegraphics[width=0.327\linewidth]{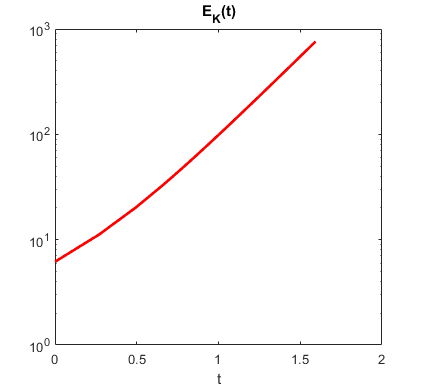}
    \caption{Growth of the kinetic energy $E_K$ over time. Left panel: $a=0.2$,  semi-log plot of $E_K$ vs. $\tau=t_c-t$ shows that $E_K$ grows slower than $\log(\tau)$ or any power of $\tau$ as $t \to t_c$. Center panel: $a=0.4$, verification of the scaling $E_K \sim \tau^{3\alpha-2}$  in \e{Eselfsim} with $I_{g,\infty }<\infty$.  Right panel: $a=1.5$, $E_K \to \infty$ exponentially as $t \to \infty$.} \label{fig:E_K(t)}
\end{figure}

However, we have to take into account that $I_{g,\xi_b}$ is multiplied by $\tau^{3\alpha-2}$ in equation \e{Eselfsim}. This means that in the limit $t\to t_c$ and for $\alpha < \frac{2}{3}$,
there is a competition between the decrease of  $\tau^{3\alpha-2}$ and the  growth of   $I_{g,\xi_b} $ as $\xi_b\to \infty.$  The scaling \e{betaa} for Eq. \e{gpower} is valid for $a\lesssim 0.4$ as seen in  Fig. \ref{fig:betta(a)} (right panel). It implies that    $I_{g,\xi_b}\propto \xi_b^{2\beta+1}= \tau^{-\alpha(2\beta+1)}x_b^{2\beta+1} $ for $a<0.265 \pm 0.001$ and $t\to t_c.$
  Then using Eqs. \e{betaa}, \e{Eselfsim} and \e{xbtau} we obtain that $E_K^{selfsim}\sim \tau^0 \sim const$. Also since the main dynamics is happening in  $x \in [-x^{}_b, x^{}_b]$ with $x^{}_b(t) \sim const,$ we  conclude that $E_K^{rest} \to const$ as $t\to t_c$, so overall the growth of $E_K(t)$ as $t\to t_c$ is very slow (i.e., slower than any power of $\tau$) for such $a$ where the scaling \e{betaa} is true.
This result is in excellent agreement with our direct calculation of  $E_K(t)$ from time-dependent simulations which shows that  for $a<0.265 \pm 0.001$ the kinetic energy grows more slowly than $\log(\tau)$ or any power of $\tau$ as $t \to t_c$; see Fig. \ref{fig:E_K(t)} (left panel) for $a=0.2$.

For $0.265 \pm 0.001<a\leq1$ the kinetic energy $E_K \to \infty$ as $t\to t_c$ (while being finite for any $t<t_c$), since $\alpha<0$ and  $E_{K} \sim\tau^{3\alpha-2} \to \infty$ as $t\to t_c$ with $I_{g,\infty } <\infty$; see Fig. \ref{fig:E_K(t)} (center panel) for a verification of this scaling when $a=0.4$.
For $a\gtrsim1.3$, which corresponds to an expanding  solution with infinite time singularity, $E_K \to \infty$ as $t\to \infty$, while being finite for any $t<\infty$; see Fig. \ref{fig:E_K(t)} (right panel) for an example with $a=1.5$.
For $a>a_c$, the above splitting of $E_K$ into two parts is no longer valid but we nevertheless verify the claims above via time-dependent numerical simulation.

\begin{figure}
\includegraphics[width=0.5\textwidth]{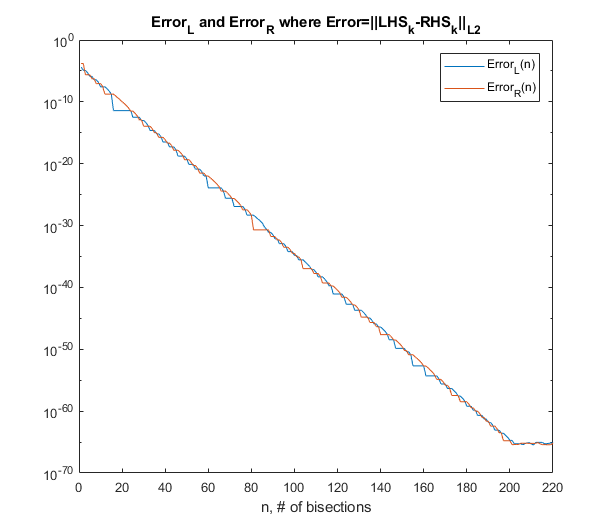}
\includegraphics[width=0.5\textwidth]{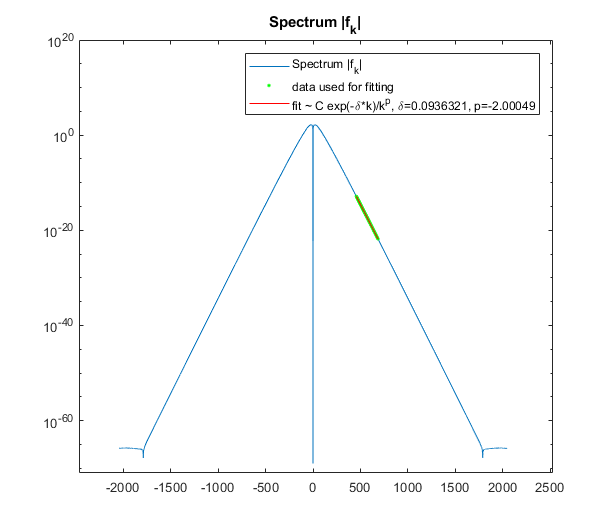}
\caption{Convergence of the residual (left panel) and spectrum of the solution (right panel)  to Eq.
\e{CLM01selfsimilar_transformed_matrix}, computed with $a=2/3$ and 68-digit
precision, and using  IC2  \e{IC2} with $V_c=1/16=0.0625$ and
$N=2048$ in the zeroth iteration.} \label{fig:a2p3_68digits}
\end{figure}

For some values of $a$ we computed  $\alpha(a)$ and nonlinear self-similar profiles with much higher precision. For example, we used 68-digit arithmetic (using commercially available Advanpix MATLAB Tolbox https://www.advanpix.com) for $a=2/3$ to find that \\ $\alpha(a)=0.0451709442203672185156916552716968964156893201125622408995729\ldots$ and to  compute $f(q)$ up to $\sim$60 digits of precision, see Fig. \ref{fig:a2p3_68digits}. High precision computations help validate  the results from double precision calculations, and allow us to obtain a good quality  analytic continuation of the solution $f(\xi)=f(q(\xi))$ from the real line $\xi\in\R$ to the complex plane $\xi\in\C$ via  the AAA-algorithm \cite{TrefethenAAA}, see Section \ref{sec:analyticalcontinuation} below.

\section{Analytical continuation into the complex plane by rational approximation and structure of singularities}\label{sec:analyticalcontinuation}

Fits of the Fourier spectrum using Eq. \e{deltakmodel}
allows us  to find only the singularity closest to the real line. A  more
powerful numerical technique of analytical continuation  based on
rational interpolants
\cite{AGH2000,DyachenkoLushnikovKorotkevichPartIStudApplMath2016,DyachenkoDyachenkoLushnikovZakharovJFM2019,TrefethenAAA}
allows us to go deeper (further away from the real line) into the
complex plane, well beyond the closest singularity. However, analytic  continuation
further  from the  real line often requires an increase in numerical
precision, even well above the standard double precision
\cite{DyachenkoLushnikovKorotkevichPartIStudApplMath2016,DyachenkoDyachenkoLushnikovZakharovJFM2019}.
In this paper we use a rational interpolation based on a modified version of the  AAA algorithm of
Ref. \cite{TrefethenAAA}.
  AAA finds an approximation $f_{AAA}(\xi)$  to a complex function $f(\xi)$ in  barycentric form by minimizing the $L_2$ error of the approximation on the real line.

The barycentric form is given
by\begin{equation}\label{AAA_barycentric_form}
f_{AAA}(\xi):=\frac{n(\xi)}{d(\xi)}=\frac{\sum_{i=1}^{m}\frac{w_i
f_i}{\xi-\xi_i}}{\sum_{i=1}^{m}\frac{w_i}{\xi-\xi_i}},
\end{equation}
where $m\geq 1$ is an integer, $\xi_i$ are a set of real distinct
\textit{support points}, $f_i$ are a set of real \textit{data values}, and
$w_i$ are a set of real \textit{weights} determined by $L_2$ error
minimization. The integer $m$ is increased until the $L_2$ error between
$f_{AAA}(\xi)$ and $f(\xi)$ on the real line is on the level of
$10^{-PR}$, where $PR$ is the current working precision. For analytic
functions the error decreases exponentially in $m$.

The Barycentric form \e{AAA_barycentric_form}  is a quotient of two
polynomials $n(\xi)$ and $d(\xi)$. A partial fraction expansion of this quotient results in a
sum of $m-1$ first order complex poles,
$f_{AAA}^{poles}(\xi)=\sum_{i=1}^{m-1}\frac{a_i}{\xi-b_i}$, with locations
$b_i$ and residues $a_i$ determined by the values of $w_i$ and $\xi_i$. The pole
locations $b_i$, which are zeros of $d(\xi)$, are determined by solving a
generalized eigenvalue problem described in Ref. \cite{TrefethenAAA}. The
values of the  residues $a_i$ can be computed using L'Hospital's rule $a_i=
res(f_{AAA},b_i)=n(b_i)/d'(b_i)$. If our data for an analytic function is
given with precision $PR$ on the real line, AAA and subsequent computations
of $b_i$ approximate the location of single poles with maximum precision
$\sim PR$, double poles with precision $\sim PR/2$, and triple poles with precision
$\sim PR/3$, etc.
The progressive loss of precision in higher order poles is due to cancellation errors.
We find we can achieve the reduced error  $|f(\xi)-f_{AAA}^{poles}(\xi)| \approx 10^{-PR} $ on the real line in
the case of higher order poles if we
increase the precision of intermediate computations in the generalized
eigenvalue problem
by a factor of
two for double poles and a factor of three for triple poles. We
additionally modified the original AAA algorithm \cite{TrefethenAAA} to
deal with odd and even functions more efficiently and output more
symmetrical sets of poles.


\begin{figure}
\includegraphics[width=0.5\textwidth]{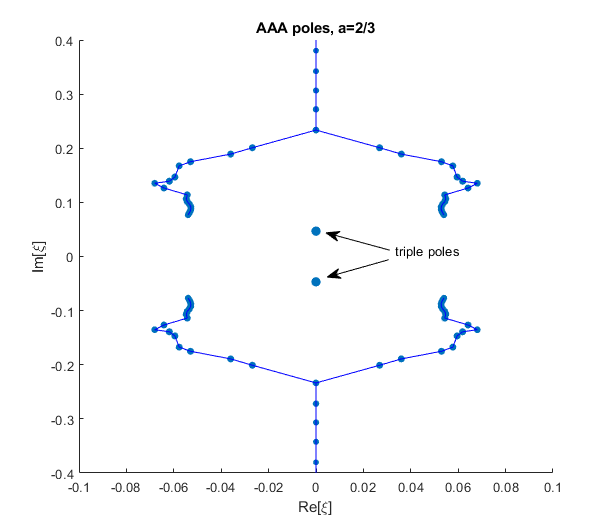}
\includegraphics[width=0.5\textwidth]{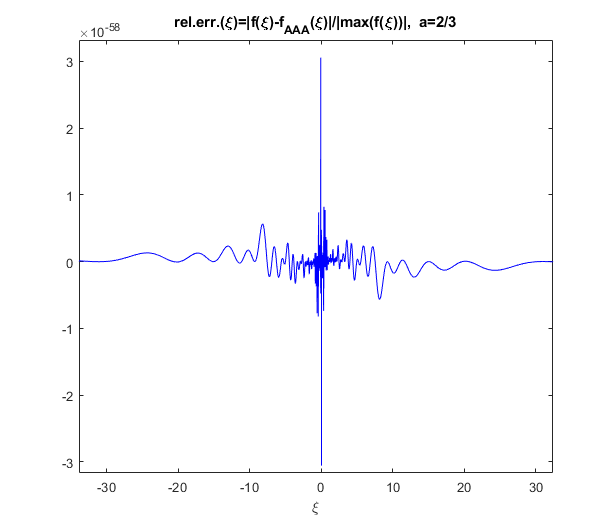}
\caption{The structure of the complex singularities of the solution from
Fig. \ref{fig:a2p3_68digits} approximated by a set of simple poles,
$f(\xi)\approx f_{AAA}^{poles}(\xi)=\sum_{i=1}^{m-1}\frac{a_i}{\xi-b_i}$
using the AAA algorithm (left panel), and the relative error on the real
line between the solution $f(\xi)$ and its approximation $f_{AAA}(\xi)$
(right panel). The simple poles are shown as dots at locations $b_i$ with
the size of dot scaled with $\log_{10}|a_i|$. The branch cuts are
approximated as lines connecting the dots. The triple poles  locations are
$\xi\approx\pm \I 0.04678$ and branch points are located at
$\xi=\xi_{branch}\approx\pm 0.05398 \pm \I 0.07674$.}
\label{fig:a2p3_AAA_68digits}
\end{figure}

In the particular case $a=2/3,$ we use  68-digit precision arithemetic for the  numerical
solution of $f(\xi)$ described at the end of  Section
\ref{sec:e_value_problem}, and incorporate this into the   AAA algorithm. This method shows  that the closest singularities to
the real line  are  a pair of the third order poles $\propto
1/(\xi\pm\I \chi_c)^3$,  in full agreement with Theorem 1  (Eq. \e{gammaa}
of Section \ref{sec:main}) and the Fourier spectrum analysis of Section
\ref{sec:numerics}. The location
 $\xi=\pm\I \chi_c$ (here $Re(\chi_c)>0 $ and $Re(\chi_c)\gg |Im(\chi_c)|$) and the third order type of these poles are automatically approximated by the AAA algorithm as three simple poles $\sum_{i=1}^3 \frac{a_i}{\xi-b_i}$ lying very close to each other ($|b_1-b_2|,|b_2-b_3|<1.54 \cdot 10^{-12})$  
 with the sum of their residues being  essentially zero ($|\sum_{i=1}^3 {a_i}|/|a_1| \approx 4.64 \cdot 10^{-47}$). 
 We  define the location of the triple pole by the average  $\I\chi_c=\sum_{i=1}^3 {b_i}/3$ and have verified that the dipole moment defined by $D:=\sum_{i=1}^3 {(b_i-\I\chi_c)a_i}$ is  negligible, $|D|\approx 1.2 \cdot 10^{-29}$.  
 In contrast, the quadrupole moment $Q:=\sum_{i=1}^3 {(b_i-\I\chi_c)^2 a_i}$ is  distinct from zero, $|Q|\approx 1.5 \cdot 10^{-4}$,
 so this multipole is  well approximated by $\frac{Q}{(\xi-\I\chi_c)^3}.$ The complex conjugate point $\xi=-\I \chi_c$ was treated in a similar way, i.e., by another set of 3 poles of AAA.

We find that  the rest of the singularities of $f(\xi)$ are branch points
with branch cuts extending from them. AAA  approximates branch cuts by
sets of poles, and  Refs.
\cite{DyachenkoLushnikovKorotkevichPartIStudApplMath2016,DyachenkoDyachenkoLushnikovZakharovJFM2019}
demonstrate how to recover branch cuts from this set of poles by increasing
the numerical precision.
The increase of numerical
precision  requires an increase in the number of poles $m$ in
rational interpolants to match the precision.  These poles, which are   located on a branch cut, become more dense with the increase in  precision and thus
recover the location of the branch cut in the continuous (infinite
precision) limit. The main motivation for using 68-digit precision in
this paper was to ensure that we robustly recover branch cuts, see Fig.
\ref{fig:a2p3_AAA_68digits} (left panel). In the particular case  $a=2/3,$
 double precision allows us to robustly see $\sim30$ poles, whereas 68-digit
precision allows us to see $\sim150$ poles. The number of poles we use for a fixed precision is determined by the minimal number of AAA poles to match the numerical precision of the solution on the real line. Increasing the number poles beyond this minimal number produces spurious poles with very small residues, which is the analog of the round-off floor in the Fourier spectrum.
We note that the exact shape of the branch cuts is not fixed analytically --
the  AAA algorithm simply provides a set of poles that corresponds to the
smallest $L_2$ error on the real axis for the given number of poles. Thus, the AAA approximation of the branch cut might move with
a change of the precision. In contrast, the branch points computed by the algorithm  are fixed. One
can see 4 branch points in Fig. \ref{fig:a2p3_AAA_68digits} (left panel),
with two branch cuts going upward and coalescing on the imaginary axis and
extending further to $+\I \infty.$ Another two branch cuts extend downwards
and merge on the imaginary axis  before going off to $-\I
\infty$.

Our investigations of complex singularities via AAA approximations
show that for any $a$, except for
$a=\frac{n-1}{n}, \quad n=1,2,3,\dots$ (which corresponds to the integer values  $\gamma=n$ in Eq. \e{gammaa}),
there is  another pair of vertical branch cuts coming out of $\xi=\pm \I
\chi_c$  and coalescing with the rest of the branch cuts on the
imaginary axis. For $a<a_c$ the side branch points are always above the
main singularity at $\xi = \pm \I \chi_c$ and their locations are
$\xi_{branch} = \pm \epsilon_1(a) \chi_c \pm \I (1+\epsilon_2(a))\chi_c, \,$ where roughly  $ \epsilon_1(a) \sim 1, \, \epsilon_2(a) \sim 1$. In particular,
$Re[\xi_{branch}]/\chi_c < 0.74$, $Im[\xi_{branch}]/\chi_c > 2$ for
$a<0.6$; $Re[\xi_{branch}]/\chi_c\approx 1.15$,
$Im[\xi_{branch}]/\chi_c\approx 1.64$ for $a=2/3$ and
$Re[\xi_{branch}]/\chi_c\approx 1.23$, $Im[\xi_{branch}]/\chi_c\approx
1.51$ near $a=a_c$.

\section{Results of time dependent simulations and Petviashvili iterations for periodic BC} \label{sec:numerics2}
Motivated by simulations of the generalized CLM equation \e{CLM01}    in Ref. \cite{Okamoto2008} for $2\pi$-periodic BC with $a=1$, we performed  simulations for a wide range of values of the parameter $a.$
For this we used  the periodic version of the Hilbert transform ${\mathcal H}^{2\pi}$  \e{HilbertHdef_periodic}  in Eq. \e{CLM01} instead of ${\mathcal H}$.

Simulations for $a<a_c$ show collapsing solutions with  $\alpha>0$, and different  types of IC  give qualitatively similar results near the collapse time $t=t_c$ as in the real line $x\in\R$ case with the same
$\alpha(a)$ (see Table \ref{Table1}). Hence  we do not describe them here.
Expanding solutions for $a>a_c$
behave differently
since the finite spatial interval  $[-\pi,\pi]$ arrested the increasing
width of the solution at large enough times. Thus we focus our discussion on $a>a_c$ and present
detailed results of our simulations,  in particular the cases of $a=0.8$ and $a=1$.

\begin{figure}
\centering
\includegraphics[width=0.49\textwidth]{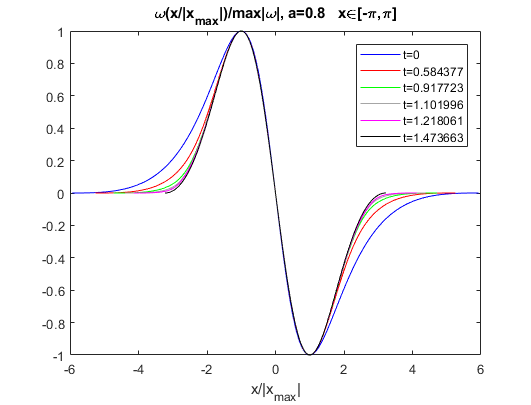}
\includegraphics[width=0.49\textwidth]{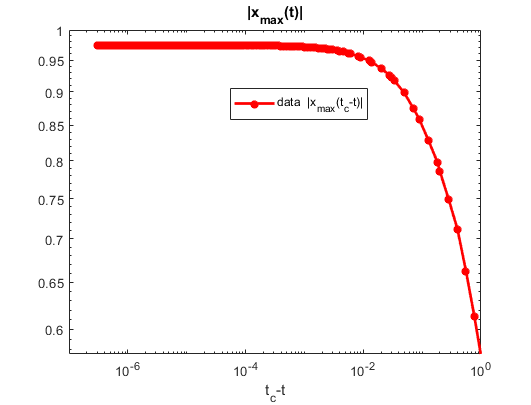}
\caption{Left panel: Convergence of time-dependent numerical solution of
Eqs. \e{CLM01} and \e{HilbertHdef_periodic} with  $a=0.8$ and IC
\e{ICperiodic} to a universal self-similar profile. The solution expands
horizontally (until arrested by the boundary condition) and extends
vertically, blowing up at $t=t_c = 1.4736630\ldots.$ The plot is scaled
vertically by $\max\limits_x|\omega|$ and  horizontally  by the location
$x_{max }(t)$ of  $\max\limits_x|\omega|$.    Right panel: Time dependence
of  $|x_{max}(t)|$, which shows slowdown and eventual arrest of the
horizontal expansion of the solution.}
\label{fig:a0p8_2pi_selfsimilarconvergence}
\end{figure}

We performed a simulation with $a=0.8$ and initial  condition
\begin{equation}\label{ICperiodic}
\omega_0(x)=-\frac{4}{3}[\sin(x)+0.5\sin(2x)]
\end{equation}
 which is qualitatively
similar to the particular case \e{IC2a} of  IC2   \e{IC2}, with $q$ replaced by $x$ and  $V_c=1,
T_c=1$. After an initial spatial expansion, the solution is arrested by the
periodic boundary conditions.
This arrest results in the qualitative change of the dynamics, see  for example the right panel of Fig.
\ref{fig:a0p8_2pi_selfsimilarconvergence} for the time dependence of the
location  $x_{max}(t)$   of $\max\limits_x|\omega(x)|.$
At  later times we still find a
finite time blow up of the solution with $\max\limits_x|\omega(x)| $ and $ \max\limits_x|u(x)|
\rightarrow{\infty}$ as $t \rightarrow{t_c}$. However, instead of Eq.
\e{self-similar1}, the solution converges to a new universal self-similar
blow-up profile given by Eq. \e{self-similar1periodic},
 %
%
as demonstrated in left panel of Fig. \ref{fig:a0p8_2pi_selfsimilarconvergence}. A comparison of Eqs. \e{self-similar1} and \e{self-similar1periodic} reveals that we can formally obtain Eq. \e{self-similar1periodic} by setting $\alpha=0$
in Eq. \e{self-similar1} (although Eq. \e{self-similar1periodic} has  periodic boundary conditions, vs. decaying BC of Eq. \e{self-similar1}). 
We note that taking the limit  $a\rightarrow{a_c^-}$ in Eq. \e{self-similar1}, we also obtain  $\alpha=0.$ However, it remains unknown if Eq. \e{self-similar1periodic} can be obtained from the continuation of  Eq. \e{self-similar1} across $a=a_c$.

\begin{figure}
\includegraphics[width=1\textwidth]{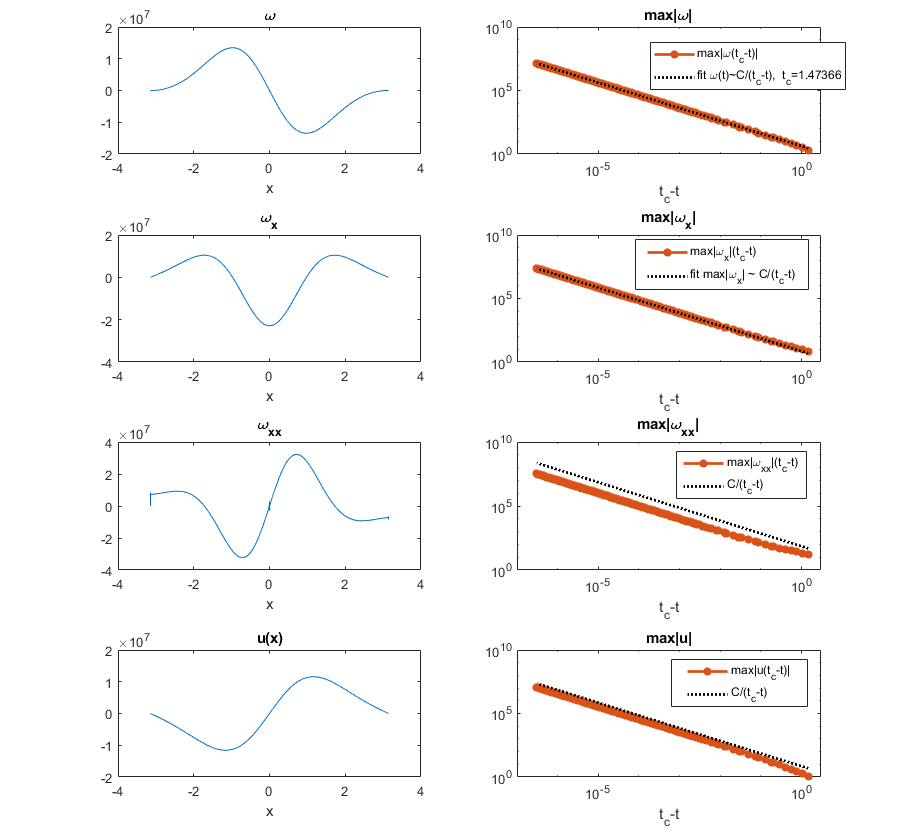}
\caption{Results of the simulation of Eqs. \e{CLM01} and \e{HilbertHdef_periodic} with  $a=0.8$ and IC \e{ICperiodic}. Left panels:   $\omega(x,t)$, its derivatives $\omega_x(x,t)$, $\omega_{xx}(x,t)$, and $u(x,t)$ at $t=1.4736627. $ Right panels: the growth of  maximum values of the corresponding quantities
over time.} \label{fig:a0p8_2pi_8pics}
\end{figure}

\begin{figure}
\includegraphics[width=0.328\textwidth]{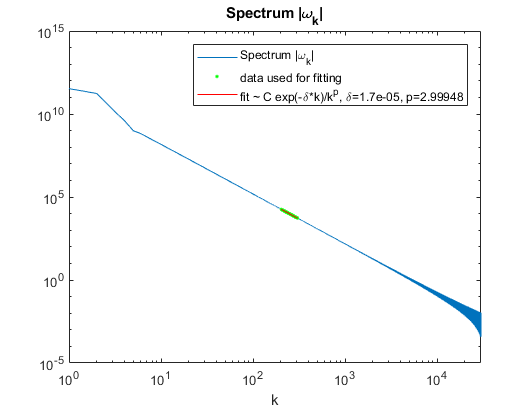}
\includegraphics[width=0.328\textwidth]{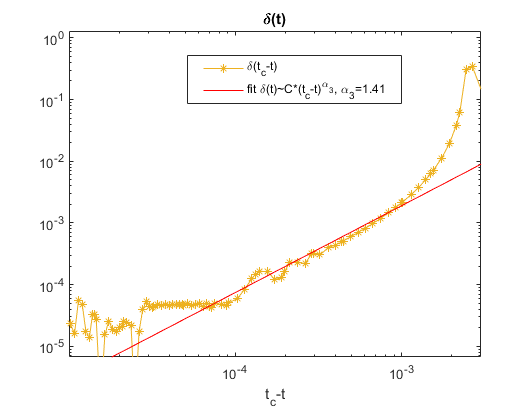}
\includegraphics[width=0.328\textwidth]{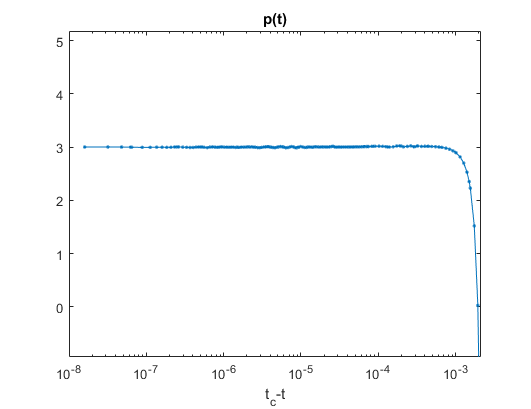}
\caption{Left panel: Log-log plot of the Fourier spectrum
$|\hat{\omega}_k|$  from Fig. \ref{fig:a0p8_2pi_8pics} at $t=1.4736627$
and $a=0.8$. The red line represents a fit to the model \e{deltakmodel}
with green line showing a portion of the  $|\hat{\omega}_k|$ used for the
fit. Center and right panels:
 $\delta(t)$ and $p(t)$ obtained from the fit of  $|\hat{\omega}_k|$  to Eq.  \e{deltakmodel} at different times. Red lines in the center panel also show a fit to  the model $\delta(t) \propto (t_c-t)^{\alpha_3}$. } \label{fig:a0p8_2pi_spectrum}
\end{figure}

The  spectrum $\hat{\omega}_k$ is initially exponentially decaying  but expands and becomes mostly algebraically decaying
(similar to Fig. \ref{fig:a0p8_Spectrum}). Finite  precision arithmetic only  ``sees" algebraic decay $|\hat{\omega}_k(x)| \sim k^{-3}$ when  $t$ is close enough to $t_c$, see Fig. \ref{fig:a0p8_2pi_spectrum}. This is  because of a jump in $\omega_{xx}$ forming at $x=\pm\pi$, see Fig. \ref{fig:a0p8_2pi_8pics} (left and middle panels). Due to the spectrum being initially oscillatory it was difficult to accurately extract values of $\delta(t)$ and $p(t)$ from a fit to Eq. \e{deltakmodel}, but using a nonoscillatory spectrum which emerges later in the simulation we were able to recover some data for $\delta(t)$ and $p(t)$ as shown in Fig. \ref{fig:a0p8_2pi_spectrum}. There,  one can see that $\delta(t)\rightarrow{0}$ and $p(t) \rightarrow{3}$ as $t\rightarrow{t_c}$.

\begin{figure}
\includegraphics[width=1\textwidth]{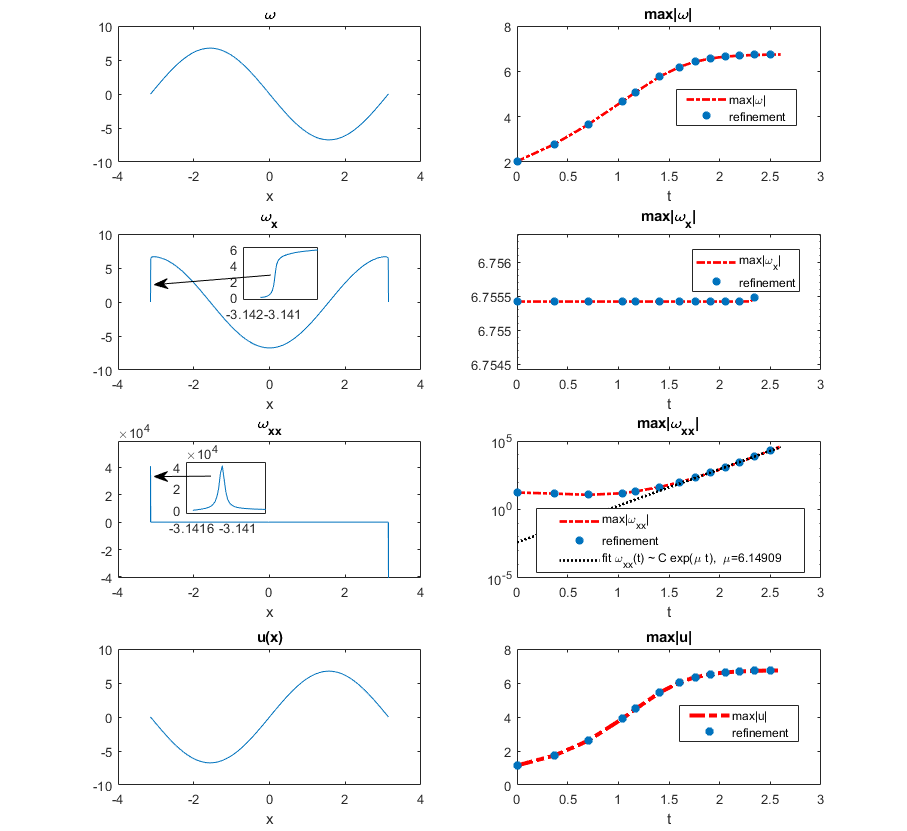}
\caption{Results of the simulation of Eqs. \e{CLM01} and \e{HilbertHdef_periodic} with  $a=1$ and IC  \e{ICperiodic}  showing  $\omega(x,t)$, its derivatives $\omega_x(x,t)$, $\omega_{xx}(x,t)$, and $u(x,t)$ at $t=2.60205$. Also shown is the growth of their maximum values as functions of time.} \label{fig:a1_2pi_8pics}
\end{figure}

For $a=1$ we considered two different types of ICs.
The first one is  IC \e{ICperiodic}, for which we observe global existence of the solution. Initially the amplitude of the solution $\omega(x)$ grows  in time, similar to the infinite domain case. But this growth slows down at later times and eventually reaches a plateau with the the same behaviour  in $u(x)$, see Fig. \ref{fig:a1_2pi_8pics}. Also $\max\limits_x|\omega_{x}|=|\omega_x(x=0)|$ remains nearly constant throughout the simulation. We observe unbounded growth of $|\omega_{xx}|$ near $x=\pm\pi$ that appears to be  exponential in time.
Due to the spectrum being oscillatory it was difficult to accurately extract values of $\delta(t)$ and $p(t)$ from a fit to Eq. \e{deltakmodel}. However, using AAA rational approximation we were able to observe two pairs of branch cuts approach the real line near $x=\pm\pi$ as $t\rightarrow{\infty}$. Replacing IC \e{ICperiodic} by the more general IC2   \e{IC2} (with $q$ replaced by $x$  and  $V_c,
T_c=1$)  is found to  only alter the transient dynamics of the expanding solution without qualitatively changing the overall behavior.

\begin{figure}
\includegraphics[width=1\textwidth]{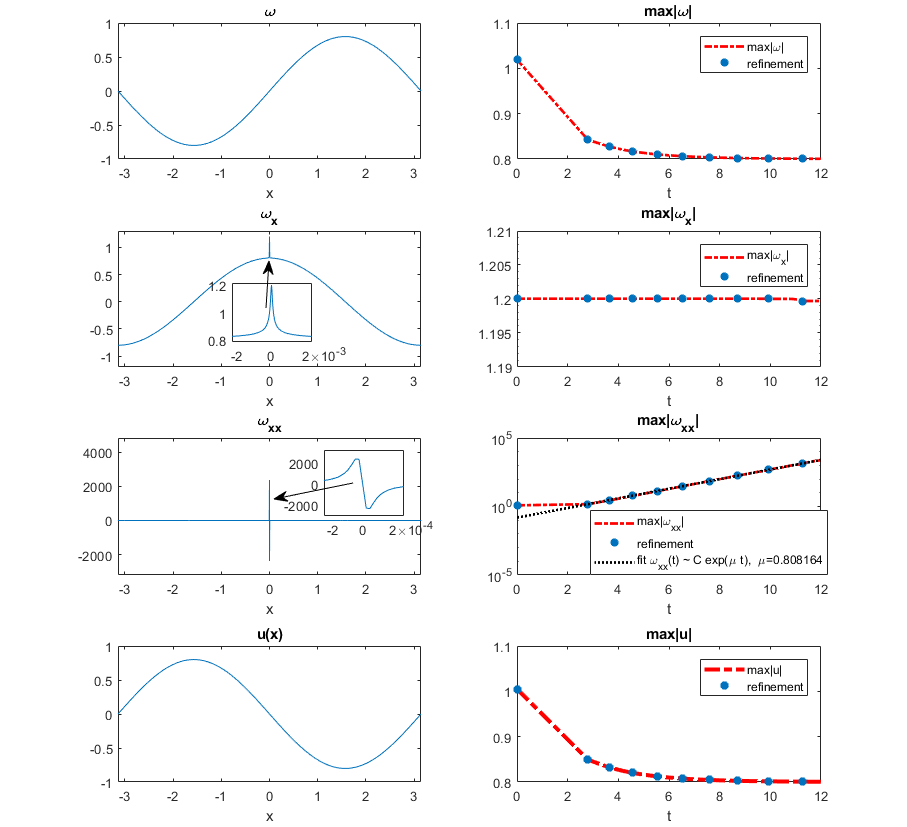}
\caption{Results of the simulation of Eqs. \e{CLM01} and \e{HilbertHdef_periodic} with  $a=1$ and IC   \e{ICOkamoto} as  in Ref. \cite{Okamoto2008} showing  $\omega(x)$, its derivatives $\omega_x(x)$, $\omega_{xx}(x)$, and $u(x)$ at $t\approx12$ and the growth of their maximum values as functions of time.} \label{fig:a1_2pi_8pics_Okamoto}
\end{figure}

\begin{figure}
\includegraphics[width=0.328\textwidth]{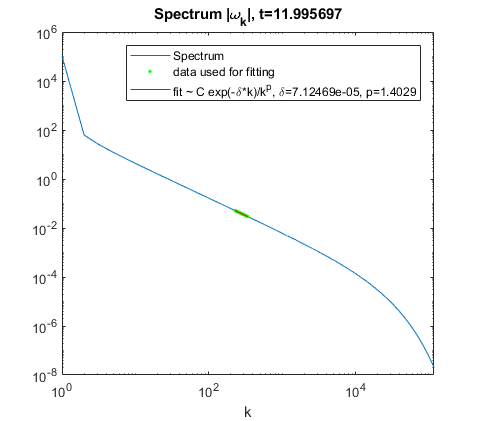}
\includegraphics[width=0.328\textwidth]{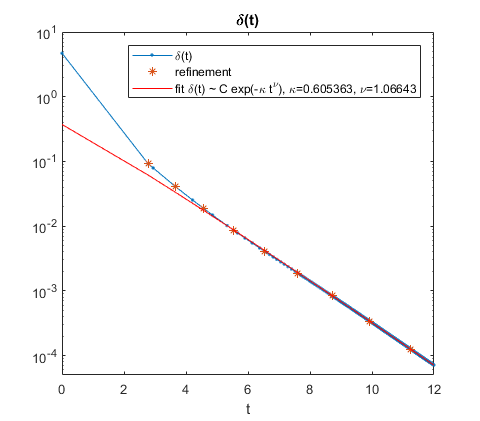}
\includegraphics[width=0.328\textwidth]{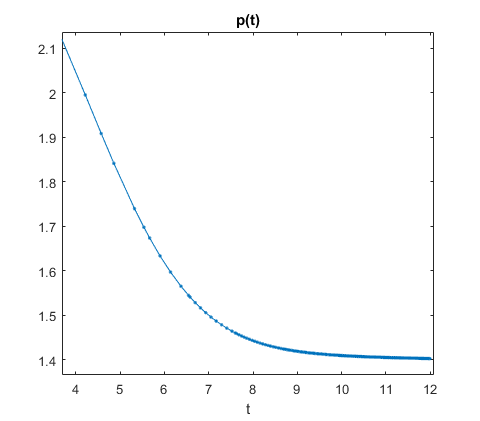}
\caption{Left panel: Log-log plot of the Fourier spectrum $|\hat{\omega}_k|$ for the solution in Fig. \ref{fig:a1_2pi_8pics_Okamoto} and a fit to the model \e{deltakmodel}.  Center and right panels: Time dependence of   $\delta(t)$ and  $p(t)$ obtained from the fit to  \e{deltakmodel}. Center panel also shows a fit of   $\delta(t)$ to  the stretched exponential  model $\delta(t) \sim e^{-\kappa t^\nu}$.} \label{fig:a1_2pi_spectrum_Okamoto} 
\end{figure}

The second type of IC we used for $a=1$  is given by%
\begin{equation}\label{ICOkamoto}
\omega_0(x)=\sin(x)+0.1\sin(2x),
\end{equation}
which is the same as in Ref. \cite{Okamoto2008}. It allows us to directly compare the results of our simulations with   Ref. \cite{Okamoto2008}. We obtain exactly the same plots as in Fig. 1 of Ref. \cite{Okamoto2008}, see Fig. \ref{fig:a1_2pi_8pics_Okamoto}.
 The difference between simulations with IC  \e{ICperiodic} and IC  \e{ICOkamoto} are seen by comparing  Figs. \ref{fig:a1_2pi_8pics} and \ref{fig:a1_2pi_8pics_Okamoto}. For example,  the spatial derivatives of $\omega$ approach  discontinuities at $x=0$ in Fig.  \ref{fig:a1_2pi_8pics} vs.  $x=\pm \pi$  in Fig. \ref{fig:a1_2pi_8pics_Okamoto}. \revc{Also,  $|\omega_{xx} |\rightarrow{\infty}$ grows exponentially in time in both cases although at different locations in $x$. The exponentially decaying spectrum $|\hat{\omega}_k|$ with  IC  \e{ICOkamoto}  widens over time and becomes mostly algebraically decaying,
 see Fig. \ref{fig:a1_2pi_spectrum_Okamoto} (left panel). In this case the only singularity is near $x=0$;} The AAA rational approximation shows an  approach of two vertical branch cuts to $x=0$ over time, so the spectrum is not oscillatory and we are able to easily recover $\delta(t)$ and $p(t)$ from the fit to Eq. \e{deltakmodel}. The fits show a  stretched-exponential in time approach of the singularity to the real line i.e.,  $\delta(t) \sim e^{-\kappa t^\nu}$, see Fig. \ref{fig:a1_2pi_spectrum_Okamoto} (middle panel). Figure \ref{fig:a1_2pi_spectrum_Okamoto} (middle and right panels) showing $\delta(t)$ and $p(t)$ can be compared with Fig. 3(a,b) of Ref. \cite{Okamoto2008}. Our values of $\delta(t)$ match those values from  Fig. 3(a) of Ref. \cite{Okamoto2008}  well, while values of $p(t)$ do not match precisely with Fig. 3(b) of Ref. \cite{Okamoto2008} because 
 they marginally depend on the particular part of spectrum $|\hat{\omega}_k|$ that is used for the fitting.

For $a > 1$ with  IC  \e{ICperiodic}, we observe  global existence of the solution. Its initial expansion in $x$-space is arrested by the periodic boundary conditions with an infinite slope forming at the boundary $x=\pm \pi$ so that $\max\limits_x|\omega_x| \rightarrow{\infty}$ as $t \rightarrow{\infty}$ (although $\max\limits_x|\omega|, \max\limits_x|u|, |\omega_x(x=0)| \rightarrow{0}$ as $t \rightarrow{\infty}$). The complex singularities approach the real line in infinite time. Their positions scale like  $x_{sing} \sim \pm \pi  \pm \I y_0 \exp{(-\kappa_2 t^{\nu_2})}$, where $y_0,\kappa_2, \nu_2>0$.
When $a\to1^+$, we observe that  $\max\limits_x|\omega|$  grows for a short time  and then decays. Unlike the  $x\in \R$ case, it is relatively  easy to  compute accurately  for $a\to1^+$  and we have been able to obtain numerical evidence of global existence for $a$ as small as 1.000001.
For IC  \e{ICOkamoto}, we also observe  global existence of the solution with  decay of $\max\limits_x|\omega|$ and unbounded growth of $|\omega_x(x=0)|$ as $t\to \infty.$  The complex singularities approach the real line like $x_{sing} \sim 0 \pm \I y_0 \exp{(-\kappa_2 t^{\nu_2})}$, where $y_0,\kappa_2,\nu_2>0$.

 We find the same behaviour of the kinetic energy for the periodic BC as in $x\in\R$ case described in Section \ref{sec:e_value_problem} for $a\le0.95$, while for $a=1$ we have that $E_K \to const$ as $t\to \infty$ (because $\max\limits_x|u|\to const$ as $t \to \infty$) and for $a>1$ we have  that $E_K\to 0$ as $t\to \infty$ (because $\max\limits_x|u|\to 0$ as $t \to \infty$).

\begin{figure}
\centering
\includegraphics[width=0.49\textwidth]{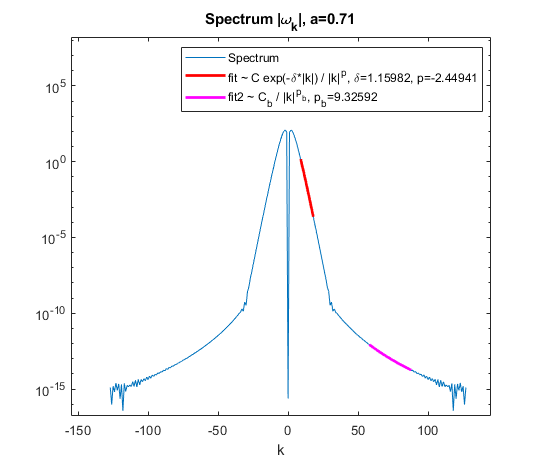}
\includegraphics[width=0.49\textwidth]{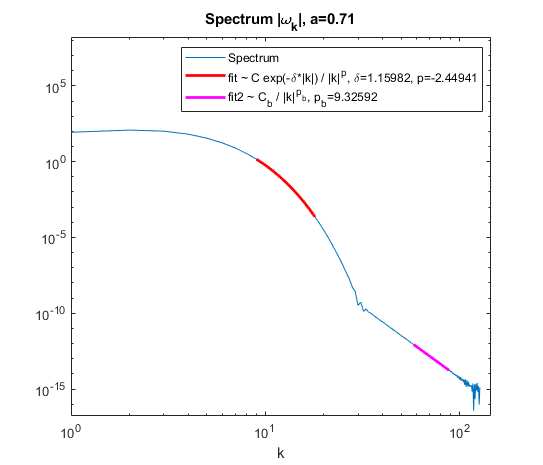}
\caption{The Fourier spectrum $|\hat\omega_k|$ of the self-similar profile  \e{self-similar1periodic}  for  $a=0.71$ obtained by GPM iterations \e{Petviashvili} of Eq.  \e{CLM01selfsimilar_nontransformed}. Two fits are  shown in different ranges of $k$ with the first fit to Eq. \e{deltakmodel}  with $\delta\ne 0$ at  intermediate $k$ and the second a power law fit $\propto |k|^{-p_b}$ for  larger $|k|$. Left panel: Log-linear plot where the first fit turns into a nearly linear function. Right panel: Log-Log plot where the second fit turns into a nearly linear function.} \label{fig:twospectrumscales}
\end{figure}

\begin{figure}
\centering
\includegraphics[width=0.49\textwidth]{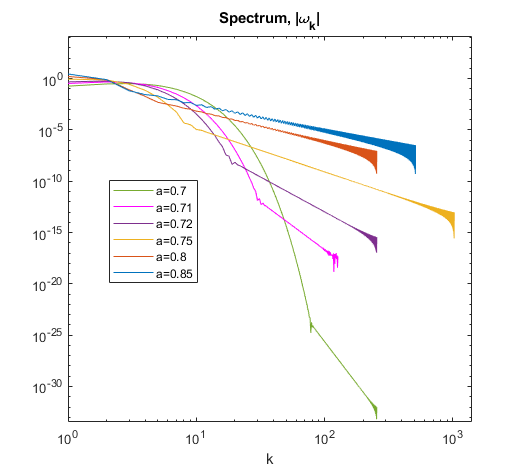}
\includegraphics[width=0.49\textwidth]{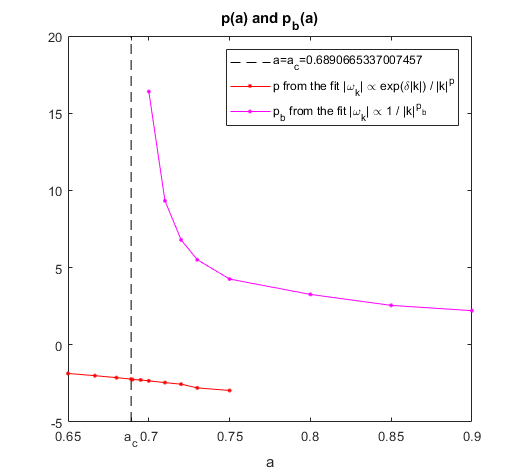}
\caption{Left panel: The Fourier spectra $|\hat\omega_k|$ of the
self-similar profile  \e{self-similar1periodic} for various values of $a$
as in the Table \ref{Table2} obtained by GPM iterations \e{Petviashvili}
of Eq.  \e{CLM01selfsimilar_nontransformed}. Right panel: $p(a)$ and
$p_b(a)$ from the Table \ref{Table2} extracted from the two fits as in
Fig. \ref{fig:twospectrumscales}.} \label{fig:GPM_spectra}
\end{figure}

{\it\ Self-similar profiles from GPM}.
We  also numerically computed the self-similar profile $f(x)$ in Eq. \e{self-similar1periodic} for $a_c < a
\leq 0.85$ using GPM described in Section \ref{sec:e_value_problem} with $\alpha=0$. In contrast to Section  \ref{sec:e_value_problem}, we do not need to use  the coordinate transformation \e{transform2} because $f(x)$ is now $2\pi$-periodic  with $\xi\equiv x$. We used
GPM to solve Eq. \e{CLM01selfsimilar} by the iteration \e{Petviashvili} with
${\mathcal M} f$ and ${\mathcal N}[f] f$ from Eq. \e{CLM01selfsimilar_transformed} replaced by
\begin{equation}\label{CLM01selfsimilar_nontransformed}
\begin{split}
&{\mathcal M} f:=f=-a g f_x + f  {\mathcal H}^{2\pi}f:={\mathcal N}[f] f  , \quad  \\
&g_x= {\mathcal H}^{2\pi}f .
\end{split}
\end{equation}
The matrix $\bf\ M$ used in Eq. \e{Petviashvili} now turns into the identity matrix. We do not need to solve the nonlinear eigenvalue problem for $\alpha$ because now $\alpha\equiv 0$.
While performing the  iteration  \e{Petviashvili},  we had to reduce $\Delta \tau$ even more than in Section  \ref{sec:e_value_problem}   to make sure the iterations converged and also had to use more Fourier modes in the spectrum, since the spectrum decay is only algebraic for these solutions. Due to these technical limitations we were unable to explore the range $0.85<a<1$,  but we fully expect that self-similar solutions exist  there because   time-dependent simulations converge to self-similar profiles, at least over the lower range $a_c<a\lesssim0.95$ (see Fig. \ref{fig:a0p8_2pi_selfsimilarconvergence}). It was not possible to obtain convergence in the upper range $0.95\lesssim a<1$  because the solution spectrum quickly widened,
and we were unable to reach the self-similar regime before the computation became prohibitively  slow.
The behavior of solutions (blow up vs. global existence) therefore remains unknown in this range. We conjecture that blow up occurs for all $a_c<a<1$ with global existence  only for $a=1$ (as demonstrated) and for larger values of $a$.

The Fourier spectrum of  $|\hat\omega_k|$ corresponding to  the self-similar profile  \e{self-similar1periodic} has two distinct domains for $|k|\gg 1$. The particular case  $a=0.71$ shown in Fig. \ref{fig:twospectrumscales}  depicts such domains. The first domain corresponds to  complex singularities  of Theorem 1 (Eq. \e{gammaa}) located at $x_{sing}=\pm \I \delta.$ This domain is well fitted by Eq. \e{deltakmodel}. From this fit  we find that $\delta=1.15982$ and $p=-2.44941$, as shown in Fig. \ref{fig:twospectrumscales}. Using Eqs. \e{gammaa} and \e{pgamma} we obtain the prediction of Theorem 1 that $p=\frac{-a}{1-a}=-2.44827\ldots$ which  agrees within an accuracy of $<0.05\%$  with the numerical fit to Eq. \e{deltakmodel}. The second domain is due to  complex singularities located at $x=\pm\pi$    and results in a discontinuity of high-order derivatives  of $\omega(x)$  at the periodic boundary.  This domain has the power law spectrum $\propto|k|^{-p_b}$(i.e., in Eq. \e{deltakmodel} it corresponds to $\delta=0$ and  $p=p_b$) which is dominant for  larger $|k|$.  In the particular case of  Fig. \ref{fig:twospectrumscales}, we obtain $p_b=9.32592\ldots.$ This implies that the 9th and  higher-order derivatives of $\omega(x)$ have a discontinuity at the periodic boundary. All these singularities can be  seen using the  AAA algorithm described in Section \ref{sec:analyticalcontinuation}.  We also find that as $a$ approaches to $a_c$ from the right, i.e. $a\rightarrow{a_c^+}$, increasingly higher order derivatives experience discontinuities at the periodic boundary, i.e.  $p_b \rightarrow{\infty}$ as $a\rightarrow{a_c^+}$, see Fig. \ref{fig:GPM_spectra} (right panel).
These solutions with  finite smoothness at the periodic boundary
can be considered the analog of the self-similar solutions with
compact support found in Sections \ref{sec:numerics} and \ref{sec:e_value_problem}, for  solutions on the real line with $a_c<a\le 1$.

Table \ref{Table2} provides the values of $\delta, p$ and $p_b$ for various values of parameter $a$ obtained from the fits described above. We note that  the symmetry \e{stretching}
is not valid for  periodic BC. Thus, the parameter $\delta$ is now fixed for each $a$, contrary to the case $x\in \R$
where it is a free parameter, cf. Section \ref{sec:e_value_problem}.


\begin{table} [h]
\centering
  \caption{Table of values of $\delta, p$ and $p_b$ extracted via a fit of spectra $|\hat{\omega}_k|$ to the model \e{deltakmodel}, obtained from eigenvalue problem simulations of Eq.  \e{CLM01selfsimilar_nontransformed} for various values of $a$, $a_c<a<1$. $\delta$ and $p$ are extracted from the fit $|\hat{\omega}_k| \propto \exp(-\delta|k|)/|k|^p$ to the central part ($k \sim 0$) of the spectrum and $p_b$ is extracted from the fit $|\hat{\omega}_k| \propto 1/|k|^{p_b}$ in the tails ($k \gg 1$) of the spectrum. Simulations with $a\geq0.71$ were performed in double precision arithmetic. To see the power law tail of the spectrum and extract $p_b$ in the case of $a=0.7$ we had to use quadruple precision. For $a_c<a\leq0.695$ the power law tail was not observable even in quadruple precision. See Fig. \ref{fig:GPM_spectra} for the spectra and plots of $p(a)$ and $p_b(a)$. The accuracy of $\delta, p$ and $p_b$ approximately corresponds to the number of digits provided in the table.
  }
\begin{tabular}{|l|l|l|l|l|}
  \hline
  $a$ & $\delta$ & $p$ & $p_b$  \\
  \hline
  0.69 & 0.2338 & -2.2446 & -    \\
  0.695& 0.5954 & -2.2787 & -    \\
  0.7  & 0.8177 & -2.3333 & 16.407    \\
  0.71 & 1.1598 & -2.4494 & 9.3259    \\
  0.72 & 1.44 & -2.55 & 6.81    \\
  0.73 & 1.73 & -2.79 & 5.51   \\
  0.75 & 2.20 & -2.96 & 4.26    \\
  0.8  &-&-& 3.26   \\
  0.85 &-&-& 2.55    \\
  0.9  &-&-& 2.21   \\
    \hline
\end{tabular} \label{Table2}
\end{table}

Here we summarize the solution behaviour of Eqs. \e{CLM01} and \e{HilbertHdef_periodic} for $x\in[-\pi,\pi]$ and generic smooth IC depending on the parameter $a$:
\begin{itemize}
\item $a < a_c:$ Behaviour of  solutions is the same at $t\to t_c$ as for the  $x\in\R$ case, with  collapse as in Eq. \e{self-similar1}.
\item $a_c < a \lesssim  0.95:$ Blow up both in $\omega$ and $u$ in  finite time $t_c$ with solution approaching  the universal self-similar profile \e{self-similar1periodic} as $t\to t_c$. That profile $f(x)$  has  discontinuities in the  high-order derivatives with complex singularities touching the real line only at $x=\pm \pi$. The number of continuous derivatives becomes infinite in the limit $a\to a_c^+$. The singularities
approach the real line as $x_{sing} \simeq \pm \pi \pm \I(t_c-t)^{\alpha_3} y_b$, where $\alpha_3(a)>0$.
\item $a = 1:$ Global existence of solution with a singularity approaching the real line exponentially in time.
For both IC \e{ICperiodic} and IC \e{ICOkamoto} we find $\max\limits_x|\omega|, \max\limits_x|u|, \max\limits_x|u_x| \rightarrow{const}$, $\max\limits_x|\omega_{x}|=|\omega_{x}(x=0)|=const$,  and $\max\limits_x|\omega_{xx}|\rightarrow{\infty}$ as $t\rightarrow{\infty}$.
\item $a>1:$ Global existence of solution with a singularity approaching the real line exponentially in time. For IC \e{ICperiodic} the singularity approaches the real line near $x=\pm\pi$ and $\max\limits_x|\omega|, \max\limits_x|u|, |\omega_x(x=0)| \rightarrow{0}$ and $\max|\omega_x| \rightarrow{\infty}$ as $t \rightarrow{\infty}$.
\end{itemize}

\section{Conclusions and discussion} \label{sec:conclusions}
We have performed a systematic sweep of the parameter $a$ in the  generalized CLM equation (\ref{CLM01}) to determine the possibility of singularity formation  and, when it occurs, its type, i.e., collapse vs. blow-up.  
We identified a new critical value
$a=a_c=0.6890665337007457\ldots$  such that for $a<a_c$  collapse occurs both on the real line $x\in \R$ and for  periodic BC.   Here,  collapse means that not only is there a finite time singularity in which  the amplitude of the solution $\omega(x,t)$ tends to infinity,  but there is also a catastrophic shrinking of the spatial extent of the solution   to zero as $t\to t_c$, described by the self-similar form \e{self-similar1}. In the intermediate range $a_c< a \le 1,$ we found there is  finite-time singularity formation    for  $x\in \R$, with the self-similar solution \e{self-similar1} experiencing an infinite rate of expansion as $t\to t_c$. This type of self-similar singularity formation, in which the spatial domain does not collapse, is termed `blow up'.  The power $\alpha$ in Eq.  \e{self-similar1}  controls  collapse (for $\alpha>0, \ a<a_c)$ vs.  blow up ($\alpha\le 0, \ a\ge a_c)$. We elucidated the dependence of $\alpha(a)$ on $a$ via both  direct numerical simulation of Eq. (\ref{CLM01}) and  the solution of a nonlinear eigenvalue problem \e{CLM01selfsimilar} using the generalized Petviashvili method   \e{Petviashvili}. We have also performed multiprecision simulations (up to 68 digits of accuracy) to demonstrate the possibility of recovering $\alpha(a)$ and the structure of self-similar solutions with any desired precision.

We show that collapsing solutions of (\ref{CLM01}) have finite energy $E_K$ up to and including the critical time $t_c$ for $a<0.265 \pm 0.001$.
  Such finite energy solutions are of interest in analogy with the problem concerning global regularity of the 3D Euler and Navier-Stokes equations with smooth initial
data,  see Refs. \cite{FeffermanMilleniumprize2006,GibbonPhysD2008}.
We found for general values of $a$  that the self-similar solution  \e{self-similar1} is real analytic for   $a<a_c$ while it  has  finite support for  $a_c< a \le 1.$

   We identified that the blow up for periodic BC with $a_c<a\le0.95$ is qualitatively different from that for   $x\in \R$,    because the periodic BC arrests or blocks  the unbounded spatial expansion of the  solution on the real line. To our  surprise, such arrest does not result in the global existence of the solution but instead leads to a new form of   self-similar blow-up  \e{self-similar1periodic}, in  which  weak singularities develop at the boundaries of the periodic domain. In the limit $a\to a_c^+,$ this self-similar solution turns into an infinitely smooth ($C^\infty$) solution.
We believe that the qualitative difference in blow up between    $x\in \R$     and periodic BC  might serve as an interesting lesson relevant to  the search for singularities in the 3D Euler equation.

Both self-similar solutions \e{self-similar1}  and  \e{self-similar1periodic} are nonlinearly stable, as follows from our simulations. Quite generic classes of IC converge to these solutions during the temporal evolution. In the case of Eq.  \e{self-similar1},  such convergence/stability is understood in the sense of convergence to a family of self-similar solutions, up to a rescaling in $x$, because of the symmetry \e{stretching} of Eq. \e{CLM01selfsimilar}.

The structure of the leading order  singularities in the complex plane $x$ (which is the analytical continuation from    $x\in \R$) is determined by Theorem 1. That result is valid  for both   $x\in \R$      and periodic BC and is in full agreement with simulations. For $a<a_c$ the leading order singularities  are the closest singularities to the real line in the complex $x$-plane. For $a>a_c$, these singularities still determine the structure of self-similar solutions near $x=0$,  while the solution near the  boundaries of  finite support in $x\in \R$  and the periodic boundaries for periodic BC are controlled by less singular terms. The self-similar solution profiles for these $a$ have been found with high accuracy by solving a nonlinear  eigenvalue problem.
We have also proved in {Theorem 3} that, except for the exact closed-form solutions for $a=0$ and $a=1/2$, the analytical structure of singularities in the complex $x$-plane goes beyond the leading order singularities.  In particular, we numerically  identified using
the  AAA algorithm the existence of additional, non-leading-order branch points for $a\ne 0, 1/2$.

We found from our simulations that quite generic IC result in the global existence of solutions for $a\gtrsim 1.3$ and $x\in \R$, while for  periodic BC  global existence is ensured for $a\ge 1.$  In the remaining gaps $1<a\lesssim 1.3$ for  $x\in \R$ and $0.95<a<1$ for the periodic case, our simulations are inconclusive and unable to distinguish between singularity formation and global existence.
We believe that more concrete results in this range of $a$ will require additional analysis and/or substantial  efforts in simulation.

We suggest that among many other issues,
the following questions would be interesting to address in  future work:

1. Analytical study of the complex singularities beyond the leading order singularities addressed in  Theorem 1.
In particular, the case $a=2/3$ might be especially interesting because the leading order singularity is very simple, namely, a third order pole.

2. Either extend GPM to the compactly supported case $a>a_c$ for $x\in \R$, or use a version of  the method in Ref.\ \cite{ChenHouHuang}  based on cubic splines. However,  splines generally lose information about the analyticity of solutions in the complex plane. One way to improve the performance of GPM in this range of $a$  might be to use a coordinate transform in the form of a conformal mapping which would simultaneously resolve the numerical grid near $x=\pm x_b$ while keeping the analyticity of the solution intact. This type of approach has been   suggested in Ref. \cite{LushnikovDyachenkoSilantyevProcRoySocA2017}.

3. Fill the gaps in our knowledge on blow up vs. global existence of solutions in the parameter regime $1<a\lesssim1.3$ for $x\in \R$ and $0.95<a<1$ for  periodic BC.

4. Look for possible analytical continuation/bifurcation at $a=a_c$ between self-similar solutions \e{self-similar1} for the case  $x\in \R$ and Eq. \e{self-similar1periodic} for periodic BC.

5. Perform an analysis of the nonlinear stability of the blow-up solutions. This could be   qualitatively similar to the stability of collapse in PDEs  such as the nonlinear Schr\"odinger equation and the
Patlak-Keller-Segel equation, see e.g. Refs.
\cite{ZakharovJETP1972,ChPe1981,SulemSulem1999,BrennerConstantinKadanoff1999,KuznetsovZakharov2007,LushnikovDyachenkoVladimirovaNLSloglogPRA2013}.

6.  Analyze the formation of singularities at the initial time $t=0^+$. This can give  information on the type of singularities which first form in the complex plane, and subsequently move toward the real line. Such an analysis has been previously performed for the evolution of  a vortex sheet in the Kelvin-Helmholtz problem \cite{CowleyBakerTanveerJFM1999}, which is also governed by a nonlocal PDE.  However, a significant difference between the current problem and the vortex sheet problem  is that here the singularities initially form at infinity in the complex plane, whereas in the vortex sheet  problem they are generated at  finite locations, due to a singularities in the kernel of the nonlocal term at these locations.

\section*{Conflict of interest}
 The authors declare that they have no conflict of interest.

\begin{acknowledgements}
P.M.L.  thanks the support of the Russian Ministry of Science and Higher
Education.
  The  work of P.M.L.  was   supported by the National
Science Foundation, grant DMS-1814619.  M.S. was supported by National Science Foundation grant DMS-1909407.
Simulations were performed  at the Texas Advanced
Computing Center using the Extreme Science and Engineering Discovery
Environment (XSEDE), supported by NSF Grant ACI-1053575.

\end{acknowledgements}

\appendix

 \section{Hilbert transform for transformed variable}
\label{sec:AppendixTransformationHilberttransform}

In this Appendix we derive the expression for 
the Hilbert transform in the auxiliary variable $q$ \e{transform} of Section \ref{sec:transform}.


%
The change of variable  \e{transform} in Eq. \e{HilbertHdef} together with \e{transform_jacobian} results in

\begin{align} \label{Hilbertchangetoq}
&{\mathcal H} f(x)=\frac{1}{\pi} \text{p.v.}
\int^{\infty}_{-\infty}\frac{f(x')}{x-x'}\D x'=\frac{1}{\pi} \text{p.v.}
\int^{\pi}_{-\pi}\frac{\tilde f(q')}{\tan{
\frac{q}{2}}-\tan{\frac{q'}{2}}}\frac{\D q'}{2\cos^2{\frac{q'}{2}}}\nonumber \\
&=\frac{1}{2\pi} \text{p.v.} \int^{\pi}_{-\pi}\frac{\tilde f(q')\left
[1+\tan{ \frac{q}{2}}\tan{ \frac{q'}{2}} -\tan{ \frac{q'}{2}}\left (\tan{
\frac{q}{2}}-\tan{\frac{q'}{2}} \right)\right ]}{\tan{
\frac{q}{2}}-\tan{\frac{q'}{2}}}\D q'\nonumber \\
&=\frac{1}{2\pi} \text{p.v.}
\int^{\pi}_{-\pi}\frac{\tilde f(q')}{\tan{\left (
\frac{q-q'}{2}\right )}}\D q'- \frac{1}{2\pi}\int\limits^{\pi}_{-\pi}{\tilde f}(q')\tan{\frac{q'}{2}}\D q'={\mathcal H}^{2\pi} f(q)+C^{2\pi}_f,
\end{align}
where we used the identities
\begin{align} 
\tan{(a-b)}=\frac{\tan{a}-\tan{b}}{1+\tan{a}\tan{b}} \quad \text{and}  \quad \frac{1}{\cos^2{\frac{q}{2}}} =\tan^2{\frac{q}{2}}+1 \nonumber
\end{align} as well as the definitions \e{HilbertHdef_periodic} and \e{C2pidef}. Eq. \e{Hilbertchangetoq} ensures that  $ \lim\limits_{q\to\pm \pi}[{\mathcal H}^{2\pi} f(q)+C^{2\pi}_f]=0.$

Also ${\mathcal H}^{2\pi} f(x)$, Eq. \e{HilbertHdef_periodic},  is the reduction of ${\mathcal H} f(x)$, Eq. \e{HilbertHdef}, to the class of $2\pi$-periodic functions.
Assuming that $f(x)$ is the periodic function with the period $2\pi$, we obtain from Eq. \e{HilbertHdef} that
\begin{equation} \label{Hilbertdefperiodic}
{\mathcal H} f(x)=\frac{1}{\pi} \sum \limits _{n=-\infty}^\infty \text{p.v.}
\int^{\pi}_{-\pi}\frac{f(x')}{x-x'+2\pi n}\D x'=\frac{1}{2\pi} \text{p.v.}
\int^{\pi}_{-\pi}\frac{f(x')}{\tan{\left (
\frac{x-x'}{2}\right )}}\D x'=:{\mathcal H}^{2\pi} f(x),
\end{equation}
where we used the definition \e{HilbertHdef_periodic} and the identity
\begin{align} \label{polessum}
\sum\limits^{\infty}_{n=-\infty}\frac{1}{x+2\pi n}=\frac{1}{2\tan{\frac{x}{2}}}.
\end{align}
%

%
 %
%






\end{document}